\documentclass[traditabstract]{aa}
\usepackage{longtable,lscape}
\usepackage{graphicx}
\DeclareGraphicsExtensions{.pdf,.png,.jpg}
\usepackage{epsfig}
\usepackage[authoryear]{natbib}

\bibpunct{(}{)}{;}{a}{}{,} 

\begin{document}

\newcommand{\Ha}{{H$\alpha$}}
\newcommand{\Hb}{{H$\beta$}}
\newcommand{\Hg}{{H$\gamma$}}
\newcommand{\Hd}{{H$\delta$}}
\newcommand{\Hep}{{H$\epsilon$}}

\newcommand{\pab}{{Pa$\beta$}}
\newcommand{\pag}{{Pa$\gamma$}}
\newcommand{\pad}{{Pa$\delta$}}
\newcommand{\brg}{{Br$\gamma$}}
\newcommand{\brd}{{Br$\delta$}}

\newcommand{\Ll}{{$L_{\rm line}$}}
\newcommand{\LHa}{{$L_{\rm H\alpha}$}}
\newcommand{\Lacc}{{$L_{\rm acc}$}}
\newcommand{\Macc}{{$\dot{M}_{\rm acc}$}}
\newcommand{\BJobs}{{BJ$_{\rm obs}$}}
\newcommand{\BJintr}{{BJ$_{\rm intr}$}}

\newcommand{\Msun}{{$M_{\odot}$}}
\newcommand{\Lsun}{{$L_{\odot}$}}
\newcommand{\Rsun}{{$R_{\odot}$}}

\newcommand{\Mstar}{{$M_{\star}$}}
\newcommand{\Lstar}{{$L_{\star}$}}
\newcommand{\Rstar}{{$R_{\star}$}}
 \newcommand{\Teff}{{$T_{\rm eff}$}}

\newcommand{\Mdisk}{{$M_{\rm disk}$}}
\newcommand{\Ldisk}{{$L_{\rm disk}$}}
\newcommand{\Rin}{{$R_{\rm in}$}}

\newcommand{\SO}{{$\sigma$-Ori}}

\title{X-Shooter spectroscopy of young stellar objects:}
\subtitle{IV -- Accretion in low-mass stars and sub-stellar objects in Lupus \thanks{Based on 
observations collected at the European Souther Observatory at
Paranal, under programs 084.C-0269(A), 085.C-0238(A), 086.C-0173(A), 087.C-0244(A) and 089.C-0143(A).}
\fnmsep\thanks{The on-line material Figures \ref{slab1} to \ref{correl6} and 
Tables \ref{optnirphot} to \ref{tab:fluxes_EWs_NaI}, are available only in electronic form.}
}

\author{
   J.M.~Alcal\'a\inst{1} 
  \and A.~Natta\inst{2,9} 
  \and C.F.~Manara\inst{7}
  \and L.~Spezzi\inst{7}
  \and B.~Stelzer\inst{3}
  \and A.~Frasca\inst{4}
  \and K.~Biazzo\inst{4,1}
  \and E.~Covino\inst{1}
  \and S.~Randich\inst{2}
  \and E.~Rigliaco\inst{8} 
  \and L.~Testi\inst{2,7,11}
  \and F.~Comer\'on\inst{7}
  \and G.~Cupani\inst{5}
  \and V.~D'Elia\inst{6,10}
}

\offprints{J.M. Alcal\'a}
\mail{alcala@oacn.inaf.it}

\institute{ 
      INAF-Osservatorio Astronomico di Capodimonte, via Moiariello 16, I-80131 Napoli, Italy
 \and INAF-Osservatorio Astrofisico di Arcetri, via Moiariello 16, Largo E. Fermi 5, I-50125 Firenze, Italy
 \and INAF-Osservatorio Astronomico di Palermo, Piazza del Parlamento 1, I-90134 Palermo, Italy
 \and INAF-Osservatorio Astrofisico di Catania, via S. Sofia 78, I-95123 Catania, Italy
 \and INAF- Osservatorio Astronomico di Trieste, via Tiepolo 11, 34143 Trieste, Italy
 \and INAF-Osservatorio Astronomico di Roma, Via Frascati 33, I-00040 Monteporzio Catone, Italy
 \and European Southern Observatory, Karl Schwarzschild Str. 2, 85748 Garching, Germany
 \and Department of Planetary Science, Lunar and Planetary Lab. University of Arizona, 1629, E. University Blvd, 85719, Tucson, AZ, USA
 \and DIAS/Dublin Institute for Advanced Studies, Burlington Road, Dublin 4, Ireland
 \and ASI-Science Data Centre, Via Galileo Galilei, I-00044 Frascati, Italy
 \and Excellence Cluster Universe, Boltsmannstr. 2, D-85748 Garching, Germany
}

\date{Received July, 7th 2013 ; accepted October, 3rd 2013}

\abstract{We present X-Shooter/VLT observations of a sample of 36 accreting low-mass stellar and 
sub-stellar objects (YSOs) in the Lupus star forming region, spanning a range in mass 
from $\sim$0.03 to $\sim$1.2\,\Msun, but mostly with 0.1\,\Msun~$<$~\Mstar~$<$~0.5\,\Msun. 
Our aim is twofold: firstly, analyse the relationship between excess-continuum and line emission 
accretion diagnostics, and, secondly, to investigate the accretion properties in terms of the physical 
properties of the central object. The accretion luminosity (\Lacc), and from it the accretion rate (\Macc), 
is derived by modelling the excess emission, from the UV to the near-IR, as the continuum emission 
of a slab of hydrogen. 
The flux and luminosity (\Ll) of a large number of emission lines of $\ion{H}{}$, $\ion{He}{}$, 
$\ion{Ca}{ii}$, etc., observed simultaneously in the range from $\sim$330\,nm to 2500\,nm, were 
computed. 
The luminosity of all the lines is well correlated with \Lacc. We provide empirical relationships
between \Lacc ~and the luminosity of 39 emission lines, which have a lower dispersion as compared 
to previous relationships in the literature. 
Our measurements extend the \pab ~ and \brg ~relationships to \Lacc ~values about two orders 
of magnitude lower than those reported in previous studies. We confirm that different methodologies 
to measure \Lacc ~and \Macc ~yield significantly different results: \Ha ~line profile modelling may 
underestimate \Macc ~by 0.6 to 0.8\,dex with respect to \Macc ~derived from continuum-excess measures. 
Such differences may explain the likely spurious bi-modal relationships between \Macc ~and 
other YSOs properties reported in the literature.
We derive \Macc ~in the range 2$\times$10$^{-12}$ -- 4$\times$10$^{-8}$\,\Msun~yr$^{-1}$ 
and conclude that \Macc~$\propto$~\Mstar$^{1.8(\pm0.2)}$, with a dispersion lower by a factor of 
about 2 than in previous studies. 
A number of properties indicate that the physical conditions of the accreting gas 
are similar over more than 5 orders of magnitude in \Macc, confirming previous 
suggestions that it is the geometry of the accretion flow that controls the rate at 
which the disc material accretes onto the central star.
}


\titlerunning{Accretion in Lupus YSOs}
\authorrunning{Alcal\'a et al.}
\maketitle

\section{Introduction} 
\label{Sec:intro}
The mass accretion rate is a key quantity for the studies of 
pre-main sequence (PMS) stellar evolution. The evolution of accretion 
discs around young low-mass (\Mstar$<$0.5\Msun) stellar and sub-stellar 
objects (YSOs) is regulated by the simultaneous effects of the 
mass accretion onto the star and the ejection of matter from 
the star-disc system \citep[][and references therein]{paatz96}. 
Understanding the evolution of accretion discs can provide 
strong constraints on theories of planet formation and evolution. 
The mass accretion rate, \Macc, sets an important constraint 
for the disc evolution models \citep{hartmann98} and disc clearing 
mechanisms \citep{alexander06, gorti09}.

All young objects, from intermediate mass stars to brown dwarfs, show 
at some stage of their evolution evidence of circumstellar discs and 
accretion.
The currently accepted paradigm that explains mass accretion in YSOs
is the magnetospheric accretion model \citep[][]{uchida85, konigl91, shu94}. 
The model assumes that matter from the inner edge of the accretion 
disc is channelled along the magnetic field lines onto the star.
The gas shocking at the stellar surface produces high temperatures ($\sim$10$^4$\,K), 
giving rise to continuum and line emission in the accretion flows. Balmer continuum 
radiation is mostly emitted in the optically thin pre-shock gas, while Paschen 
continuum emission is generally produced in optically thick post-shock gas.  
Emission in the Balmer lines as well as in other lines of $\ion{Ca}{ii}$ and 
$\ion{He}{i}$ originates mainly in the accretion columns 
\citep{hartmann94, calvetgulbring98, muzerolle01, kurosawa06}.
The accretion luminosity, \Lacc, which measures the radiative losses, 
integrated over the whole spectrum, can be converted into mass accretion rate, \Macc, 
if the YSOs radius and mass are known \citep{gullbring98, hart98}. 
 
In the past, \Lacc ~has been calculated using veiling measurements 
in high-resolution spectra 
\citep[e.g.][and references therein]{hartigan91, hartigan03, white04}, 
or by modelling the Balmer and Paschen continuum excess emission 
with a plane-parallel slab of hydrogen in local thermodynamic 
equilibrium (LTE) \citep[e.g.][]{valenti93, gullbring98, HH08, 
rigliaco12}. More recently, the continuum excess emission has been 
measured by incorporating in the models multiple accretion components
with a large range of energy fluxes \citep{ingleby13}. 

First direct measurements of UV excess emission in very low-mass 
stars and sub-stellar objects were done by \citet{HH08, herczeg09} 
and \citet{rigliaco12}. Prior to theses studies, the difficulty
of detecting the Balmer continuum emission in brown dwarfs 
called for the necessity of developing magnetospheric accretion 
models to reproduce the \Ha ~line profile \citep{muzerolle01, 
muzerolle03, muzerolle05}, more easily measured in these objects. 
These studies additionally showed that the size of the accretion 
flows and the mass accretion rate are correlated, suggesting that 
it is the geometry of the accretion flows that controls the rate 
at which the disc material accretes onto the central star 
\citep{muzerolle01}.

It has been shown that \Lacc, hence \Macc ~is well correlated with the 
line luminosity, \Ll, of several emission lines of the Balmer series, 
$\ion{He}{i}$ and $\ion{Ca}{ii}$ lines \citep[e.g][and references therein]{HH08, rigliaco12},
as well as with hydrogen recombination lines in the near infrared 
\citep{muzerolle98, calvet04}. This underlines the importance of these 
emission features as accretion diagnostics.  The \Macc -- \Ll ~relationships
were then extended down to the sub-stellar regime \citep{natta04, mohanty05, HH08}, 
by combining measures of \Macc ~ using different methodologies for different 
mass regimes (spectral veiling determination and modelling of the \Ha ~line profile 
for the stars with \Mstar ~greater and lower than 0.3\Msun, respectively).
A two-mode relation, with respect to YSO mass, between \Macc ~and 
the surface flux of the $\ion{Ca}{ii}$ infrared triplet (IRT) line 
at ~866.2\,nm was suggested \citep{mohanty05}, but the origin of 
the bi-modality was unclear. A relationship between \Macc ~and 
the width of the \Ha ~line at 10\% of the line peak was also established, 
though with a large scatter \citep{natta04}. 

Researchers have been using emission lines as proxies to measure 
\Lacc ~\citep[][and references therein]{muzerolle98, 
muzerolle00, muzerolle01, muzerolle05, natta04, mohanty05, gatti06, 
gatti08, antoniucci11, biazzo12}. When applied to large samples 
of YSOs, the \Lacc -- \Ll ~relations provided a correlation between 
accretion rate and mass of the central object, i.e. 
\Macc~$\propto$~\Mstar$^\alpha$ ~with $\alpha \approx 2$  
\citep{muzerolle05, natta06}, but with a  scatter of more 
than 2\,dex in $\log{}$\Macc, at a given YSO mass.
Such a large scatter is not easily explained in terms of YSO variability
\citep{costigan12, costigan13} or different methodological approach used to 
derive \Lacc ~\citep{HH08}.
 
The main caveat of the \Lacc -- \Ll ~relations is the non-simultaneity 
of photometry and spectral line measurements, complicated by the fact 
that emission lines can trace strong winds as well as mass accretion. 
Additionally, emission lines are produced not only by accretion, but also 
in other processes such as chromospheric activity. This adds more 
uncertainty on the \Lacc ~determinations from lines. The uncertainty 
depends on the level of the accretion rate and on the stellar properties 
\citep[see][]{manara13}. 

Deriving precise \Lacc -- \Ll ~relations for accretion requires well-calibrated 
observations of both the continuum and line fluxes over a wide range of wavelengths, 
and over the widest possible range of YSO physical parameters, i.e., \Lacc, \Macc, 
and \Mstar. 
Furthermore, an accurate knowledge of both the photospheric and chromospheric 
spectrum of the YSO, which need to be subtracted from the observed flux to 
isolate the accretion emission, is required. On the other hand, a large number 
of emission line diagnostics over a wide wavelength  need to be studied 
simultaneously to avoid problems related to YSO variability, in order to 
probe gas at different excitation conditions.
These requirements can be fulfilled with the use of broad spectral 
range spectrographs such as X-Shooter \citep{vernet11} at the VLT. 
X-shooter is an ideal instrument for this purpose, as its 
wide wavelength range (300-2500 nm), covered in a single shot, 
allows comparing many different accretion diagnostics simultaneously.

In this paper we present the analysis and results of X-Shooter observations 
of a sample of 36 accreting YSOs in the Lupus star forming region, spanning 
a range in mass from $\sim$0.03 to 1.2\,\Msun, but the majority with \Mstar~$<$~0.5\Msun. 
We use the continuum UV-excess emission as a measure of \Lacc, hence of 
\Macc,  and provide revised relationships between \Lacc ~and the luminosity, 
\Ll, of an unprecedentedly large number of emission lines. The accretion 
properties of the sample are also analysed. 

The sample and observations are presented in Section~\ref{Sec:spectra},
while in Section~\ref{params} the sample characterisation is provided. 
The determinations of \Lacc, obtained by fitting the Balmer and Paschen
continua with a model of a slab of hydrogen, and derivation of \Macc, 
as well as the luminosity of a number of permitted emission lines  
are presented in Section~\ref{accretiondiagnostics}. 
The relationships between \Lacc ~and \Ll ~are derived in Section~\ref{correlations}, 
while the accretion properties of the sample are examined in Section~\ref{accprop}. 
The results are then discussed in Sections~\ref{discussion} and our main conclusions 
are summarised in Section~\ref{summary}.

\vspace{1cm}

\section{Sample, observations and data reduction} 
\label{Sec:spectra}

As part of the consortium that built X-Shooter, INAF was granted 46 nights 
of guaranteed time observations (GTO) at the VLT-UT2 (Kueyen) spread over a period 
of three years, starting on ESO period 84. Eight of these nights were devoted to young 
stellar objects in nearby star forming regions ($\sim$3, 3 and 2, nights for the 
$\sigma$-Ori cluster, the Lupus clouds and the TW~Hya association, respectively). 
The data used in this paper were acquired within the context of the X-Shooter 
INAF/GTO \citep{alcala11} in  April 06/07, 2010, April 23, 2011 and April 18, 2012. 

\subsection{The sample}
We selected YSOs in the Lupus star forming region.
The Lupus cloud complex is one of the low-mass star forming regions 
located closest (d$<$200\,pc) to the Sun \citep[see][for a review]{comeron08}. 
Similarly to other regions (e.g., Taurus, Chamaeleon, and $\rho$-Oph), 
a large variety of objects in various stages of evolution are present 
in Lupus. Several sub-stellar objects have also been discovered in the 
region \citep{comeron08,lopezmarti05}. 

Our sample comprises 36 YSOs mainly in the Lupus~I and III clouds 
satisfying as close as possible the following criteria: 
i) YSOs with infrared Class-II characteristics 
and low extinction to take full advantage of the broad X-Shooter 
spectral range, from UV to near-IR; ii) targets well surveyed in 
as many photometric bands as possible: mainly Spitzer (IRAC \& MIPS) 
surveys and complementary Wide-field Infrared Survey Explorer \citep[WISE;][]{wright10}
data, as well as optical photometry available; iii) mostly very 
low-mass (VLM, \Mstar~ $<$ 0.3\,\Msun) objects, but also a number 
of more massive (\Mstar~ $<$ 1\,\Msun) stars, in order to explore a wider 
range of accretion luminosity. 

The two main bibliographic sources from which our sample was compiled
are \citet{allen07} and \citet{merin08}. The former reports several new VLM 
young stellar and sub-stellar objects with Spitzer colors, while the latter 
provides a well characterised sample in terms of spectral energy 
distribution (SED) and SED spectral index, based on the Spitzer 
c2d criteria \citep{evans09}. 
Additional Class-II  YSOs in Lupus-III that extend the sample to a broader 
mass range, and eventually to higher accretion luminosity, were selected 
from \citet{hughes94}, following as close as possible criteria i) and ii) 
above, although several of these targets do not possess Spitzer fluxes 
because they were not covered by the c2d or other Spitzer surveys. 
Thus, the available WISE data were used.

Among the selected objects, there are two visual binaries, namely Sz\,88 
and Sz\,123, where both the components were observed by us. In other 8 
of our YSOs the Spitzer images revealed objects with separation between
2\farcs0 and  10\farcs0 (see Table\ref{targets}), among these 6 at separations 
larger than 4\farcs0 (see Table~1 in Mer\'{i}n et al. 2008 and Table~9 in
Comer\'{o}n 2008). The spatial resolution of X-Shooter is 
sufficiently high, allowing the observation of all our targets without light 
pollution from any of those nearby objects.
To our knowledge, none of our targets has been reported as a spectroscopic
binary in previous investigations using high-resolution spectroscopy \citep[e.g.][]{melo03, guenther07}.
Our sample also includes Par-Lup3-4, one of the lowest mass young stellar 
objects in Lupus known to host an outflow \citep{comeron03}.  Although 
separate X-Shooter papers are being devoted to studying  outflows  
\citep{bacciotti11, whelan13, giannini13}, we include  Par-Lup3-4 here to 
investigate its accretion properties in the same way as the other targets.

It must be stressed that, although the sample covers rather well 
the mass range between $\sim$0.05\,\Msun~ and $\sim$1\,\Msun ~(40\% with 
\Mstar~$<$~0.2\,\Msun, 35\% with \Mstar ~in the range 0.2-0.5\,\Msun, 
and 25\% with \Mstar~$>$0.5\Msun), it is incomplete at each mass 
bin. However, our sample represents about 50\% of the total 
population of Class-II YSOs in the Lupus~I and Lupus~III clouds
\citep[see Table~6 by][for the statistics of the different YSO Classes]{merin08}.

Table~\ref{targets} provides the list of the targets. Ancillary photometric 
data, in the optical, near- and mid-IR, were collected from \citet{allen07}, 
\citet{merin08}, the WISE All-Sky Source Catalog  and from an 
unpublished catalog by \citet{comeron09}. Such data are reported in the 
on-line Tables~\ref{optnirphot} and \ref{mirphot} (published in electronic 
form only). Although not simultaneous with the X-Shooter spectroscopy, 
these are the only photometric data available that can be used for 
comparisons with the spectroscopic fluxes.

\subsection{X-Shooter spectroscopy}
With its three spectrograph arms, X-Shooter provides simultaneous 
wavelength coverage from $\sim$300 to $\sim$2480 nm. For most of the targets 
slits of 1.0/0.9/0.9~arc-sec were used in the UVB/VIS/NIR arms, 
respectively, yielding resolving powers of 5100/8800/5600. 
Only the two brightest objects in the sample (Sz\,74 and Sz\,83) 
were observed through the 0.5/0.4/0.4 arc-sec slits in the UVB/VIS/NIR 
arms, respectively, yielding resolving powers  of 9100/17400/10500. 
Table~\ref{targets} includes the log of observations. 
Most of the targets were observed in one cycle using the A-B nodding mode, 
while three (Lup\,706, Par-Lup3-4, and 2MASS~J16085953-3856275) were
observed in two cycles using the A-B-B-A nodding mode. 
Several Class-III YSOs, indistinctly quoted here as Class-III YSOs 
or Class-III templates, were also observed throughout the various 
Italian-GTO star formation runs, and their properties were published in  
separate papers  \citep{manara13, stelzer13b}. The spectra of such stars 
are used as templates for the modelling of Balmer and Paschen continua 
(Section~\ref{bjmodelling}) and to estimate extinction and revisit the 
spectral types (Section~\ref{params}) of the Class-II YSOs.
The Class-III templates cover rather well the spectral range from K5 to M9 
\citep[see][]{manara13}. 

Except for a few minutes of thin cirrus when the seeing was
$\sim$1.3\,arc-sec (only for Sz\,130) and some clouds at the
end of the night on April 06, 2010 
\citep[but only for the Class-III YSOs Par-Lup3-1 and Par-Lup3-2, see][]{manara13}, 
the weather conditions were mostly photometric with sub-arcsecond 
seeing.

With a few exceptions explained next, all the  targets were observed 
at parallactic angle. The components of Sz\,88 and Sz\,123, with 
separation 1\farcs49 and 1\farcs70, respectively, were observed simultaneously 
by aligning the 11\,arc-sec slit at their position angle. 
Likewise, in the case of Par-Lup3-4 the slit was aligned at the 
position angle of the outflow \citep[][]{bacciotti11, whelan13}. 

\setlength{\tabcolsep}{3pt}
\begin{table*}
\caption[ ]{\label{targets} Selected YSOs and observing log.}
\begin{tabular}{l|cc|cc|ccc|ccc|c}
\hline \hline
 Object/Other name &  RA(2000)  & DEC(2000)  & Obs. Date &  MJD       & \multicolumn{3}{c}{ \underline {$T_{\rm exp}$ (sec)}}  & \multicolumn{3}{c}{\underline{$S/N$ ratio}} $\ddag$ &  Lupus  \\ 
      &  h \, :m \, :s & $^\circ$ \, ' \, ''   & YY-MM-DD  & (+2400000) &   UVB & VIS & NIR                                   &      UVB & VIS & NIR                     &  cloud \\ 	       
                   &            &            &	        &	      &	      &     &                                       &          &     &                         &         \\ 	       
\hline
  Sz66$^\dag$               	    & 15:39:28.28 & $-$34:46:18.0  & 2012-04-18 &  56035.2516 &  2x300 & 2x250  & 2x100 & 40 & 140 & 55 & I   \\ 
  AKC2006-19            	    & 15:44:57.90 & $-$34:23:39.5  & 2011-04-23 &  55674.1966 &  2x900 & 2x900  & 2x900 &  7 &  50 & 40 & I   \\ 
  Sz69            /    HW Lup$^\dag$ & 15:45:17.42 & $-$34:18:28.5  & 2011-04-23 &  55674.0991 &  2x300 & 2x300  & 2x300 & 12 &  70 & 60 & I   \\ 
  Sz71            /      GW Lup     & 15:46:44.73 & $-$34:30:35.5  & 2012-04-18 &  56035.0949 &  2x300 & 2x250  & 2x100 & 40 & 110 & 52 & I   \\ 
  Sz72            /      HM Lup     & 15:47:50.63 & $-$35:28:35.4  & 2012-04-18 &  56035.1926 &  2x300 & 2x250  & 2x100 & 33 &  95 & 39 & I   \\ 
  Sz73                  	    & 15:47:56.94 & $-$35:14:34.8  & 2012-04-18 &  56035.2923 &  2x300 & 2x250  & 2x100 & 20 & 115 & 55 & I   \\ 
  Sz74            /      HN Lup     & 15:48:05.23 & $-$35:15:52.8  & 2012-04-18 &  56035.1695 &  2x150 & 2x100  & 2x50  & 22 & 135 & 45 & I   \\ 
  Sz83            /      RU~Lup     & 15:56:42.31 & $-$37:49:15.5  & 2012-04-18 &  56035.1794 &  2x100 & 2x50	& 2x30  & 45 & 130 & 50 & I   \\ 
  Sz84                  	    & 15:58:02.53 & $-$37:36:02.7  & 2012-04-18 &  56035.1083 &  2x350 & 2x300  & 2x115 & 40 & 120 & 80 & I   \\ 
  Sz130                 	    & 16:00:31.05 & $-$41:43:37.2  & 2010-04-07 &  55293.3609 &  2x300 & 2x300  & 2x300 & 40 &  90 & 45 & IV  \\ 
  Sz88A (SW)      /      HO Lup (SW)& 16:07:00.54 & $-$39:02:19.3  & 2012-04-18 &  56035.3342 &  2x300 & 2x250  & 2x100 & 60 & 110 & 50 & I   \\ 
  Sz88B (NE)      /      HO Lup (NE)& 16:07:00.62 & $-$39:02:18.1  & 2012-04-18 &  56035.3342 &  2x300 & 2x250  & 2x100 & 18 &  75 & 48 & III \\ 
  Sz91                  	    & 16:07:11.61 & $-$39:03:47.1  & 2012-04-18 &  56035.3213 &  2x300 & 2x250  & 2x300 & 40 &  70 & 35 & III \\ 
  Lup713$^\dag$               	    & 16:07:37.72 & $-$39:21:38.8  & 2010-04-06 &  55292.1950 &  2x900 & 2x900  & 2x900 &  8 &  50 & 45 & III \\ 
  Lup604s               	    & 16:08:00.20 & $-$39:02:59.7  & 2010-04-06 &  55292.2374 &  2x450 & 2x450  & 2x450 &  5 &  60 & 40 & III \\ 
  Sz97                  	    & 16:08:21.79 & $-$39:04:21.5  & 2011-04-23 &  55674.1139 &  2x300 & 2x300  & 2x300 & 15 & 120 & 55 & III \\ 
  Sz99                  	    & 16:08:24.04 & $-$39:05:49.4  & 2010-04-07 &  55293.3452 &  2x300 & 2x300  & 2x300 & 14 &  60 & 35 & III \\ 
  Sz100$^\dag$                 	    & 16:08:25.76 & $-$39:06:01.1  & 2011-04-23 &  55674.1391 &  2x300 & 2x300  & 2x300 & 10 &  50 & 55 & III \\ 
  Sz103                 	    & 16:08:30.26 & $-$39:06:11.1  & 2011-04-23 &  55674.1542 &  2x300 & 2x300  & 2x300 & 15 &  80 & 50 & III \\ 
  Sz104                 	    & 16:08:30.81 & $-$39:05:48.8  & 2010-04-06 &  55292.2549 &  2x300 & 2x300  & 2x300 & 10 &  70 & 50 & III \\ 
  Lup706                	    & 16:08:37.30 & $-$39:23:10.8  & 2010-04-06 &  55292.2796 &  4x900 & 4x900  & 4x900 &  6 &  20 & 30 & III \\ 
  Sz106                 	    & 16:08:39.76 & $-$39:06:25.3  & 2012-04-18 &  56035.3051 &  2x450 & 2x400  & 2x450 & 36 &  85 & 50 & III \\ 
  Par-Lup3-3            	    & 16:08:49.40 & $-$39:05:39.3  & 2010-04-06 &  55292.3586 &  2x300 & 2x300  & 2x300 &  5 &  50 & 30 & III \\ 
  Par-Lup3-4$^\dag$            	    & 16:08:51.43 & $-$39:05:30.4  & 2010-04-07 &  55293.1763 &  4x900 & 4x900  & 4x900 &  7 &  30 & 45 & III \\ 
  Sz110           /      V1193 Sco  & 16:08:51.57 & $-$39:03:17.7  & 2011-04-23 &  55674.3413 &  2x300 & 2x300  & 2x300 & 35 & 103 & 55 & III \\ 
  Sz111           /      Hen 3-1145 & 16:08:54.69 & $-$39:37:43.1  & 2012-04-18 &  56035.1229 &  2x300 & 2x250  & 2x100 & 50 &  90 & 45 & III \\ 
  Sz112                 	    & 16:08:55.52 & $-$39:02:33.9  & 2012-04-18 &  56035.2644 &  2x350 & 2x300  & 2x350 & 20 &  40 & 35 & III \\ 
  Sz113                 	    & 16:08:57.80 & $-$39:02:22.7  & 2011-04-23 &  55674.3566 &  2x900 & 2x900  & 2x900 & 15 &  70 & 38 & III \\ 
2MASS J16085953-3856275$^\dag$ 	    & 16:08:59.53 & $-$38:56:27.6  & 2011-04-23 &  55674.2808 &  4x900 & 4x900  & 4x900 &  5 &  18 & 20 & III \\ 
SSTc2d160901.4-392512   	    & 16:09:01.40 & $-$39:25:11.9  & 2011-04-23 &  55674.2260 &  2x450 & 2x450  & 2x450 & 20 &  80 & 45 & III \\ 
  Sz114           /      V908 Sco   & 16:09:01.84 & $-$39:05:12.5  & 2011-04-23 &  55674.3850 &  2x300 & 2x300  & 2x300 & 30 &  90 & 30 & III \\ 
  Sz115                 	    & 16:09:06.21 & $-$39:08:51.8  & 2012-04-18 &  56035.3469 &  2x350 & 2x300  & 2x350 & 20 &  70 & 45 & III \\ 
  Lup818s$^\dag$               	    & 16:09:56.29 & $-$38:59:51.7  & 2011-04-23 &  55674.2442 &  2x900 & 2x900  & 2x900 &  5 &  25 & 33 & III \\ 
  Sz123A (S)            	    & 16:10:51.34 & $-$38:53:14.6  & 2012-04-18 &  56035.2784 &  2x700 & 2x600  & 2x350 & 25 &  60 & 45 & III \\ 
  Sz123B (N)            	    & 16:10:51.31 & $-$38:53:12.8  & 2012-04-18 &  56035.2784 &  2x700 & 2x600  & 2x350 & 25 &  70 & 35 & III \\ 
  SST-Lup3-1$^\dag$            	    & 16:11:59.81 & $-$38:23:38.5  & 2010-04-06 &  55292.3781 &  2x450 & 2x450  & 2x450 &  6 &  35 & 40 & III \\ 

\hline

\end{tabular}
\tablefoot{ $\dag$ : nearby (2\farcs0 $<$ d $<$ 10\farcs0) object detected in the Spitzer images \citep[see][]{merin08, comeron08}.  \\
          $\ddag$ : the S/N ratio refers to the central wavelength of each spectrograph arm.
}

\end{table*}


Several telluric standard stars were observed with the same
instrumental set-up and at very similar airmass as the targets. 
Normally two flux standards per night were observed through 
a 5\,arcsec slit for flux calibration purposes.

\subsection{Data reduction}
\label{datared}
The data reduction was done independently for each spectrograph arm 
using two versions of the X-Shooter pipeline \citep{modigliani10}, 
depending on the period in which the data were acquired: 
version 1.0.0 was used for the April 2010 data, while for 
the data acquired in April 2011 and April 2012 version 1.3.7 was 
used. The standard steps of processing included bias or dark 
subtraction, flat fielding, optimal extraction, wavelength calibration, 
and sky subtraction. Since version 1.0.0 did not include flux calibration,
a MIDAS\footnote{The Munich Image Data Analysis System (MIDAS) 
provides general tools for image processing and data reduction. 
It is developed and mantained by the ESO.} routine \citep[see details in][]{alcala11}
was used for the April 2010 data. Wavelength shifts due to instrumental 
flexure were corrected using the {\em flexcomp} package within the 
pipeline. The precision in wavelength calibration is better than 0.01\,pix
in the UVB and VIS arms, corresponding to 0.002\,nm, but errors can be 
as large as $\sim$0.006\,nm in the NIR arm.
The flux calibration for the April 2011 and April 2012 data was done using 
the pipeline. A test data-set was processed with both the MIDAS procedure 
and the pipeline, confirming that the flux-calibrated spectra resulting
from both procedures are identical.
The final extracted one-dimensional flux-calibrated spectra, 
from the MIDAS procedure or the pipeline, are not corrected for telluric 
absorption bands. The telluric correction was performed independently 
in the VIS and NIR spectra as explained in Appendix~\ref{appendix1}. 
The X-Shooter scale of $\sim$0.16 arc-sec/pix along the slit direction
allowed to resolve the components of the binaries, making possible the 
extraction of the spectra of the individual components, without any 
light contamination. 

By comparing the response function of different flux standards observed 
during the same night, we estimate an intrinsic precision on the flux 
calibration of $\sim$5\%. This does not consider, however, errors due to 
flux losses induced by seeing variations during the science exposures. Such 
errors are expected to be small (less than 10\%), as the seeing was always 
sub-arcsecond during the science exposures. In the 
case of Sz\,74 and Sz\,83, both observed with the narrow slits, the seeing 
was $\sim$0.5\,arc-sec. Therefore, $\sim$10\% can be considered as a robust 
relative error for the flux calibration.

The overlap of $\sim$40nm between the UVB-VIS and VIS-NIR spectra
allows us to verify the consistency in flux between the different arms. 
While the flux match between the UVB and VIS spectra is always perfect, 
the NIR flux in some cases is low with respect to 
the VIS flux by a factor $<$ 1.2\footnote{Some flux losses in the NIR
may result due to slit vignetting caused by a slight misalignment of the 
NIR slit with respect to the VIS and UVB arms.}. 
Thus, an additional correction was applied to the NIR spectrum to match 
the flux in the VIS arm.

Finally, the ancillary photometric data were used to compare the 
spectroscopic fluxes with the photometric ones.
The spectra follow the corresponding SED shape very well, with most of 
them matching the photometric fluxes within a factor of less than 1.5, 
whereas in some cases of objects well known to be strongly variable 
(c.f. Par-Lup3-4, Sz\,74, Sz\,83, Sz\,106, Sz\,113) the ratio may 
be as large as 2.5. In these latter cases the difference between 
photometric and spectroscopic fluxes can be ascribed to variability 
of the object.

\section{Stellar and sub-stellar properties}
\label{params}

Estimates of the physical parameters exist in the literature for some of
the YSOs in our sample, but they are unknown for the majority of our 
targets.
The quality and performance of the X-Shooter spectra allow us to derive 
many physical quantities \citep[c.f.][]{manara13b}. In this paper, however, 
we focus only on those parameters strictly needed to study the accretion 
properties of our sample.
Examples of full exploitation of the X-Shooter spectra for brown dwarfs 
and YSOs can be found in \citet{alcala11}, \citet{stelzer12}, \citet{manara13} 
and \citet{stelzer13, stelzer13b}. 

\subsection{Spectral type}
\label{spt}

For the M-type stars various spectral indices were calculated following 
\citet{riddick07} for optical wavelengths, and the H2O-K2 index from 
\citet{rojasayala12} for the near-IR spectra. The \citet{riddick07} 
spectral indices in the VIS are almost extinction independent\footnote{Note 
that the spectral indices may be affected by high extinction ($A_{\rm V}~>$5\,mag). 
Worth to say that none of our targets has such high $A_{\rm V}$.}. 
The spectral type assigned to a given 
object was estimated as the average spectral type resulting from 
the various indices in the VIS. From the dispersion over the average
of all the available spectral indices we obtain an uncertainty of half 
a spectral subclass.

The NIR indices provide spectral types which are consistent with the 
VIS results typically within one spectral sub-class. 
Thus, the spectral types derived from the VIS are adopted for the analysis, 
consistently with the spectral type assignment for the Class-III templates 
\citep[][]{manara13}. The spectral types are listed in Table~\ref{pars}.
For both the earliest-type stars in our sample (Sz73 and Sz83), an accurate 
spectral type of K7 is reported in the literature \citep[see][]{HH08, comeron08}. 
We revisit, however, the determination using the procedures described in 
Section~\ref{extinction}.

\begin{figure}[h]
\resizebox{1.0\hsize}{!}{\includegraphics[]{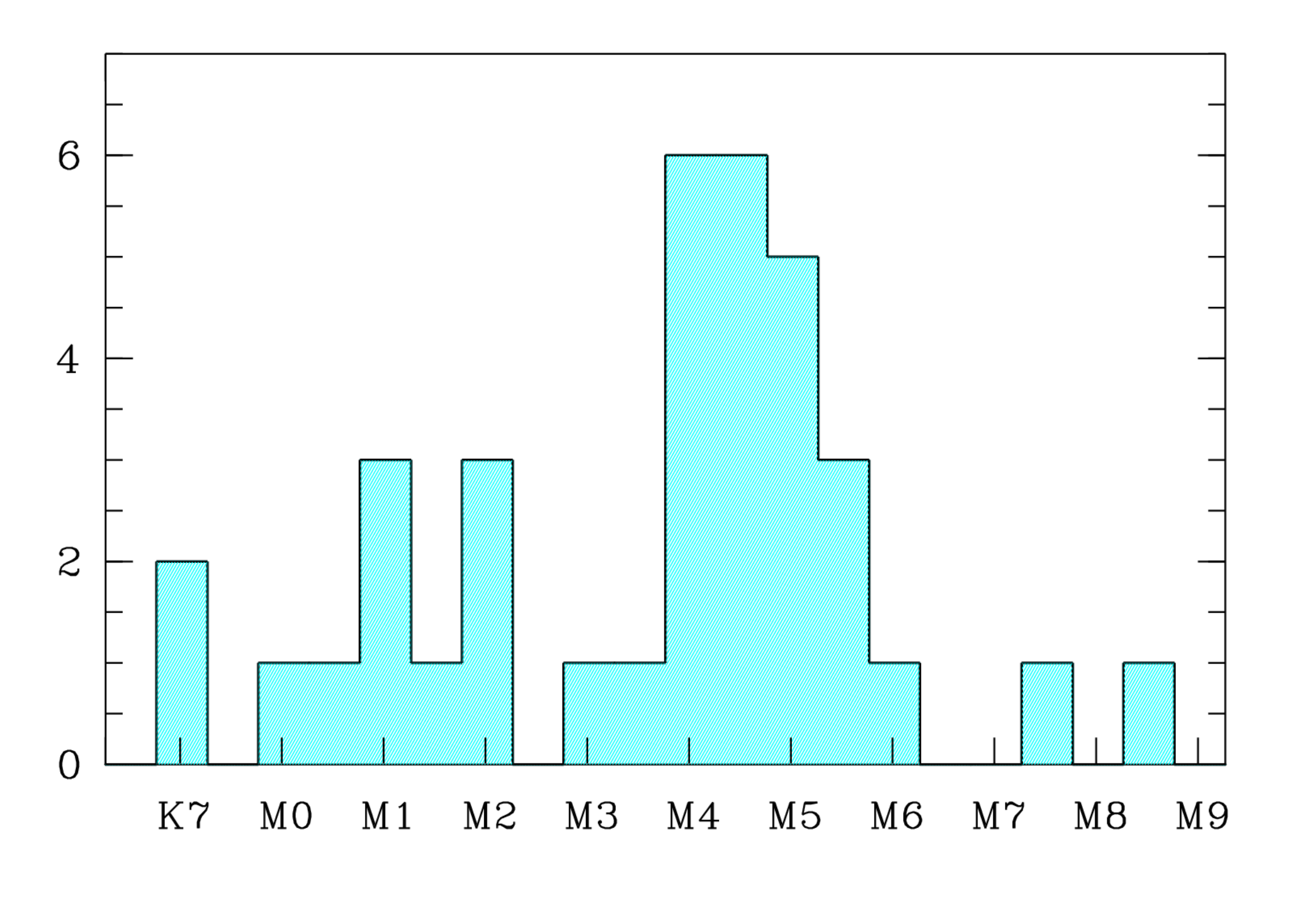}}
\caption{Distribution of spectral types of the YSO sample.
    \label{histspt}}
\end{figure}

Some difference in spectral type determinations in the literature can be 
ascribed to the different spectral ranges used in the different investigations
\citep{comeron03, hughes94, mortier11}. 
With the wide spectral range of X-Shooter we can overcome this problem. 
In Appendix~\ref{appendix2} we investigate the impact of veiling on the 
spectral indices, and conclude that for all  M-type YSOs veiling is 
estimated to influence spectral type classification to less than 0.3 
spectral subclasses, i.e. well within the uncertainty of spectral type
determinations using spectral indices.
Generally, our spectral types are consistent within $\pm$0.5 sub-class
with those in the literature, with a few exceptions
that are discussed at the end of the next section. 
The range of spectral types of our sample is from K7 to M8, with an 
over-abundance of M4-M5 objects (see Figure~\ref{histspt}).

\subsection{Extinction}
\label{extinction}
To derive the extinction, $A_{\rm V}$, for a given Class-II YSO, 
its VIS spectrum was compared with the Class-III templates best 
matching the Class-II spectral type. 
All our Class-III templates have low extinction ($A_{\rm V} <$ 0.5) 
\citep[see][]{manara13}. The templates were then artificially reddened 
by $A_{\rm V}=$0...4.0\,mag, in steps of 0.25\,mag, until the best 
match to the Class-II YSO was found. To redden the spectra we
used the extinction law by \citet{weingartner01}, which  covers a
wide range in wavelength, from the UV to the mid-IR, and has also 
been adopted for the c2d investigations \citep[see][]{evans09}. 
The procedure simultaneously provided a further test for the correct 
assignment of the template to derive the accretion luminosity 
(see Section~\ref{bjmodelling}). 
The $A_{\rm V}$ values derived in this way are listed in Table~\ref{pars}. 
We confirm that the majority of the targets possess zero extinction, 
as they were selected with this criterion. The highest values, 
2.2\,mag and 3.5\,mag, are found for Par-Lup3-3 and Sz73, 
respectively.

The main sources of uncertainty on $A_{\rm V}$ are the errors 
in spectral type when associating a template to a given YSO
and the error in the extinction of the template. 
The combined effect leads to an error of $<$0.5\,mag. However, 
another important source of uncertainty  may be introduced by strong 
veiling, which makes the YSOs spectra intrinsically bluer than the 
templates. In Appendix~\ref{appendix2} we investigate this effect 
using the object with  the strongest veiling among the M-type YSOs 
in our sample, i.e. Sz\,113 (see Section~\ref{bjmodelling}) and conclude 
that in this case we may underestimate $A_{\rm V}$ by less than 
$\sim$0.5\,mag. 
Thus, the extinction as derived above is not severely affected by 
the veiling in our sample, but in order to minimize the impact of 
veiling the extinction was derived only from the red portion of the 
spectra starting at 600\,nm.
For earlier type stars ($<$K7) with much higher levels of veiling 
than those in our sample, other methods to derive extinction 
must be used \citep[see for instance][]{manara13b}.

To check the self-consistency of the extinction derived in another spectral 
range we repeated the same procedure on the NIR spectra. The result is 
that the $A_{\rm V}$ values are consistent within the 0.5\,mag uncertainty, 
but affected by a larger error ($\sim$0.75\,mag). 
The latter is mainly  due to the higher uncertainty in the spectral type 
provided by the spectral indices in the NIR than in the VIS.
Thus, the values derived from the VIS are adopted for the following analysis.
Another obvious reason for preferring the extinction in the VIS is that 
the extinction in the NIR is small and one needs to multiply it (and its 
uncertainty) by a large factor to derive $A_{\rm V}$.

Our spectral type and extinction determinations are in good agreement with the 
literature values, except in a few cases. For Sz\,69, \citet{hughes94} 
give a spectral type M1 with a visual extinction of 3.20\,mag, while 
in our case the M4 template with zero extinction fits much better the 
entire X-Shooter spectrum. For Sz\,110, \citet{hughes94} give M2, while 
\citet{mortier11} claim M3, more consistent with our M4 determination. 
In the case of Sz\,113, the M4 spectral type reported by \citet{hughes94} 
agrees with our M4.5 determination, while \citet{mortier11} 
report M1 and \citet{comeron03} M6. 
The visual extinction values in the literature for Par-Lup3-4 range 
from 2.4 to 5.6\,mag \citep{comeron03}. The confirmed under-luminosity 
and edge-on geometry of this object \citep{comeron03, huelamo10} suggest 
that our zero extinction can be interpreted as wavelength-independent, 
i.e. gray extinction, rather than as a null extinction \citep{whelan13}. 
Interestingly, a null extinction is consistent with the value 
derived off-source using the [Fe\,{\sc ii}] lines at 1.27, 1.32, 
1.64\,$\mu$m \citep{bacciotti11, giannini13}, that trace jet emission 
\citep{nisini05}.

On the other hand, it is worth mentioning that the zero extinction we 
find for Sz\,83 agrees with the value derived by 
\citet{herczeg05} from the profile of the Ly$\alpha$ line. 
Our procedure also confirms the spectral type of this YSO, despite 
its strong veiling (see Section~\ref{bjmodelling}). Also, our $A_{\rm V}$ 
determination for Sz\,113, the most veiled among the M-type YSOs in 
our sample, agrees with the one by \citet{hughes94}.

\begin{figure}[h]
\resizebox{1.0\hsize}{!}{\includegraphics[]{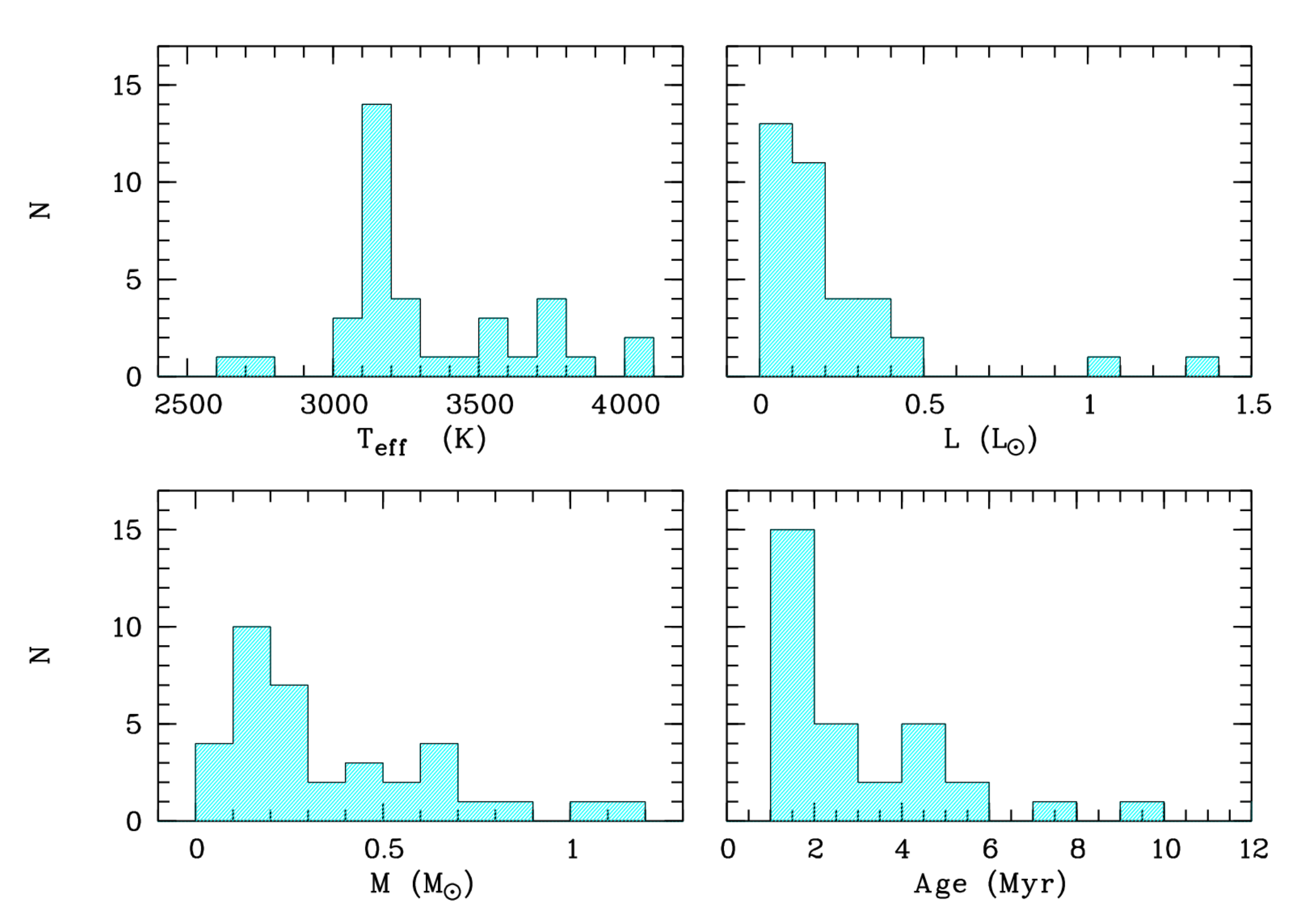}}
\caption{Histograms of YSOs properties. The YSOs in the sample are cooler 
      than 4000\,K with 70\% cooler than 3500\,K. The vast majority 
      have a luminosity of less than 0.5\,\Lsun, while more 
      than 70\% have a mass lower than 0.5\,\Msun. Two YSOs (Lup\,713 and 
      Lup\,818s) have a mass just at the hydrogen burning limit, while other 
      two (Lup\,706 and SST\,c2d160901.4-392512) are definitely sub-stellar.
    \label{histprop}}
\end{figure}

\subsection{YSOs physical parameters}
\label{physpar}

The effective temperature, $T_{\rm eff}$, was derived using the temperature 
scales given by \citet{Ken95} for the two K-type stars, and by \citet{luhman03} 
for the M-type YSOs. The latter scale is intermediate between the dwarf and 
giant temperature scales, and more appropriate for young objects than
temperature scales derived for field M dwarfs \citep[e.g.][]{testi09, rajpurohit13}.
The stellar luminosity and radius were computed using the methods 
described in \citet{spezzi08}, adopting the extinction and distance values
given in Table~\ref{pars}. The stellar radius has also been determined using 
the flux-calibrated X-Shooter spectra as explained in \citet{alcala11}. 
The good agreement between the radius calculated with the two methods 
\citep[c.f. Figure~5 in][]{alcala11} further confirms the reliability of the
flux calibration of the spectra. Mass and age were derived by comparison 
with theoretical PMS evolutionary tracks \citep{baraffe98, chabrier00} on 
the HR diagramme. The physical parameters of the targets are listed 
in Table~\ref{pars} and summarised in Figure~\ref{histprop}. 
Uncertainties in luminosity, radius and mass take into account the error 
propagation of about half spectral sub-class in spectral typing, as well 
as errors in the photometry and uncertainty on extinction.

\begin{figure}[h]
\resizebox{1.0\hsize}{!}{\includegraphics[]{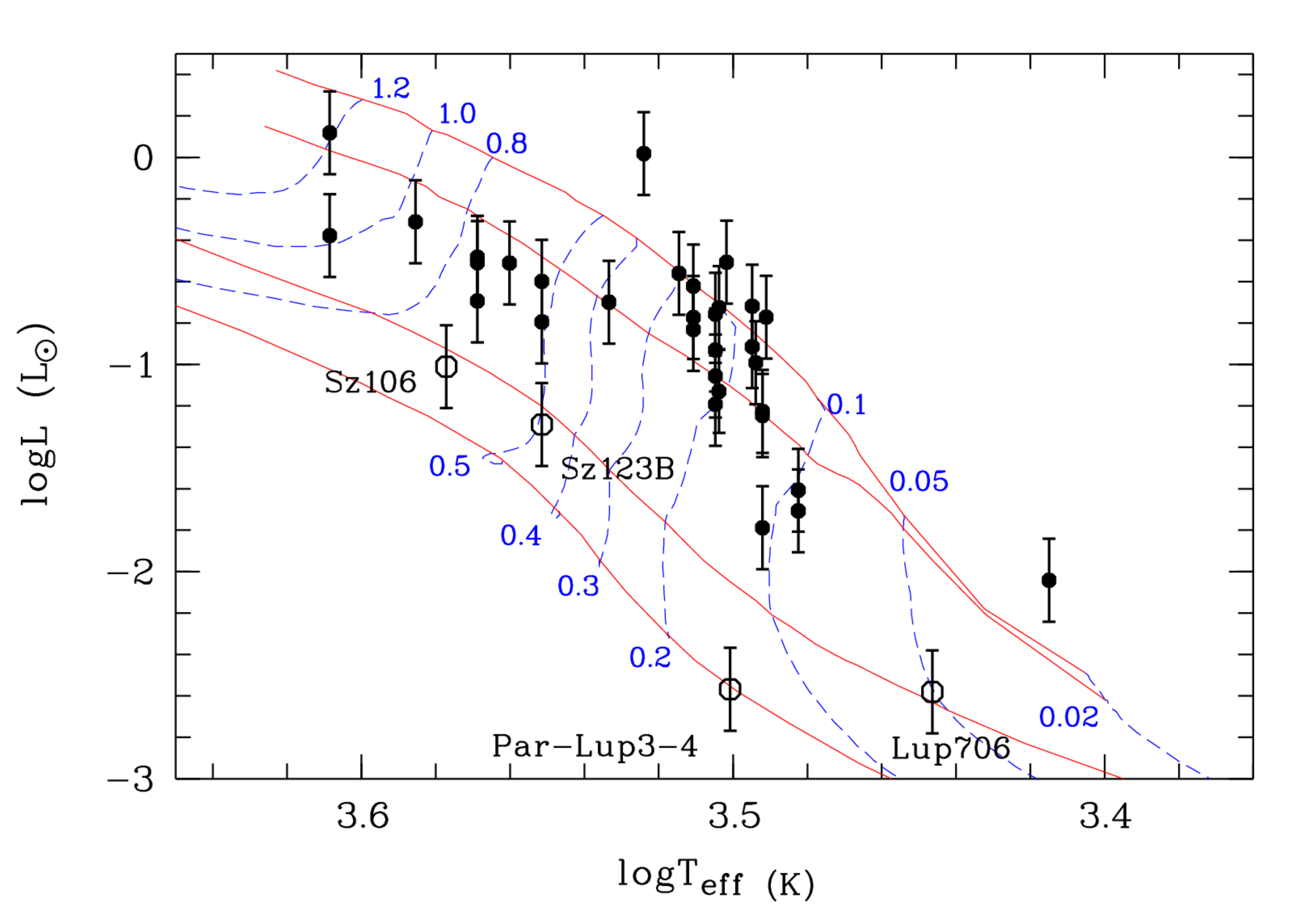}}
\caption{Hertzsprung-Russell diagramme for the Lupus sample. The
       four sub-luminous objects described in the text are 
       represented with open circles as labelled. The continuous
       lines show the 1\,Myr, 3\,Myr, 30\,Myr and 10\,Gyr isochrones 
       by \citet{baraffe98}, while the dashed lines show the 
       low-mass Pre-Main Sequence evolutionary tracks by the 
       same authors as labelled.
    \label{HRD}}
\end{figure}

The luminosity of four objects, namely Par-Lup3-4, Lup706, Sz\,123B, 
and Sz\,106, is significantly lower than for the other YSOs of similar 
spectral type or mass, hence their age results apparently older than 15\,Myr.
The sub-luminosity of these objects in comparison with the others 
is evident in Figure~\ref{HRD}, where the HR-diagramme for the sample is 
shown. It is not entirely clear whether the relatively low-luminosity of 
these objects is due to evolution or to obscuration effects because of a 
particular disc geometry. Sub-luminosity has been reported for Sz\,106
and Par-Lup3-4 \citep{comeron03}, and for the latter it has been shown that 
the disc is edge-on \citep{huelamo10}. No evidence of subluminosity
for the other two objects is found in the literature. 
At the end of the paper (Section~\ref{subluminous}) we provide arguments 
suggesting that the sub-luminosity of these objects is most likely due 
to geometrical effects. 

The average age of 3$\pm$2\,Myr for the sample, excluding the sub-luminous 
objects\footnote{Note that these objects are not included in the age hitogramme
in Figure~\ref{histprop}.}, is consistent with previous age estimates for Lupus YSOs
\citep[][and references therein]{comeron08}. 
We do not detect statistically significant differences among the 
stellar parameters of the Lupus-I and Lupus-III YSOs because our number 
statistics is small and this work deals only with Class-II YSOs. Note, 
however, that a recent investigation \citep{rygl13} based on Herschel 
data of significant samples of all classes of YSOs, concludes that there 
are differences in the star formation rate between the various Lupus clouds.

Finally, the Li~I $\lambda$670.78nm absorption line is detected in all 
the spectra, and it is identified for the first time in nine of our targets, 
two of which are part of the Class-III sample \citep{manara13}. 
Among these nine, three objects previously considered only as candidates 
were confirmed to be truly YSOs. We also stress that for the stars 
Sz\,105 and Sz\,94, originally included in the X-Shooter observations, no 
lithium absorption was detected, despite the high S/N of their spectra. 
Therefore, they are not considered in our analysis.
The nature of the Class-III source Sz\,94 has been discussed in 
\citet{manara13} and \citet{stelzer13b}, whereas the YSO nature of 
Sz\,105 is unclear, but will be discussed in a future work.

\setlength{\tabcolsep}{8pt}
\begin{table*}
\caption[ ]{\label{pars} Spectral types, extinction, and physical parameters} 
\begin{tabular}{lcccccccrl}
\hline \hline
Object  &   SpT  & $T_{\rm eff}$  & $A_{\rm V}$  & $d$    &  \Lstar         &    \Rstar       & \Mstar         & Age & Notes \\
        &        & [K]           & [mag.]       & [pc] &  [\Lsun] & [\Rsun]  & [\Msun] & [Myr] &  \\
        &        &               &              &      &                 &   		 &              &        & \\
\hline
  Sz66                  & M3.0 &  3415 &  1.00 &  150  & 0.200$\pm$0.092 &  1.29$\pm$0.30 &  0.45$_{-0.15}^{+0.05}$  &   4   &   \\ 
  AKC2006-19            & M5.0 &  3125 &  0.00 &  150  & 0.016$\pm$0.008 &  0.44$\pm$0.10 &  0.10$_{-0.02}^{+0.03}$  &  13   & {\bf 1, 2} \\ 
  Sz69                  & M4.5 &  3197 &  0.00 &  150  & 0.088$\pm$0.041 &  0.97$\pm$0.22 &  0.20$_{-0.03}^{+0.00}$  &   3   &   \\ 
  Sz71                  & M1.5 &  3632 &  0.50 &  150  & 0.309$\pm$0.142 &  1.43$\pm$0.33 &  0.62$_{-0.17}^{+0.02}$  &   4   &   \\ 
  Sz72                  & M2.0 &  3560 &  0.75 &  150  & 0.252$\pm$0.116 &  1.29$\pm$0.30 &  0.45$_{-0.00}^{+0.12}$  &   3   &   \\ 
  Sz73                  & K7   &  4060 &  3.50 &  150  & 0.419$\pm$0.193 &  1.35$\pm$0.31 &  1.00$_{-0.00}^{+0.00}$  &   9   &   \\ 
  Sz74                  & M3.5 &  3342 &  1.50 &  150  & 1.043$\pm$0.480 &  3.13$\pm$0.72 &  0.50$_{-0.10}^{+0.10}$  &   1   &   \\ 
  Sz83                  & K7   &  4060 &  0.00 &  150  & 1.313$\pm$0.605 &  2.39$\pm$0.55 &  1.15$_{-0.05}^{+0.25}$  &   2   &   \\ 
  Sz84                  & M5.0 &  3125 &  0.00 &  150  & 0.122$\pm$0.056 &  1.21$\pm$0.28 &  0.17$_{-0.02}^{+0.08}$  &   1   &   \\ 
  Sz130                 & M2.0 &  3560 &  0.00 &  150  & 0.160$\pm$0.074 &  1.03$\pm$0.24 &  0.45$_{-0.00}^{+0.05}$  &   6   &   \\ 
  Sz88A (SW)            & M0   &  3850 &  0.25 &  200  & 0.488$\pm$0.225 &  1.61$\pm$0.37 &  0.85$_{-0.10}^{+0.10}$  &   4   &   \\ 
  Sz88B (NE)            & M4.5 &  3197 &  0.00 &  200  & 0.118$\pm$0.054 &  1.12$\pm$0.26 &  0.20$_{-0.03}^{+0.05}$  &   2   &   \\ 
  Sz91                  & M1   &  3705 &  1.20 &  200  & 0.311$\pm$0.143 &  1.36$\pm$0.31 &  0.62$_{-0.08}^{+0.13}$  &   4   &   \\ 
  Lup713                & M5.5 &  3057 &  0.00 &  200  & 0.020$\pm$0.009 &  0.52$\pm$0.12 &  0.08$_{-0.00}^{+0.05}$  &   4   & {\bf 1} \\ 
  Lup604s               & M5.5 &  3057 &  0.00 &  200  & 0.057$\pm$0.026 &  0.83$\pm$0.19 &  0.11$_{-0.02}^{+0.04}$  &   2   &    \\ 
  Sz97                  & M4.0 &  3270 &  0.00 &  200  & 0.169$\pm$0.078 &  1.34$\pm$0.28 &  0.25$_{-0.00}^{+0.05}$  &   2   &   \\ 
  Sz99                  & M4.0 &  3270 &  0.00 &  200  & 0.074$\pm$0.034 &  0.89$\pm$0.20 &  0.17$_{-0.00}^{+0.08}$  &   3   &   \\ 
  Sz100                 & M5.5 &  3057 &  0.00 &  200  & 0.169$\pm$0.078 &  1.43$\pm$0.33 &  0.17$_{-0.04}^{+0.00}$  &   1   &   \\ 
  Sz103                 & M4.0 &  3270 &  0.70 &  200  & 0.188$\pm$0.087 &  1.41$\pm$0.30 &  0.25$_{-0.00}^{+0.05}$  &   1   &   \\ 
  Sz104                 & M5.0 &  3125 &  0.00 &  200  & 0.102$\pm$0.047 &  1.11$\pm$0.26 &  0.15$_{-0.02}^{+0.02}$  &   1   &   \\ 
  Lup706                & M7.5 &  2795 &  0.00 &  200  & 0.003$\pm$0.001 &  0.22$\pm$0.05 &  0.06$_{-0.02}^{+0.03}$  &  32   & {\bf 1, 3 }\\ 
  Sz106                 & M0.5 &  3777 &  1.00 &  200  & 0.098$\pm$0.045 &  0.72$\pm$0.17 &  0.62$_{-0.05}^{+0.00}$  &  32   & {\bf 3} \\ 
  Par-Lup3-3            & M4.0 &  3270 &  2.20 &  200  & 0.240$\pm$0.110 &  1.59$\pm$0.37 &  0.25$_{-0.05}^{+0.05}$  &   1   & {\bf 1} \\ 
  Par-Lup3-4            & M4.5 &  3197 &  0.00 &  200  & 0.003$\pm$0.001 &  0.17$\pm$0.04 &  0.13$_{-0.00}^{+0.02}$  & $>$50 & {\bf 1, 3} \\ 
  Sz110                 & M4.0 &  3270 &  0.00 &  200  & 0.276$\pm$0.127 &  1.61$\pm$0.37 &  0.35$_{-0.05}^{+0.05}$  &   1   &   \\ 
  Sz111                 & M1   &  3705 &  0.00 &  200  & 0.330$\pm$0.152 &  1.40$\pm$0.32 &  0.75$_{-0.13}^{+0.05}$  &   6   &   \\ 
  Sz112                 & M5.0 &  3125 &  0.00 &  200  & 0.191$\pm$0.088 &  1.52$\pm$0.35 &  0.25$_{-0.08}^{+0.00}$  &   1   &   \\ 
  Sz113                 & M4.5 &  3197 &  1.00 &  200  & 0.064$\pm$0.030 &  0.83$\pm$0.19 &  0.17$_{-0.04}^{+0.03}$  &   3   &   \\ 
2MASS J16085953-3856275 & M8.5 &  2600 &  0.00 &  200  & 0.009$\pm$0.004 &  0.47$\pm$0.11 &  0.03$_{-0.01}^{+0.01}$  &   1   & {\bf 1, 2}\\ 
SSTc2d160901.4-392512   & M4.0 &  3270 &  0.50 &  200  & 0.148$\pm$0.068 &  1.25$\pm$0.29 &  0.20$_{-0.05}^{+0.10}$  &   1   &   \\ 
  Sz114                 & M4.8 &  3175 &  0.30 &  200  & 0.312$\pm$0.144 &  1.82$\pm$0.42 &  0.30$_{-0.10}^{+0.05}$  &   1   &   \\ 
  Sz115                 & M4.5 &  3197 &  0.50 &  200  & 0.175$\pm$0.080 &  1.36$\pm$0.31 &  0.17$_{-0.08}^{+0.08}$  &   1   &   \\ 
  Lup818s               & M6.0 &  2990 &  0.00 &  200  & 0.025$\pm$0.011 &  0.58$\pm$0.13 &  0.08$_{-0.02}^{+0.02}$  &   3   & {\bf 1, 2}\\ 
  Sz123A  (S)           & M1   &  3705 &  1.25 &  200  & 0.203$\pm$0.093 &  1.10$\pm$0.25 &  0.60$_{-0.03}^{+0.20}$  &   7   &   \\ 
  Sz123B  (N)           & M2.0 &  3560 &  0.00 &  200  & 0.051$\pm$0.024 &  0.58$\pm$0.13 &  0.50$_{-0.10}^{+0.00}$  &  40   & {\bf 3} \\ 
  SST-Lup3-1            & M5.0 &  3125 &  0.00 &  200  & 0.059$\pm$0.027 &  0.85$\pm$0.19 &  0.13$_{-0.04}^{+0.02}$  &   2   &   \\ 
 
\hline
\end{tabular}
\tablefoot{
\\
{\bf 1}: Li~I $\lambda$670.78nm absorption line detected for the first time. 
 The other two targets for which the line is seen for the first time are 
 the Class-III YSOs Par-Lup3-1 and Par-Lup3-2 \citep[see][]{manara13}; 
{\bf 2}: YSO nature confirmed; 
{\bf 3}: sub-luminous YSO.
}
\end{table*}

\section{Accretion diagnostics}
\label{accretiondiagnostics}
The energy loss per unit time of the accretion energy, or accretion luminosity \Lacc, 
shows up as continuum and line emission over a wide spectral range. 
The contribution of the lines to \Lacc ~is generally ignored because the 
continuum emission is normally larger, but as will be seen in
Section~\ref{Llines_Lacc}, significant energy losses  also occur in line 
emission. 
In this section we derive the accretion luminosity of each YSO in our 
sample from the continuum-excess measurements, that we describe as the continuum 
emission of a slab of hydrogen. Previous determinations 
of \Lacc ~in the literature have also excluded line emission from estimates of 
the accretion excess emission, making our results directly comparable to those 
earlier studies. The slab modelling accounts also for the excess emission 
short-ward 330nm, which is about the shortest wavelength into which our 
spectra are useful, and allows us to describe the excess emission in the Paschen
continuum, where it is measured at very few wavelengths only, as line veiling. 

\subsection{Continuum emission}
In accreting YSOs the continuum excess emission is most easily detected as
Balmer continuum emission \citep[see][and references therein]{valenti93, gullbring98, 
HH08, rigliaco12}. 
Except for two very low-mass objects (Lup\,604 and Par-Lup3-3),
with rather noisy UVB spectra, Balmer continuum emission is 
evident in all the YSOs of our sample (see examples in Figure~\ref{slab_fits}). 
The observed Balmer jump, \BJobs, i.e.  the ratio of the flux at 360\,nm 
to the flux at 400\,nm, ranges from 0.4 to 3.7 (c.f. Table~\ref{accretion}),
with only three objects (Lup\,604, Par-Lup3-3 and Sz\,112) having \BJobs$\le$0.5.
This satisfies the criterion suggested by \citet{HH08} 
that any mid M-type dwarf with an observed Balmer jump greater than 0.5 
should be considered an accretor.
The intrinsic Balmer jump, \BJintr, which is the ratio of the flux at 360\,nm 
to the flux at 400\,nm after extinction correction and subtraction of the 
photospheric emission, was calculated using the photospheric spectrum of 
the same Class-III template as in the slab modelling described below. 

Paschen continuum emission (in the wavelength range from 364nm to 820nm) 
is dimmer than the Balmer emission in the UVB range, but is detected as veiling 
of the photospheric lines. Examples of such filling-in of the Ca\,{\sc i} 
$\lambda$422.7\,nm absorption line in some of the YSOs in $\sigma$-Ori 
are shown in Figure~3 by \citet{rigliaco12} \citep[see also Figure~2 in][]{HH08}. 
The filling-in of this line may be partially due to the  Fe\,{\sc i} 
emission lines at 421.6, 422.7 and 423.3\,nm.
The continuum emission at wavelengths longer than about 700\,nm  is rather weak 
in very low-mass YSOs and is hardly detected in many of our targets.

\subsubsection{Calculation of continuum accretion luminosity}
\label{bjmodelling}
\begin{figure}
 \advance\leftskip-0.9cm
 \includegraphics[width=11.0cm, height=12.5cm]{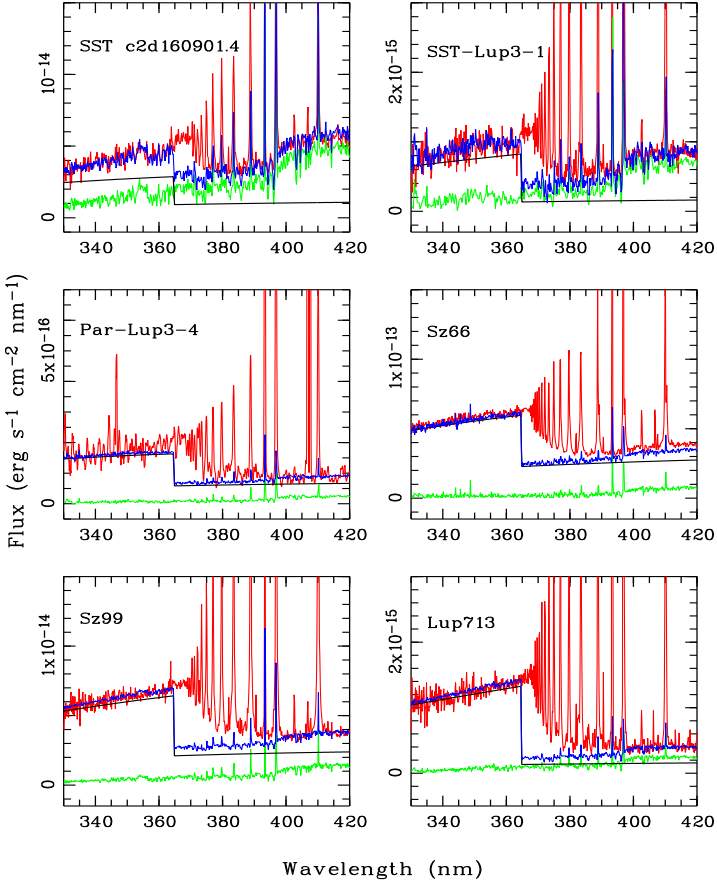}{\centering}
\caption{Examples of X-Shooter spectra of Class-II YSOs in the region of the 
       Balmer jump (red lines). The spectrum of the adopted Class-III templates 
       are over-plotted with green lines. The continuum emission from the slab is
       shown by the black continuous line. The best fit with the
       emission  predicted from the slab model added to the template is 
       given by the blue lines.
     \label{slab_fits}}
 \end{figure}

Following \citet{valenti93}, \citet{HH08} and \citet{rigliaco12}, the spectrum 
of each Class-II YSO is fitted as the sum of the photospheric spectrum  
and the emission of a slab of hydrogen;the accretion luminosity is given by 
the luminosity emitted by the slab. We take the photospheric spectrum to be 
that of the Class-III template best matching the spectral type of the 
Class-II YSO reported in Table~\ref{pars}.
The input spectra are extinction corrected using the extinction 
values reported in Table~\ref{pars}. 
The hydrogen slab emission includes bound-free and free-free emission from 
$\ion{H}{}$  and $\ion{H}{}^-$; it is computed assuming LTE conditions and 
is described by three parameters, namely the electron temperature and density 
and the slab length, which is related to the optical depth at 300\,nm. 
These three parameters determine the wavelength dependence 
of the excess emission. In addition, there are two "normalization" parameters 
one for the slab emission and one for the Class-III template flux. We  vary 
these 5 parameters over a wide range of values to find the model that best 
matches the observed continuum of the target spectrum. More specifically,
we consider a number of spectral features, namely the observed Balmer jump, 
the slope of the Balmer continuum measured between $\sim$340 and 360\,nm, 
the slope of the Paschen continuum between $\sim$400 and $\sim$475\,nm, 
the value of the observed fluxes at $\sim$360\,nm, $\sim$460\,nm, and $\sim$710\,nm. 
This includes spectral regions where the emission is likely to be dominated 
by the accretion shock emission and regions where the photospheric emission 
is likely the dominant component. The best-fit procedure is described
in \citet{manara13b}.
After the best-fitting model has been found, we then check if it reproduces 
well other spectral features, in particular the Ca\,{\sc i} $\lambda$422.7\,nm 
absorption line. 
Examples of slab modelling are shown in Figure~\ref{slab_fits} 
and the complete set of plots showing the fits for all targets are provided 
in electronic form only (Figures from \ref{slab1} to \ref{slab4}).

The adopted Class-III templates, and the \BJobs, \BJintr, and \Lacc ~ values 
corresponding to each Class-II YSO are reported in Table~\ref{accretion}. In this 
table we also give the ratio of the excess emission over the photospheric one at 
710\,nm as in the best-fit model.
The uncertainties on \Lacc ~are dominated by the uncertainty in the extinction 
and by the choice of the Class-III template \citep{manara13b}.
An additional, non-negligible uncertainty for low values of \Lacc, comes from 
the uncertainty on the Paschen continuum excess emission, especially in objects 
with poor signal-to-noise \citep[e.g.][]{rigliaco12}. In general, we estimate 
that the uncertainty on \Lacc ~is $\sim$0.2\,dex.

The Balmer and Paschen continua, as well as the
Balmer jump, are well reproduced by our fits. On the other hand, we
stress that we did not attempt to fit the hydrogen emission lines but 
only the continuum emission. For this reason the region on the longer 
wavelength side of the Balmer jump ($\lambda\sim$ 365-370 nm) is not well
reproduced by our fits, as the emission in this region originates from
a superposition of unresolved hydrogen emission lines and not from 
continuum emission. In some cases, even when the lines are clearly resolved, 
the overlapping of their wings produce a pseudo-continuum that we do not 
account for. Such line blending shifts the apparent Balmer jump to $\sim$370\,nm, 
while the actual jump is at 364.6\,nm. In Section~\ref{Llines_Lacc} we estimate 
the contribution of this effect on the total budget of the Balmer continuum 
emission. 

Only in one case namely Sz123B, the slope of the Balmer continuum is not
exactly reproduced by any combination of our free parameters. We checked 
that this is not due to data reduction problems (e.g. flat-field correction 
or incorrect spectrum extraction), but perhaps to slit-loss effects. 
Small differences between the observed and best-fit spectra in the Paschen 
continuum are present in some objects (e.g. AKC2006-19, Sz115), but 
differences are small compared to the excess emission in the Balmer 
continuum region.
We were always able to fit our observations with a single slab
model, without the necessity of multiple accretion components \citep{ingleby13}, 
and our fits reproduce well the observed spectra on a large wavelength 
interval, from $\sim$330 nm to $\sim$720\,nm.

In about 50\% of the objects, the excess emission in the Paschen
continuum accounts for more than 50\% of \Lacc, and in all the targets at
least 30\% of the total excess is emitted in that region.

\setlength{\tabcolsep}{4pt}
\begin{table*}
\caption[ ]{\label{accretion} Accretion properties of Lupus YSOs.} 
\begin{tabular}{llcrccrcc}
\hline \hline
Object  &   Template  & BJ$_{\rm obs}$ & BJ$_{\rm intr}$ & log\Lacc       & veiling & log\Macc   & $L_{\rm all\_lines}$/\Lacc  & W\Ha(10\%)   \\
        &             &               &                 & [\Lsun] & at 710nm   & [\Msun yr$^{-1}$] &           & [km s$^{-1}$]  \\
        &             &               &                 &        &            &                 &           &                \\
\hline
   Sz66 		  &  SO797	&  1.50   &   1.90  & $-$1.8    & 0.45 &  $-$8.73  &  0.09 & 460 \\
   AKC2006-19   	  &  SO641	&  0.60   &  14.67  & $-$4.1    & 0.04 & $-$10.85  &  0.16 & 228 \\
   Sz69 		  &  SO797	&  2.83   &   4.50  & $-$2.8    & 0.18 &  $-$9.50  &  0.55 & 403 \\
   Sz71 		  &  TWA15A	&  1.19   &   2.65  & $-$2.2    & 0.08 &  $-$9.23  &  0.24 & 350 \\
   Sz72 		  &  TWA9B	&  2.73   &   3.70  & $-$1.8    & 0.23 &  $-$8.73  &  0.28 & 455 \\
   Sz73 		  &  SO879	&  1.05   &   2.08  & $-$1.0    & 0.22 &  $-$8.26  &  0.05 & 504 \\
   Sz74 		  &  TWA15A	&  0.90   &   1.46  & $-$1.5    & 0.10 &  $-$8.09  &  0.07 & 401 \\
   Sz83 		  &  SO879	&  2.28   &   2.27  & $-$0.3    & 1.88 &  $-$7.37  &  0.14 & 604 \\
   Sz84 		  &  SO641	&  1.44   &   2.65  & $-$2.7    & 0.10 &  $-$9.24  &  0.28 & 456 \\
   Sz130		  &  TWA2A	&  1.50   &   2.43  & $-$2.2    & 0.15 &  $-$9.23  &  0.20 & 266 \\
   Sz88A (SW) 	          &  TWA25	&  2.08   &   3.03  & $-$1.2    & 0.32 &  $-$8.31  &  0.18 & 597 \\
   Sz88B (NE)		  &  SO797	&  0.87   &   3.34  & $-$3.1    & 0.04 &  $-$9.74  &  0.18 & 405 \\
   Sz91 		  &  TWA13A	&  1.12   &   2.14  & $-$1.8    & 0.12 &  $-$8.85  &  0.22 & 374 \\
   Lup713		  &  Par-Lup3-2 &  3.50   &   6.50  & $-$3.5    & 0.12 & $-$10.08  &  0.54 & 378 \\
   Lup604s		  &  SO925	&  0.50   &   8.50  & $-$3.7    & 0.05 & $-$10.21  &  0.13 & 264 \\
   Sz97 		  &  Sz94	&  1.00   &   3.70  & $-$2.9    & 0.03 &  $-$9.56  &  0.38 & 452 \\
   Sz99 		  &  TWA9B	&  1.90   &   2.65  & $-$2.6    & 0.26 &  $-$9.27  &  0.19 & 373 \\
   Sz100		  &  SO641	&  0.87   &   2.16  & $-$3.0    & 0.09 &  $-$9.47  &  0.31 & 251 \\
   Sz103		  &  Sz94	&  0.74   &   1.87  & $-$2.4    & 0.14 &  $-$9.04  &  0.09 & 426 \\
   Sz104		  &  SO641	&  0.65   &   1.90  & $-$3.2    & 0.06 &  $-$9.72  &  0.23 & 201 \\
   Lup706		  &  TWA26	&  2.40   &  13.50  & $-$4.8    & 0.11 & $-$11.63  &  0.51 & 328 \\
   Sz106		  &  TWA25	&  0.97   &   1.87  & $-$2.5    & 0.14 &  $-$9.83  &  0.06 & 459 \\
   Par-Lup3-3		  &  TWA15A	&  0.50   &   1.10  & $-$2.9    & 0.01 &  $-$9.49  &  0.19 & 240 \\
   Par-Lup3-4		  &  SO641	&  2.00   &   3.00  & $-$4.1    & 0.25 & $-$11.37  &  0.46 & 393 \\
   Sz110		  &  Sz94	&  1.43   &   2.14  & $-$2.0    & 0.26 &  $-$8.73  &  0.20 & 498 \\
   Sz111		  &  TWA13A	&  1.36   &   9.60  & $-$2.2    & 0.04 &  $-$9.32  &  0.41 & 455 \\
   Sz112		  &  SO641	&  0.40   &   1.40  & $-$3.2    & 0.03 &  $-$9.81  &  0.15 & 160 \\
   Sz113		  &  SO797	&  1.86   &   2.29  & $-$2.1    & 0.56 &  $-$8.80  &  0.20 & 392 \\
 2MASS J16085953-3856275  &  TWA26	&  3.69   &  15.00  & $-$4.6    & 0.08 & $-$10.80  &  0.39 & 147 \\
 SSTc2d160901.4-392512    &  Sz94	&  0.84   &   3.85  & $-$3.0    & 0.04 &  $-$9.59  &  0.35 & 447 \\
   Sz114		  &  Sz94	&  0.67   &   2.07  & $-$2.5    & 0.05 &  $-$9.11  &  0.26 & 222 \\
   Sz115		  &  SO797	&  0.53   &   1.00  & $-$2.7    & 0.10 &  $-$9.19  &  0.07 & 338 \\
   Lup818s		  &  SO925	&  1.06   &   4.00  & $-$4.1    & 0.08 & $-$10.63  &  0.48 & 200 \\
   Sz123A  (S)            &  TWA2A	&  1.71   &   2.71  & $-$1.8    & 0.22 &  $-$8.93  &  0.30 & 487 \\
   Sz123B  (N)            &  TWA15B	&  1.36   &   2.45  & $-$2.7    & 0.09 & $-$10.03  &  0.36 & 519 \\
   SST-Lup3-1		  &  SO641	&  1.23   &   6.00  & $-$3.6    & 0.03 & $-$10.17  &  0.48 & 254 \\
\hline
\end{tabular}
\end{table*}

\subsection{Mass accretion rate}
The accretion luminosities were converted into mass accretion rates, \Macc,
using the relation:

{\setlength{\mathindent}{0pt}
\begin{equation}
\label{Macc}
\dot{M}_{acc} = ( 1 - \frac{R_{\star}}{R_{\rm in}} )^{-1} ~ \frac{L_{acc} R_{\star}}{G M_{\star}}
 \approx 1.25 ~ \frac{L_{acc} R_{\star}}{G M_{\star}}
 \label{Macc}
\end{equation}

\noindent
where $R_{\star}$ and $M_{\star}$ are the YSO radius and mass reported in 
Table~\ref{pars}, respectively, and $R_{\rm in}$ is the YSO inner-disc 
radius \citep{gullbring98, hart98}. 
The latter corresponds to the distance from the star at which the disc 
is truncated -- due to the stellar magnetosphere -- and from which the 
disc gas is accreted, channelled by the magnetic field lines. 
It has been found that $R_{\rm in}$ ranges from 3$R_{\star}$ to 
10$R_{\star}$ \citep{johns-krull07}.
For consistency  with previous studies \citep[e.g.][]{gullbring98, HH08, rigliaco12},
here we assume $R_{\rm in}$ to be $5\,R_\star$. The results on \Macc 
~ are listed in column~7 of Table~\ref{accretion}.

We calculate \Macc ~values from 2$\times$10$^{-12}$\,\Msun~yr$^{-1}$ to 
4$\times$10$^{-8}$\,\Msun~yr$^{-1}$.
The sources of error in \Macc ~are the uncertainties on \Lacc, 
stellar mass and radius (see Table~\ref{pars}). Propagating these, we estimate 
an average error of $\sim$0.35\,dex in \Macc. 
However, additional errors on the aforementioned quantities come from the 
uncertainty in distance, as well as from differences in evolutionary 
tracks. 
The uncertainty on the Lupus YSOs distance is estimated to be $\sim$20\% 
\citep[see][and references therein]{comeron08}, yielding a relative
 uncertainty of about 0.26\,dex in the mass accretion rate
\footnote{Note that \Macc~$\propto$~d$^3$, as \Lacc~$\propto$~d$^2$ and 
\Rstar~$\propto$~d.}.
On the other hand, using the \citet{DM97} tracks we obtain a difference
in mass from 10\% to 70\% (with an average of 30\%) with respect to the 
\citet{baraffe98} tracks, leading to uncertainties of 0.04\,dex to 0.3\,dex 
in \Macc. We estimate that the cumulative relative uncertainty in \Macc ~is 
about 0.5\,dex.

With \Macc$=$3.4$\times$10$^{-8}$\,\Msun~yr$^{-1}$ , the strongest accretor 
in our sample is Sz\,83. A variety of \Macc ~estimates for this star exist
in the literature that range from 10$^{-7}$ to a few 10$^{-8}$\,\Msun~yr$^{-1}$, 
and may be as high as 10$^{-6}$ \citep{comeron08}.
Our \Macc ~estimate is in very good agreement with that calculated by 
\citet{HH08} (1.8$\times$10$^{-8}$\,\Msun~yr$^{-1}$). 

There are large discrepancies between our \Macc determinations and those 
derived by \citet{comeron03} for Sz\,100,  Sz\,106, Sz\,113 and Par-Lup3-4. 
The \citet{comeron03} estimates, which are based on the flux of the 
$\ion{Ca}{ii}$ $\lambda$854.2\,nm line, are larger by up to 1\,dex. Although 
part of the discrepancies may be in principle ascribed to variable accretion, 
such variability must be enormous over timescales of years to explain the 
differences. \citet{costigan12} and \citet{costigan13} have observed variable 
accretion over years, but their results show that it is very rare 
to have YSOs that vary \Macc ~by large factors. Most of the variability 
they find occurs on rotational timescales, suggesting asymmetric, 
rather than strongly variable accretion flows. 

\subsection{Emission lines}
 \label{emisslines}
A large number of permitted and forbidden emission lines, displaying 
a variety of profiles, are detected. The analysis of forbidden emission 
lines is deferred to a forthcoming paper (Natta et al. 2013, in prep). 
The emission lines studied here are listed in Table~\ref{linfits}.
The detection/non-detection of such lines depends on instrumental sensitivity 
and on the accretion rate of the individual YSO. The number of detections 
of each  line is given in column 5 of Table~\ref{linfits}, labelled as 
number of points used for the linear fits in Section~\ref{correlations}.

The detected emission lines include several from the Balmer and 
Paschen series, the \brg~ line, as well as several helium and calcium 
lines. Examples of permitted emission lines in X-Shooter spectra from 
our GTO programme have been published in previous papers 
\citep[e.g.][]{alcala11, bacciotti11, rigliaco11c, rigliaco12, 
manara13, stelzer13}.

Balmer lines are detected up to H25 in six objects (Lup\,713, Sz\,113, 
Sz\,69, Sz\,72, Sz\,83, Sz\,88A). One among these (Lup\,713) is at the 
hydrogen burning limit, with its spectrum resembling that of the young 
brown dwarf J\,053825.4$-$024241 reported in \citet{rigliaco11b}. 
In Sz\,88A Balmer line emission is detected up to H27 at the 2$\sigma$ 
level.
The Pa\,8, Pa\,9 and Pa\,10 are  located in spectral regions of dense 
telluric absorption bands. Although the telluric correction has been 
performed as accurately as possible, some residuals from the correction still 
remain. Thus, the detection and analysis of these three Paschen lines 
is more uncertain, leading to larger errors, in particular for Pa\,8. 

In this work we consider the nine He\,{\sc i} lines with the highest 
transition strength. Among these, the He\,{\sc i} $\lambda$1082.9nm 
has been found to be also related to winds/outflows \citep{edwards06}. 
Thus, the line may include both the accretion and wind contributions. 
In most cases the He\,{\sc i} $\lambda$492.2\,nm is blended with 
the Fe\,{\sc i} $\lambda$492.1\,nm line and we did not attempt any 
de-blending.

The Ca\,{\sc ii} H \& K lines are detected in all YSOs. 
The Ca\,{\sc ii}~H-line is partially blended with \Hep.  
The Ca\,{\sc ii}~IRT $\lambda\lambda\lambda$ 849.8, 854.2, 866.2\,nm 
and the D-lines of the $\ion{Na}{i}$ $\lambda\lambda$ 589.0,589.6\,nm 
doublet are very well resolved in all our spectra.
In several objects both the Ca\,{\sc ii}~IRT and the $\ion{Na}{i}$ 
lines show up as an emission reversal superposed on the broad 
photospheric absorption lines. Thus, the strength of these lines must be 
corrected for the photospheric contribution. This was done for 
the complete sample (see Section~\ref{linefluxes}).

Finally, the two O\,{\sc i} lines at 777.3\,nm and 844.6\,nm are well 
detected in 14 and 18 YSOs, respectively. These lines are seen in the 
objects with the strongest Balmer, He\,{\sc i} and Ca\,{\sc ii} lines.

\subsection{Line fluxes and equivalent widths}
\label{linefluxes}
The flux in permitted lines was computed by direct integration of 
the flux calibrated and extinction corrected spectra, using the {\em splot} 
package under IRAF\footnote{IRAF is distributed by the National Optical Astronomy 
Observatory, which is operated by the Association of the Universities for 
Research in Astronomy, inc. (AURA) under cooperative agreement with the
National Science Fundation}. 
Three independent measurements per line were done, 
considering the lowest, highest, and middle position of the local 
continuum, depending on the local noise level of the spectra. The flux and 
its error were then computed as the average and standard deviation of 
the three independent measurements, respectively. 
The extinction-corrected fluxes, equivalent widths, and their errors 
are reported in several tables provided in electronic form only 
(from Table \ref{tab:fluxes_EWs_Hae} 
to \ref{tab:fluxes_EWs_NaI}) \footnote{The flux errors reported in these 
tables are those resulting  from the uncertainty in continuum placement. 
The estimated $\sim$10\% uncertainty of flux calibration (see Section~\ref{datared}) 
should be added in quadrature.}. In the cases where the lines were not 
seen, 3$\sigma$ upper limits were estimated using the relationship 
${\rm 3 \times F_{\rm noise} \times \Delta\lambda}$, where F$_{\rm noise}$  
is the rms flux-noise in the region of the line and $\Delta\lambda$ 
is the expected average line width, assumed to be 0.2\,nm.

In the case of the H$\epsilon$ line, blended with the Ca\,{\sc ii}~H 
line, measurements of both lines were attempted by a de-blending procedure
using IRAF. In many cases, the X-Shooter resolution allows us to resolve both 
lines almost entirely. However, 11 YSOs display very broad lines,  
making the de-blending measurements unreliable. These objects are flagged 
in the electronic Tables~\ref{tab:fluxes_EWs_Hae} and \ref{tab:fluxes_EWs_CaII}.

The contribution of the photospheric absorption lines of the $\ion{Na}{i}$~D 
lines and the Ca\,{\sc ii}~IRT, strongest in the late-K and early-to-mid 
M-type objects, have been removed in all the spectra by using 
the synthetic BT-Settl spectra by \citet[][]{allard10} of the same 
\Teff ~and $\log{g}$ ~as the YSOs, binned at the same resolution of X-Shooter, 
and rotationally broadened at the same $v\sin{i}$ as the YSOs. For this purpose, 
the ROTFIT code \citep{frasca06}, specifically modified for X-Shooter data 
\citep[See][for details]{stelzer13b}, was applied. 
In some objects with very broad emission lines, the Ca\,{\sc ii}~IRT 
lines are fully blended with the Paschen lines Pa\,13 ($\lambda$866.502\,nm), 
Pa\,15 ($\lambda$854.538\,nm) and Pa\,16 ($\lambda$849.249\,nm). These 
objects are flagged in the electronic Table~\ref{tab:fluxes_EWs_CaII}. 

We note that the ratio of the Ca\,{\sc ii}~IR triplet lines is always
very close to  1:1:1 (see the electronic Table~\ref{tab:fluxes_EWs_CaII}), 
i.e. consistent with optically thick gas conditions 
\citep[][and references therein]{herbig80, hamann92},  suggesting that 
the lines are formed in a high density region near the surface 
of the central YSO, rather than in a low-density outflow environment 
\citep[see][]{reipurth86, graham88, fernandez01}.

Although several objects display hydrogen recombination lines with high 
quantum numbers, here the analysis is restricted to Balmer lines up to 
H15, Paschen lines up to Pa\,10 and the \brg ~line, as well as the helium, 
calcium, sodium, and oxygen lines listed in Table~\ref{linfits}.

The luminosity of the different emission lines was computed as
\Ll ~$ = 4 \pi  d^2 \cdot f_{\rm line}$, where $d$ is the YSO distance 
in Table~\ref{pars} and $f_{\rm line}$ is the extinction-corrected 
absolute flux of the lines reported in the electronic 
Tables~\ref{tab:fluxes_EWs_Hae} to \ref{tab:fluxes_EWs_NaI}.
 
\subsection{Lines versus continuum losses}
\label{Llines_Lacc}
We have calculated the total line luminosity, $L_{\rm all\_lines}$, as the 
sum of the luminosity of all the emission lines detected in every YSO. 
In the integrated luminosity of the Balmer lines we also account for the 
flux of the pseudo-continuum produced by  line blending close to the 
Balmer Jump. The latter was measured by subtracting the flux of the best-fit 
model to the extinction-corrected spectrum of the YSO. 
On the average, more than 70\% of the total line luminosity comes from
the Balmer lines. In most cases (90\%), the integrated Balmer line luminosity 
amounts to more than 60\% of $L_{\rm all\_lines}$, while for a few objects 
with very strong lines (e.g. Sz\,72, Sz\,83, Sz\,88\,A, Sz\,113) the emission 
in other lines may be as high as 55\% of $L_{\rm all\_lines}$.

The integrated line luminosity is strongly correlated with \Lacc ~
(Figure~\ref{Lall_lines_Lacc}). A linear fit yields:
$\log{(L_{\rm all\_lines}/L_{\odot})}  = 0.86 \cdot \log($\Lacc/\Lsun)~$- 1.05$,
with a standard deviation $\sigma=$0.25. This can be expressed as 
$\log({L_{\rm all\_lines}}$/\Lacc)~=$ - 0.14 \cdot \log$(\Lacc/\Lsun)~$- 1.05$.
From the latter equation and considering the scatter of the relation
we calculate that objects with \Lacc~$<$10$^{-4}$\Lsun ~(only five YSOs) may 
have $L_{\rm all\_lines}$/\Lacc ~ratios as high as 0.55, while the ratio is lower 
for high \Lacc ~and is $<$0.25 for \Lacc~$>$~10$^{-3}$\Lsun.
Column 8 of Table~\ref{accretion} lists the $L_{\rm all\_lines}$/\Lacc ~ratios.
Although these numbers show that $L_{\rm all\_lines}$ ~is a fraction of 
\Lacc, there is considerable emission also in the lines.

\begin{figure}[h]
\resizebox{1.0\hsize}{!}{\includegraphics[]{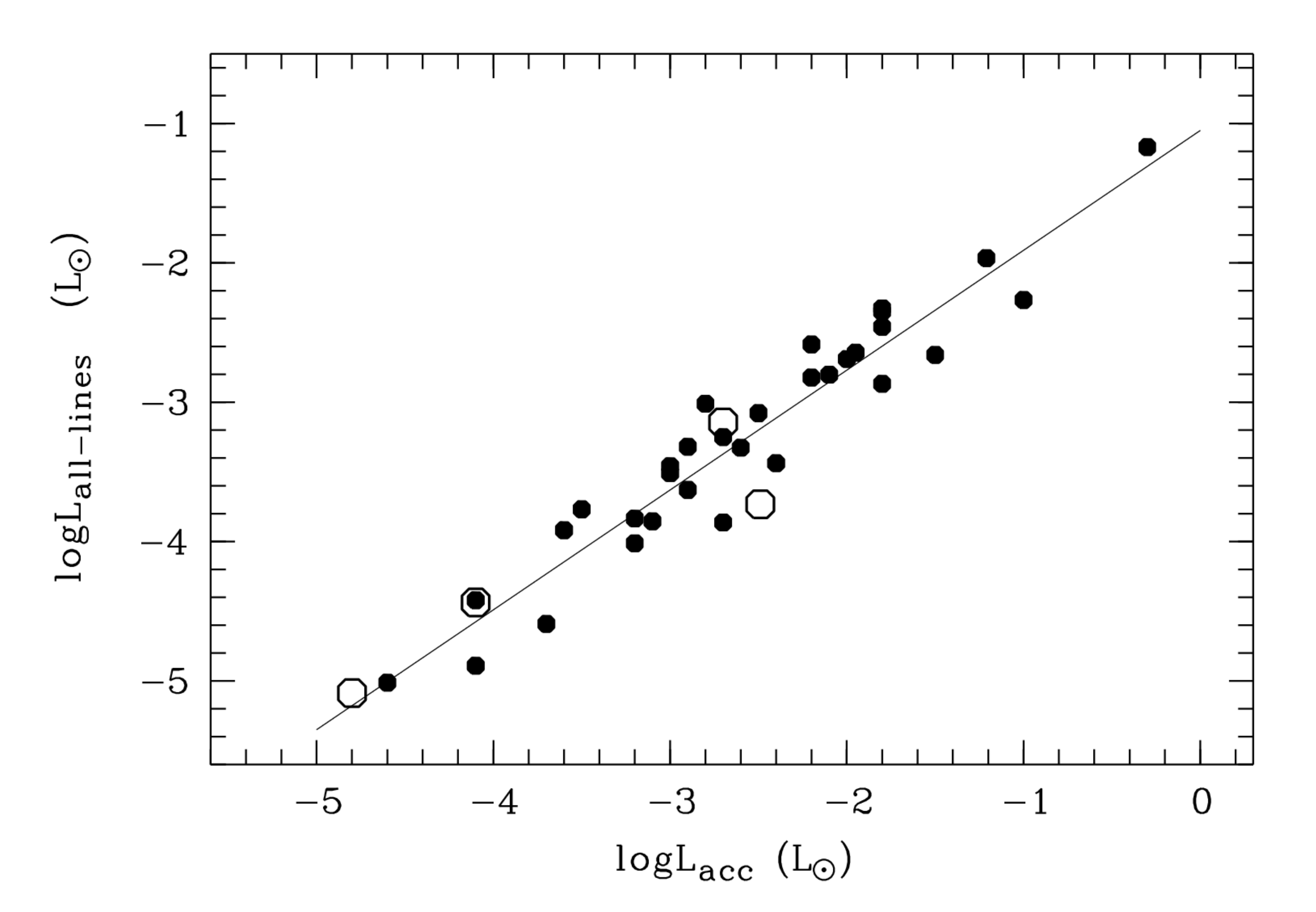}}
\caption{Total emission line luminosity as a function of \Lacc.
The \Lacc ~values were derived from the slab model in Section~\ref{bjmodelling}.
Symbols are as in Figure~\ref{HRD}. The continuous line represents the linear
fit discussed in the text.
    \label{Lall_lines_Lacc}} 
\end{figure} 

\setlength{\tabcolsep}{6pt}
\begin{table*}
\begin{center}
\caption[ ]{\label{linfits} Results of the \Lacc -- \Ll ~linear fits$^\dag$. } 
\begin{tabular}{l|r|c|c|l|r|c|l}
\hline \hline

Diagnostic   &   $\lambda$ & $a$ ($\pm$err) & $b$ ($\pm$err) & N$_{\rm points}^\ddag$ & $\sigma ^{\star}$  &  No. upper & Comments \\
             &     [nm]   &               &               &                  &             &   limits       &   \\
\hline     
        &            &               &               &                  &            &         &  \\

H3 (\Ha)  & 656.2800 &  1.12 (0.07)  &  1.50 (0.26)  & 36 + 6  &  0.36  & 0  &  \\
H4 (\Hb)  & 486.1325 &  1.11 (0.05)  &  2.31 (0.23)  & 36 + 6  &  0.27  & 0  & ~~$\bullet$ \\
H5 (\Hg)  & 434.0464 &  1.09 (0.05)  &  2.50 (0.25)  & 36 + 6  &  0.27  & 0  & ~~$\bullet$ \\
H6 (\Hd)  & 410.1734 &  1.06 (0.06)  &  2.50 (0.28)  & 36      &  0.30  & 0  & ~~$\bullet$ \\
H7 (\Hep) & 397.0072 &  1.07 (0.06)  &  2.64 (0.29)  & 36      &  0.30  & 0  &  {\bf 1 } \\
H8        & 388.9049 &  1.04 (0.06)  &  2.55 (0.29)  & 36      &  0.30  & 0  & ~~$\bullet$ \\
H9        & 383.5384 &  1.01 (0.05)  &  2.53 (0.27)  & 36      &  0.29  & 0  & ~~$\bullet$ \\
H10       & 379.7898 &  1.00 (0.05)  &  2.58 (0.27)  & 35      &  0.29  & 1  & ~~$\bullet$ \\
H11       & 377.0630 &  1.02 (0.05)  &  2.74 (0.25)  & 35 + 6  &  0.27  & 1  & ~~$\bullet$ \\
H12       & 375.0151 &  0.99 (0.05)  &  2.73 (0.25)  & 35      &  0.26  & 1  & ~~$\bullet$ \\
H13       & 373.4368 &  1.00 (0.05)  &  2.85 (0.25)  & 34      &  0.25  & 2  & ~~$\bullet$ \\
H14       & 372.1938 &  1.02 (0.06)  &  3.09 (0.31)  & 31      &  0.25  & 5  & ~~$\bullet$ \\
H15       & 371.1977 &  1.02 (0.06)  &  3.13 (0.31)  & 31      &  0.25  & 5  & ~~$\bullet$ \\

\hline     
         &          &               &              &                &        &     &   \\
Pa5  (\pab) & 1281.8070  &   1.04 (0.08)  &   2.45 (0.39)  &  29 + 6 &  0.38  & 7  &  ~~$\bullet$ \\ 
Pa6  (\pag) & 1093.8086  &   1.18 (0.06)  &   3.17 (0.31)  &  33 + 6 &  0.29  & 3  &  ~~$\bullet$ \\
Pa7  (\pad) & 1004.9368  &   1.18 (0.10)  &   3.33 (0.47)  &  25     &  0.36  & 9  &  ~~$\bullet$ \\
Pa8         &  954.5969  &   1.11 (0.12)  &   3.19 (0.58)  &  17     &  0.39  & 16 &   \\
Pa9         &  922.9014  &   1.13 (0.09)  &   3.40 (0.47)  &  27     &  0.35  & 7  &    \\
Pa10        &  901.4909  &   1.03 (0.09)  &   2.99 (0.49)  &  26     &  0.39  & 8  &    \\

\hline
        &            &               &               &                  &       &    &  \\

Br7  (\brg) & 2166.1210  &  1.16 (0.07) &  3.60 (0.38) &  19   &    0.28  & 14 &  ~~$\bullet$ \\

\hline
        &            &               &               &                  &     &      &  \\

He\,{\sc i}	       &  402.6191  &   1.04 (0.06)  &  3.62 (0.37)  &  31      &  0.27 & 5  & ~~$\bullet$ \\
He\,{\sc i}	       &  447.1480  &   1.05 (0.06)  &  3.45 (0.35)  &  33      &  0.30 & 3  & ~~$\bullet$ \\
He\,{\sc i}	       &  471.3146  &   1.04 (0.11)  &  4.25 (0.69)  &  16      &  0.29 & 12 & ~~$\bullet$ \\
He\,{\sc i}Fe\,{\sc i} &  492.1931  &   0.90 (0.06)  &  2.65 (0.36)  &  32      &  0.31 & 4  & {\bf 2}   \\
He\,{\sc i}	       &  501.5678  &   0.98 (0.06)  &  3.36 (0.38)  &  30      &  0.27 & 4  & ~~$\bullet$ \\
He\,{\sc i}	       &  587.5621  &   1.13 (0.06)  &  3.51 (0.30)  &  36 + 6  &  0.28 & 0  & ~~$\bullet$ \\
He\,{\sc i}	       &  667.8151  &   1.16 (0.08)  &  4.12 (0.45)  &  36      &  0.36 & 0  & ~~$\bullet$ \\
He\,{\sc i}	       &  706.5190  &   1.14 (0.07)  &  4.16 (0.39)  &  36      &  0.32 & 0  & ~~$\bullet$ \\
He\,{\sc i}	       & 1082.9091  &   1.11 (0.12)  &  2.62 (0.57)  &  30      &  0.44 & 4  & {\bf 3}   \\
He\,{\sc ii}	       &  468.5804  &   1.10 (0.09)  &  4.21 (0.57)  &  28      &  0.36 & 2  &           \\

\hline
        &            &               &               &                  &        &   &  \\

Ca\,{\sc ii} (K)     &  393.3660  &  0.96 (0.05)  &  2.06 (0.27)  &  36  &  0.31  & 0 & ~~$\bullet$ \\
Ca\,{\sc ii} (H)     &  396.8470  &  1.02 (0.05)  &  2.37 (0.23)  &  36  &  0.25  & 0 & {\bf 4} \\
Ca\,{\sc ii}         &  849.8020  &  0.95 (0.07)  &  2.18 (0.38)  &  34  &  0.41  & 2 & \\
Ca\,{\sc ii}         &  854.2090  &  0.95 (0.08)  &  2.13 (0.42)  &  32  &  0.44  & 2 & \\
Ca\,{\sc ii}         &  866.2140  &  0.95 (0.09)  &  2.20 (0.43)  &  29  &  0.43  & 4 & \\

\hline
        &            &               &               &                  &        &   &   \\
Na\,{\sc i} &  588.995   & 0.93 (0.06)  & 2.56 (0.32)  & 36	&  0.33  & 0      &      \\
Na\,{\sc i} &  589.592   & 0.90 (0.06)  & 2.56 (0.37)  & 36	&  0.38  & 0      &      \\
\hline
        &            &               &               &                  &        &   &    \\
O\,{\sc i}  &  777.3055  & 1.16 (0.09)  & 3.91 (0.51)  & 14	 &  0.36 & 6 & {\bf 5}    \\
O\,{\sc i}  &  844.6360  & 1.06 (0.18)  & 3.06 (0.90)  & 18	 &  0.61 & 3 &            \\

\hline
\end{tabular}
\tablefoot{
\\
$\dag$: the relations  are of the form ~~~
 $\log{(L_{\rm acc}/L_{\odot})} = a\cdot\log{(L_{\rm line}/L_{\odot})} ~+~ b $.\\
\vspace{0.1cm}
{\large $\ddag$}: number of points for the fit. The fits in which the six YSOs in $\sigma$-Ori 
  \citep{rigliaco12} were included are indicated with ''+6~''. Although measurements of 
  the Ca\,{\sc ii}~IRT are also available in \citet{rigliaco12}, we did not include 
  them because those measurements were not corrected for the photospheric 
  contribution. The number of points for  Lupus is the number of YSOs in which
  the corresponding line was detected.\\
\vspace{0.0cm}
$\star$ : standard deviation from linear fit\\
Comments in last column: 
{\bf (1)}~partially blended with  Ca\,{\sc ii}~H ;
{\bf (2)}~He\,{\sc i} + Fe\,{\sc i} blend ;
{\bf (3)}~this line is also produced in winds/outflows \citep{edwards06}; the relationship must be used with caution.;
{\bf (4)}~partially blended with  \Hep ;
{\bf (5)}~O\,{\sc i} $\lambda \lambda$ 777.194, 777.417nm doublet. ~~
$\bullet$: Suggested relations for deriving \Lacc ~from the line luminosity.}
\end{center}
\end{table*}

\section{Relationships between continuum excess luminosity 
        and emission line luminosity}
\label{correlations}
In units of \Lsun, the dynamical range of \Lacc ~ for our sample covers 
more than four orders of magnitude, while the luminosity of the line 
diagnostics discussed in the previous section spans over more than
five orders of magnitude. This allows us to investigate the
relationships between continuum excess emission and the emission in 
individual permitted lines.
 
\subsection{Continuum vs. line emission relationships}
Electronic Figures~\ref{correl1} to \ref{correl6} (on-line material only) 
show the relationships between \Lacc ~~and the luminosity of all the  
permitted emission lines discussed in Section~\ref{emisslines}. 
When available, values of \Lacc ~and \Ll ~from previous investigations 
of YSOs in Taurus  \citep[c.f.][and references therein]{HH08} 
and the $\sigma$-Ori cluster \citep[][]{rigliaco12} are overlaid.  

Linear fits of the  $\log{}$\Lacc ~vs. $\log{}$\Ll ~ relationships have been 
calculated using the package ASURV \citep[][]{feigelson85} under the IRAF 
environment. ASURV includes censoring of upper or lower limits in the fits.
In our various relationships the upper limits in the independent variable 
\Ll ~are well consistent with the trends seen in the \Lacc ~vs. \Ll ~ plots.
The results of the fits (c.f. Table~\ref{linfits}) with and without the 
inclusion of upper limits are consistent within the errors. The total number 
of points and the standard deviation of the fits are given in the fifth and 
sixth columns of Table~\ref{linfits}, respectively. The errors in the 
computed relationships also account for upper limits when included.
No fits were calculated for the Br\,8 (\brd) relation, as the number of 
upper limits is larger than the number of detections, and the relationship 
is very scattered (see electronic Figure~\ref{correl3}).

The trends in our  \Lacc ~vs. \Ll ~relationships  generally agree with 
those found in previous investigations 
\citep[][]{muzerolle98, calvet04, natta04, HH08, ingleby13}  (see
electronic Figures~\ref{correl1} to \ref{correl6}). However,
because of the different methodologies adopted to derive \Lacc ~  
(\Ha ~line profile modelling, veiling in the FUV, UV and VIS, etc.), 
systematic differences may exist at different mass regimes  
\citep[see][and next Section]{HH08}. Therefore, except for the YSOs 
in $\sigma$-Ori \citep{rigliaco12}, whose \Lacc ~and \Ll ~values  
were computed in the same way as here, we do not combine other literature 
data to derive \Lacc -- \Ll ~relationships. Note also that our 
sample comprises \Lacc ~$\le$ 1\,\Lsun, while literature data 
extend to higher accretion luminosities. 

While the accretion luminosity is well correlated with the luminosity 
of all the emission lines, the scatter in the correlations differs for 
the various lines (see standard deviation from the fits in Table~\ref{linfits}).

\subsection{Comparison with previous relationships}
\label{comparison_previous_rels}
The electronic figures \ref{correl1} to \ref{correl6} show that the 
\Lacc -- \Ll ~~relations for the Lupus YSOs are fairly consistent 
with those in previous investigations of continuum-excess in YSOs 
in Taurus \citep[c.f.][and references therein]{HH08}, and $\sigma$-Ori 
\citep[][]{rigliaco12}. The slopes and zero points of the \Lacc ~vs. \Ll 
~relations derived here are consistent within the errors with those 
reported in \citet{HH08} (see their Table~16\footnote{Note that the slope 
and zero points in the \citet{HH08} relationships are swapped in their 
Table~16.}). A comparison of the \Lacc ~values derived here from the 
slab model with the average accretion luminosity drawn from different 
line diagnostics (e.g. \Ha, \Hb, \Hg, \Hd, the He\,{\sc i} lines at 501.6nm 
and 587.6nm, the He\,{\sc ii} line at 468.5nm and the Ca\,{\sc ii}~K line) 
 as measured from the X-Shooter spectra (fluxes in electronic 
Tables~\ref{tab:fluxes_EWs_Hae}-\ref{tab:fluxes_EWs_NaI}) and using the 
\citet{HH08} relationships, leads to a rms difference $<$0.3\,dex. 
This is significantly less than the error drawn from the application 
of single-diagnostic relationships \citep[see also][]{rigliaco12}. 
We also note that our relationships for the \Ha, \Hb, and Ca\,{\sc ii}~K 
lines are practically identical to those of \citet{ingleby13}, which were 
derived by fitting UV and optical spectra with multiple accretion 
components.

Our \Lacc -$L_{\rm Pa\beta}$ ~and  \Lacc -$L_{\rm Br\gamma}$ ~relations are 
also similar to those in previous works by \citet{muzerolle98}, \citet{natta04}, 
\citet{calvet00} and \citet{calvet04}, but extend to much lower values of \Lacc, 
toward the very low-mass regime (see electronic Figure~\ref{correl3}). 
As mentioned earlier, however, systematic differences may arise due to the 
different methodologies to derive \Lacc. For instance, the difference in 
$\log{L_{\rm acc}}$ when using our \brg ~relationship and \citet{muzerolle98}'s 
may be up to $\sim$0.5\,dex for a typical 0.5\,\Msun ~T~Tauri star with  
$L_{\rm Br\gamma}$~$\sim$10$^{-3}$\Lsun.

\subsection{Impact of chromospheric emission on \Lacc ~estimates}
\label{chrom_contrib}

Previous studies \citep[][]{ingleby11,rigliaco12,manara13} have stressed 
the impact of chromospheric emission on \Lacc ~at low levels of 
accretion. Our estimates of \Lacc ~ are not influenced by chromospheric
line emission, as they are derived from the continuum excess 
emission. The typical continuum emission of chromospheric origin, if present,
is automatically corrected for by using as templates the Class-III 
stars, rather than field dwarfs.

\begin{figure}[h]
\advance\leftskip-0.6cm
\includegraphics[width=10.0cm, height=11.5cm]{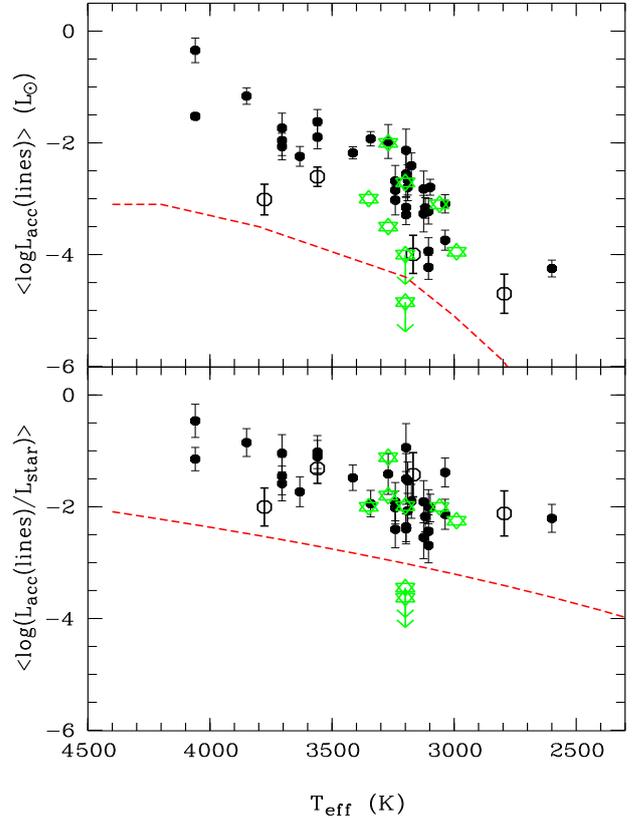}{\centering}
\caption{The average accretion luminosity  $\langle$log\Lacc(lines)$\rangle$ derived 
       from 15 emission lines  as described in the text (upper panel) 
       and the $\langle$log\Lacc(lines)/\Lstar $\rangle$ ratio (lower panel) 
       as a function of effective temperature for Lupus and $\sigma$-Ori 
       (open stars) YSOs. Plotting symbols for the Lupus YSOs are as in 
       Figure~\ref{HRD}. The dashed lines in both panels mark the locus 
       below which chromospheric emission is important in comparison with \Lacc.
       The vertical error bars represent the standard
       deviation over the average of 15 emission line diagnostics.
    \label{Lacc_Teff}}
\end{figure}

In order to investigate the possible effects of chromospheric line emission 
in our sample, we derived the accretion luminosity, \Lacc(lines), 
using emission line diagnostics and the \Lacc--\Ll ~relations in 
Table~\ref{linfits}. 
We calculate $\langle$log\Lacc(lines)$\rangle$ as the average over 
fifteen diagnostics discussed in \citet{manara13}. 
In Figure~\ref{Lacc_Teff} the $\langle$log\Lacc(lines)$\rangle$ values 
and the $\langle$log\Lacc(lines)/\Lstar $\rangle$ ratio are plotted as 
a function of \Teff. The dashed lines in the figure show the level 
of chromospheric noise as determined by \citet{manara13}. 
Those lines represent the locus below which the contribution of 
chromospheric emission starts to be important in comparison with 
energy losses due to accretion. 
The accretion level of all the Lupus YSOs studied here is well above the 
chromospheric noise. Therefore, we conclude that the chromospheric
contribution to \Lacc ~is influential, even at the lowest values of \Lacc. 
Our \Lacc--\Ll ~relationships are hence calculated for \Lacc ~values
well above the chromospheric threshold.
The two objects of $\sigma$-Ori indicated with upper limits correspond 
to SO\,587 and SO\,1266. For these two objects the fraction of luminosity 
in the Balmer lines with respect to the upper limit in \Lacc ~is higher 
than one. \citet{rigliaco11a, rigliaco12} show that in SO\,587 the strong 
permitted lines probably originate in a photo-evaporation wind, while in
SO\,1266 they are dominated by chromospheric emission. We do not consider 
these two objects in the following plots and analysis.

\section{Accretion properties}
\label{accprop}

\subsection{Accretion luminosity versus YSO luminosity}
\label{Lacc_corr}
Previous investigations \citep[e.g.][ and references therein]{natta06, rigliaco11a} 
have shown that \Lacc ~ and stellar luminosity in Class-II YSOs are correlated, 
although with significant scatter at a given YSO luminosity. \citet{clarke06} 
pointed out that the distribution of points in the \Lacc -- \Lstar ~plane more or 
less fills a region which is bounded by the \Lacc = \Lstar ~relation at high 
\Lacc, but which is dominated by detection biases at low values of \Lacc, roughly 
following a power-law \Lacc~$\propto$~\Lstar$^{1.6}$.
This agrees with the relationships they derive from the data (detections and
upper limits) in \citet{natta06}. 
\citet{tilling08} presented simplified stellar evolution calculations for stars subject 
to time-dependent accretion history, and derived evolutionary tracks on the \Lacc -- \Lstar 
~diagramme for a variety of fractional disc masses, $f_{disc}\equiv M_{disc}$/\Mstar, 
and YSO masses. Using the \citet{DM97} models, they assumed that the mass accretion 
rate declines with time as $t^{-\eta}$, with $\eta=1.5$.

\begin{figure}[h]
 \advance\leftskip-0.3cm
 \includegraphics[width=9.5cm, height=7.5cm]{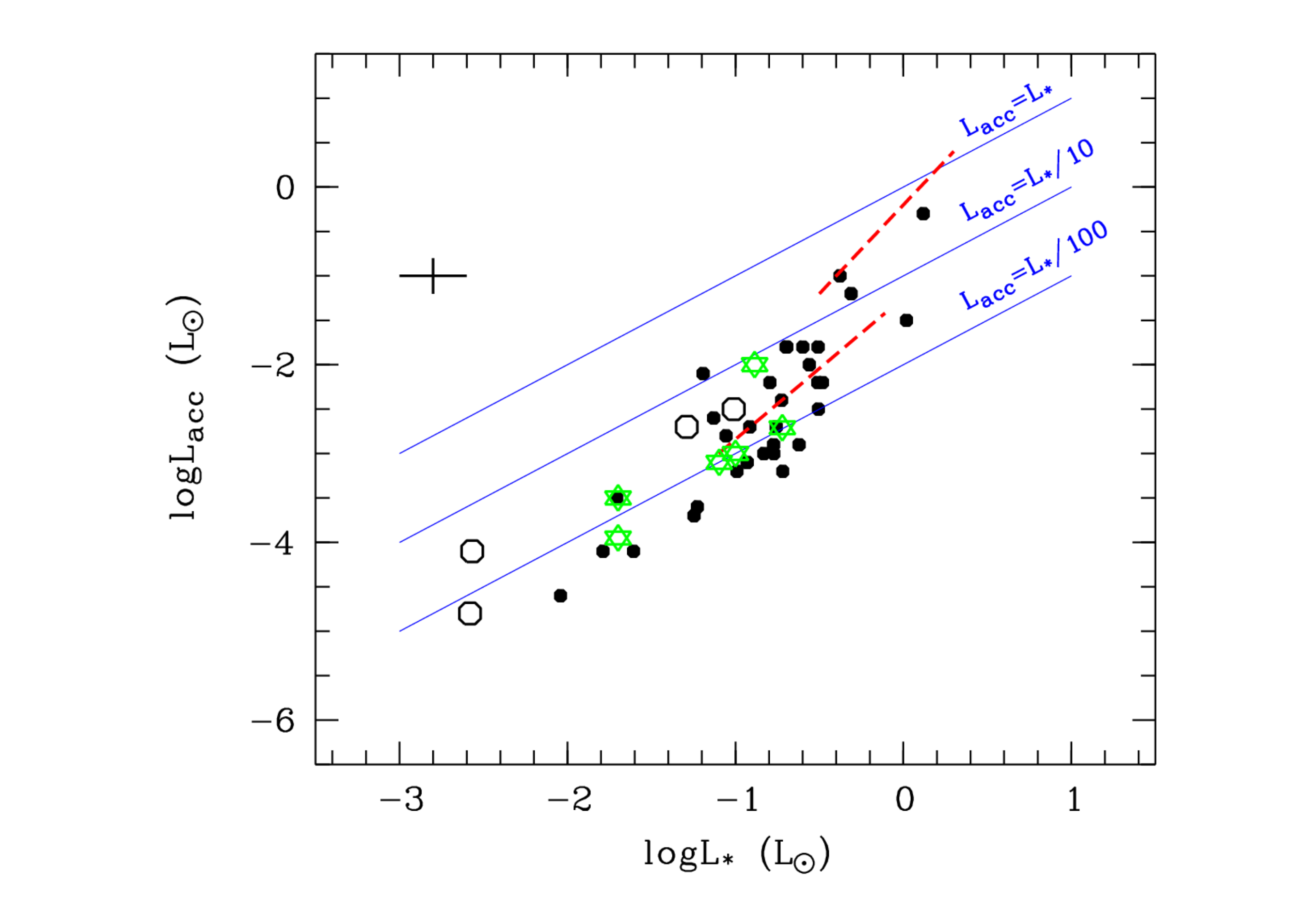}{\centering}
  \caption{Accretion luminosity as a function of stellar luminosity for
       Lupus and $\sigma$-Ori YSOs. Symbols are as in Figure~\ref{Lacc_Teff}.
       The two Lupus YSOs with the lowest luminosities
       are Par-Lup3-4 and Lup\,706. The continuous lines represent 
       the three \Lacc ~vs. \Lstar ~relations as labelled. 
       Average error bars are shown in the upper left.
       The red dashed lines are the model tracks by \citet{tilling08}
       as follows: the upper line is for 1.0\Msun YSOs and $f_{disc}=$0.2, 
       while the lower one is for 0.4\Msun ~YSOs with $f_{disc}=$0.014.
     \label{Lacc_vs_Lstar}}
\end{figure}

As shown in Figure~\ref{Lacc_vs_Lstar}, the Lupus YSOs also fall below the \Lacc=\Lstar 
~boundary, with a small number of objects between 0.1 and 1\,\Lsun, and many with 
\Lacc/\Lstar $<$0.01. The data points are apparently less scattered than those of 
previous samples.
Our sample lacks YSOs with \Lacc$>$~1\Lsun, this being most likely the reason 
why we do not populate the region between \Lacc=0.1\Lstar ~and \Lacc=\Lstar ~on 
the diagramme. The slope for the data in Figure~\ref{Lacc_vs_Lstar} 
is steeper than the \Lacc/\Lstar~$=$constant lines, more or less following 
the slope of the  \citet{tilling08} tracks.
A linear fit to the data yields \Lacc~$\propto$~\Lstar$^{1.7}$, i.e. 
very similar to the claim by \citet{clarke06} for the lower envelope of 
the \Lacc -- \Lstar ~relation, but based on detections only. Note that 
\citet{manara12} find an almost identical power-law (\Lacc~$\propto$~\Lstar$^{1.68}$), 
using a complete sample in the Orion Nebula Cluster, but with \Lacc 
~determinations based on deep photometry.

According to the \citet{tilling08} model and by a qualitative comparison 
between their evolutionary tracks and our data set on the \Lacc--\Lstar 
~diagramme (Figure~\ref{Lacc_vs_Lstar}), one would expect the discs of the 
lowest mass YSOs to have masses lower than 0.014 \Mstar. However, within 
the uncertainties of the measurements by \citet{ricci10}, there is no evidence 
of a scaling between the disc mass and the stellar mass, or the mass accretion 
rate. Note also that a recent compilation for a wide range of masses suggests 
that the fractional disc mass is compatible with an uniform distribution 
around the value $f_{disc}\approx$0.01 \citep[][]{olofsson13}.
 
\subsection{Accretion rate versus mass}
\label{Macc_corr}

Previous investigations \citep[][and references therein]{muzerolle03,mohanty05,
natta06,HH08,rigliaco11a,antoniucci11, biazzo12}, have found that 
\Macc~ goes roughly as the square of \Mstar ~although with a 
significant scatter (up to 3\,dex) in \Macc ~for a given YSO mass. 

\begin{figure}[h]
\resizebox{1.0\hsize}{!}{\includegraphics[]{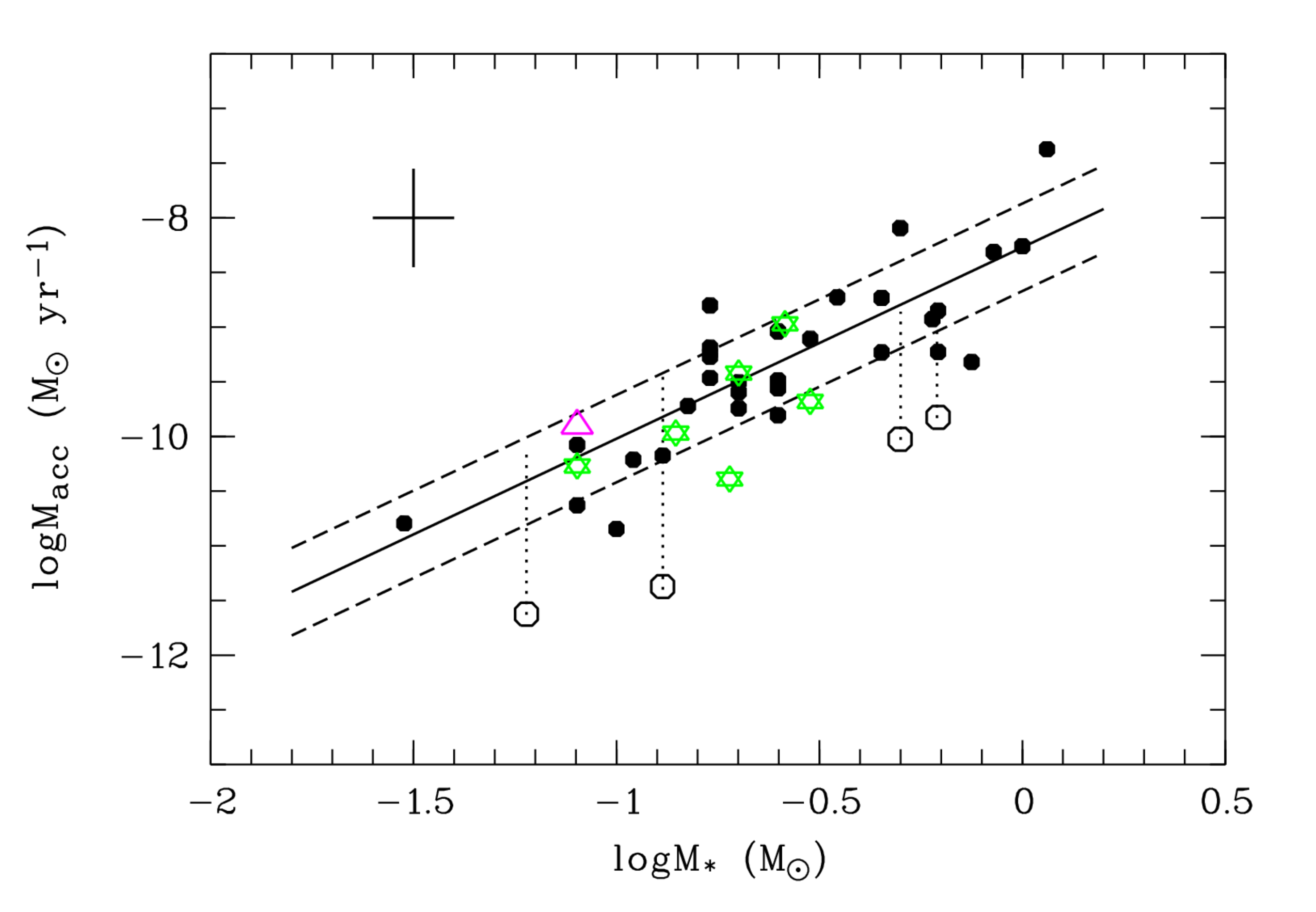}}
\caption{Mass accretion rate \Macc ~as a function of mass. Symbols are as in 
 Figure~\ref{Lacc_Teff}. The vertical shifts in \Macc ~for the low-luminosity 
 YSOs (open circles), after correction for obscuration as explained in 
 Section~\ref{subluminous}, are shown by the dotted lines.
 The big triangle represents the young brown dwarf FU~Tau\,A, for which the 
 \Macc ~and \Mstar ~values were derived by \citet{stelzer13}. The  
 continuous line represents the linear fit of Equation~\ref{fit_without_slysos},
 i.e. without including the sub-luminous objects. The dashed lines represent
 the 1\,$\sigma$ deviation from the fit. 
 Average error bars are shown in the upper left corner.
     \label{Macc_Mstar}}
\end{figure}

Figure~\ref{Macc_Mstar} ~shows the \Macc ~versus \Mstar ~diagramme for the Lupus
YSOs studied here. The position of the $\sigma$-Ori stars in this plot
is consistent with Lupus and the young brown dwarf FU~Tau\,A,
also investigated with X-Shooter \citep{stelzer13}, follows the trend as 
well. The scatter is significantly increased by the four sub-luminous 
YSOs. A linear fit to the complete Lupus sample yields:

 \begin{equation}
 \log{\dot M_{\rm acc}}  = 1.89 (\pm0.26) \cdot \log{M_{\star}}  - 8.35 (\pm0.18) 
 \label{fit_with_slysos}
 \end{equation} 

\noindent 
with a standard deviation of 0.6. The scatter decreases if the sub-luminous 
objects are excluded from the fit, yielding  

 \begin{equation}
 \log{\dot M_{\rm acc}}  = 1.81 (\pm0.20) \cdot \log{M_{\star}}  - 8.25 (\pm0.14) 
 \label{fit_without_slysos}
 \end{equation} 

\noindent
with a standard deviation of 0.4. The slope and zero point of the \Macc -- \Mstar
~fit do not change significantly in either cases because the sub-luminous 
objects represent only 11\% of our sample. It is thus reasonable to conclude that 
for our sample \Macc~$\propto$~\Mstar$^{1.8(\pm0.2)}$, which is in agreement
with previous studies \citep{natta06, muzerolle05, HH08, rigliaco11a, antoniucci11, 
biazzo12, manara12}, but inconsistent with the results by \citet{fang09} 
(\Macc~$\propto$~\Mstar$^{3}$) for their sub-solar mass sample in the 
Lynds~1630N and 1641 clouds in Orion. This inconsistency is more likely to 
be related to the different methodologies used to derive both \Macc ~ and \Mstar ~
rather than to different environmental conditions.

\section{Discussion}
\label{discussion}

\subsection{Emission lines as tracers of accretion}
Fitting the UV excess emission, and in general continuum excess emission,
is the most reliable and accurate method to derive accretion luminosity in 
low-extinction YSOs \citep{muzerolle03, HH08, rigliaco12, ingleby13}. 
In the absence of UV spectra, \Lacc, hence \Macc, can be calculated 
using emission line diagnostics. While the relationships by \citet{HH08} 
are based on simultaneous low-resolution UV and optical spectroscopy, ours 
encompass and extend simultaneous observations of the diagnostics from 
$\sim$330\,nm up to $\sim$2500\,nm, at intermediate spectral 
resolution.

Being based on almost twice the number of points as in
previous works, the \Lacc--\Ll ~relations computed here have 
in general a lower dispersion than those found in the literature applying 
similar methodologies ~\citep[e.g.][]{HH08, rigliaco12, ingleby13}. 
As in those works, each point in the relationships represents 
an instantaneous snapshot of \Lacc ~and \Ll. However, results of 
temporal monitoring of several YSOs indicate variability in 
optically thick line fluxes, without significant changes in 
the corresponding continuum accretion rate \citep[e.g.][]{gahm08, herczeg09},
so that some dispersion may still arise from variability even
when the observations are simultaneous. 
Long-term spectrophotometric monitoring of YSOs over a range of
masses is still required to shed light on the magnitude of this
effect.

The strongest line in optical spectra of YSOs is the \Ha ~line. 
Nevertheless, similarly as for the Taurus sample \citep{HH08}, the 
\Lacc--\LHa ~relation for the Lupus YSOs is the most scattered 
among the Balmer lines relationships (see Table~\ref{linfits}). 
Not surprisingly, as it is well known that several other processes 
(e.g. outflows, hot spots, chromospheric activity, complex magnetic 
field topology and geometry, stellar rotation) besides accretion may 
contribute to the strength of the line. All these processes have an 
important impact on the line profile, in particular on its width. 
Previous investigations have used the full width of the \Ha ~line at 10\% 
of the line peak, W\Ha(10\%) expressed in km~s$^{-1}$, to investigate 
accretion \citep[see][and references therein]{whitebasri03, natta04}.
Since W\Ha(10\%) is easily gathered from optical spectra, many authors 
have used it to estimate \Macc. Nevertheless, as discussed by many 
authors, the relationship has a very large scatter and its use is 
discouraged when reliable measurements of the line luminosity are 
possible.
In \citet{natta04} the \Macc ~values used for low-mass stars ($<$0.3\Msun) 
are based on modelling of the \Ha ~line profile 
\citep[e.g.][]{muzerolle01,muzerolle03}, whereas those for higher mass 
stars are mainly based on spectral veiling measurements \citep[e.g.][]{gullbring98}.

\begin{figure}[h]
\resizebox{1.0\hsize}{!}{\includegraphics[]{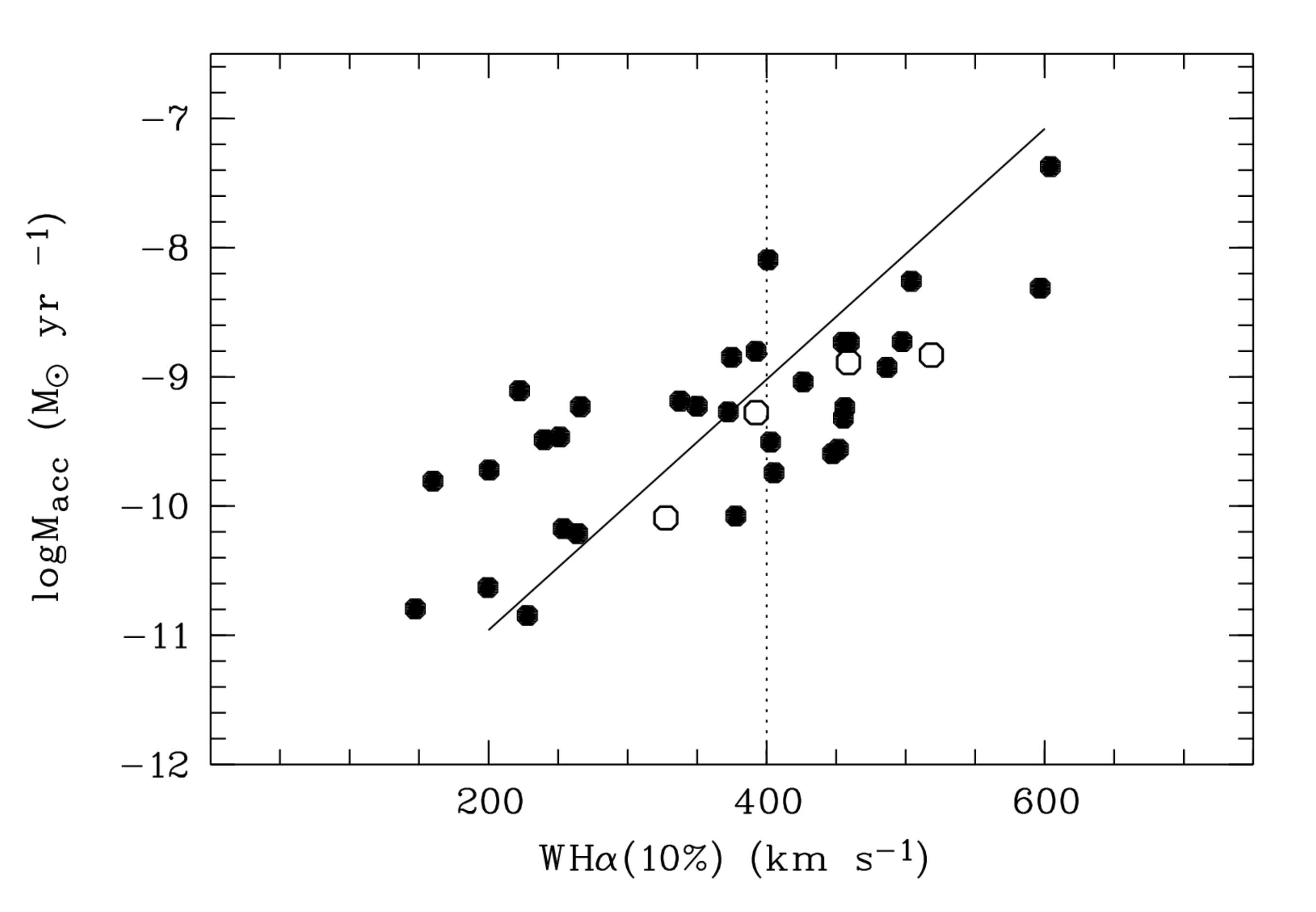}}
\caption{The \Macc ~vs. W\Ha(10\%) plot for the Lupus  YSOs. 
       Symbols are as in Figure~\ref{HRD}. The W\Ha(10\%) measures are 
       reported in Table~\ref{accretion}. The \Macc ~values of the sub-luminous
       objects were corrected, as discussed in Section~\ref{subluminous}. 
       The continuous line represents the relationship 
        $ \log{(\dot M_{\rm acc})} = 9.7 \times 10^{-3} \cdot {\rm WH\alpha(10\%)}~- 12.89$ 
       by  \citep{natta04}.
       \label{Macc_Wha10pc}}
\end{figure}

In Figure~\ref{Macc_Wha10pc} the \Macc--W\Ha(10\%) scatter plot for 
the Lupus sample is compared with the \citet{natta04} relationship, 
represented with a continuous line. 
For objects with W\Ha(10\%)\,$<$\,400\,km s$^{-1}$, the \citet{natta04} 
relation tends to underestimate \Macc ~with respect to our
determinations from continuum-excess modelling by 
$\sim$0.6\,dex in $\log{\dot{M}_{\rm acc}}$, on the average, but the 
differences may be up to about an order of magnitude. Similar 
differences were seen in the \citet{HH08}'s Taurus sample and 
in the \citet[][see their Figure~45]{fang13} sample in the L1641 cloud.
For objects with W\Ha(10\%)\,$>$\,400\,km s$^{-1}$, the slope of
the Lupus correlation seems rather consistent with the 
\citet{natta04} relationship, but the $\log{\dot M_{\rm acc}}$ 
values are systematically lower by  $\sim$0.5-0.7\,dex.
According to the \citet{natta04} relationship, YSOs with W\Ha(10\%)\,$<$\,400\,km s$^{-1}$ 
~should have \Macc\,$<\,10^{-9}$\,\Msun\,${\rm yr^{-1}}$, and from 
Figure~\ref{Macc_Mstar}, these correspond to objects with \Mstar\,$<$\,0.3\,\Msun, 
i.e. those for which the \Macc's come from the modelling of the \Ha ~line 
profile in \citet{natta04}. Therefore, the differences we observe between 
the \Macc's calculated from continuum excess and those derived from
the \citet{natta04} relationship, can be attributed to the different 
methodologies adopted to measure \Macc. 
In conclusion, although the \Ha ~line is the strongest in optical spectra 
of YSOs, special attention should be paid when using it to estimate 
mass accretion rates from the line width. 

The least scattered \Lacc--\Ll ~relations are those of the Balmer lines 
with n$>$3, the \brg ~line and the $\ion{He}{i}$ lines. The \pab ~and 
the \brg ~relations are recommended because these lines are the least
affected from chromospheric emission. In contrast, the Ca\,{\sc ii}~IRT
relations are the most scattered. Even after correction for the 
photospheric absorption, the  Ca\,{\sc ii}~IRT relations appear more 
scattered than any of the Balmer lines, mainly because in some YSOs 
with very broad lines (e.g. Sz\,83, Sz\,72, Sz\,69 and Lup\,713)
the blending with the Pa\,13, Pa\,15 and Pa\,16 lines contributes 
to the integrated flux, increasing the line luminosity. Without correction 
for the photospheric contribution the scatter in the CaII~IRT relations 
would be even larger. As in \citet{mohanty05}, we also investigated relations 
between the surface flux of the Ca\,{\sc ii}~IRT lines, $F_{\rm \ion{Ca}{ii}~IRT}$, 
and \Macc, but the scatter remains rather large, on the order of 0.6\,dex. 
In addition to the problem of blending with the Paschen lines, uncertainties 
in stellar radius and distance make the surface flux relations very scattered 
and uncertain. Moreover, no evidence is found in our data for a two-mode 
$F_{\rm \ion{Ca}{ii}~IRT}$-\Macc ~relation depending on the mass range, 
in contrast to the suggestion by \citet{mohanty05}.
This casts some doubts on whether the two \citet{mohanty05} relations 
may be produced by the different methodologies used to calculate \Macc.
In fact, the \Macc ~values for six of the eight objects used by \citet{mohanty05} 
to derive the "low-mass" relation come from modelling of the \Ha ~line profile, 
while their  "high-mass" relation is entirely based on veiling estimates. 
Note that \citet{mohanty05} briefly mention this as a possible reason for 
their different \Macc's at their low and high mass regimes.

Another important aspect to be considered when determining accretion rates 
from emission lines and \Lacc--\Ll ~relationships regards the contribution of 
chromospheric emission. The relative importance of (hydrogen) line emission with 
respect to \Lacc ~is higher for small \Lacc ~values, and chromospheric emission 
may be the dominant process in the lines. Based on the luminosity of several 
chromospheric emission lines in the Class-III templates, \citet{manara13} 
determined a threshold below which chromospheric emission dominates line 
luminosities.  The threshold depends on YSO effective temperature and age. 
Line luminosities yielding \Lacc ~values below or just above that threshold 
should not be considered as accretion diagnostics.

Finally, as in \citet{rigliaco12}, we stress that the average \Lacc ~derived 
from several lines, measured simultaneously, has a much reduced error.

\subsection{Discrepancies with magnetospheric accretion models}

The most extensive calculations of line emission in the context of
magnetospheric accretion models remain those of \citet{muzerolle01}; 
they were performed for the stellar parameters typical of T~Tauri 
stars (\Mstar$=0.5$\,\Msun, \Rstar$=2$\,\Rsun), mass accretion rates between 
10$^{-6}$ and 10$^{-9}$\,\Msun/yr, different disc truncation radii and a wide 
range of gas temperatures.  Their predictions of the dependence of the 
line luminosity and line ratios are at odds with the observed trends: 
namely, while the 
observed line luminosities increase roughly linearly with \Macc, the models, 
which include constraints on the gas temperature,  predict that the line 
luminosity will stay constant above \Macc~$\sim$~10$^{-8}$\,\Msun/yr. 
Muzerolle et al. propose that the observations can only be understood if 
both \Macc ~ and the line fluxes are controlled by the size of the accretion 
flow, rather than reflecting the physical conditions of the accreting gas. 
In other words, if the flow occurs along discrete separate magnetic flux 
tubes, the gas physical conditions in each tube are similar, but the number 
of them can increase by an order of magnitude from object to object.
We confirm the trend and show that it extends over a very large
range of \Macc.

The similarity of the physical conditions in the accreting gas for
objects with very different \Macc ~is suggested by a number of other 
properties as well. One is the fact that the hydrogen line ratios remain 
quite constant over a large range of \Macc, and that there is no indication 
that optical depth effects play a significant role. See, e.g., the \pab/\brg 
~ratio in Figure~\ref{PabBrg_Macc}: 
over a mass accretion range of 6 orders of magnitude, this ratio is 
constant within the uncertainties in the range  3--5, with no evidence 
for the lines to become optically thin at low \Macc ~\citep[e.g.][]{muzerolle01}. 
We note that in our sample there 
are no objects with \pab/\brg ~ $\sim$2, as found in some $\rho$-Oph 
brown dwarfs by \citet{gatti06}. It is possible that the physical conditions 
in the younger and brighter BDs in $\rho$-Oph differ from those in Lupus;
however, it would be worthwhile to confirm the \citet{gatti06} results using
spectra of the same quality of the ones used here. 

\begin{figure}[h]
\resizebox{1.0\hsize}{!}{\includegraphics[]{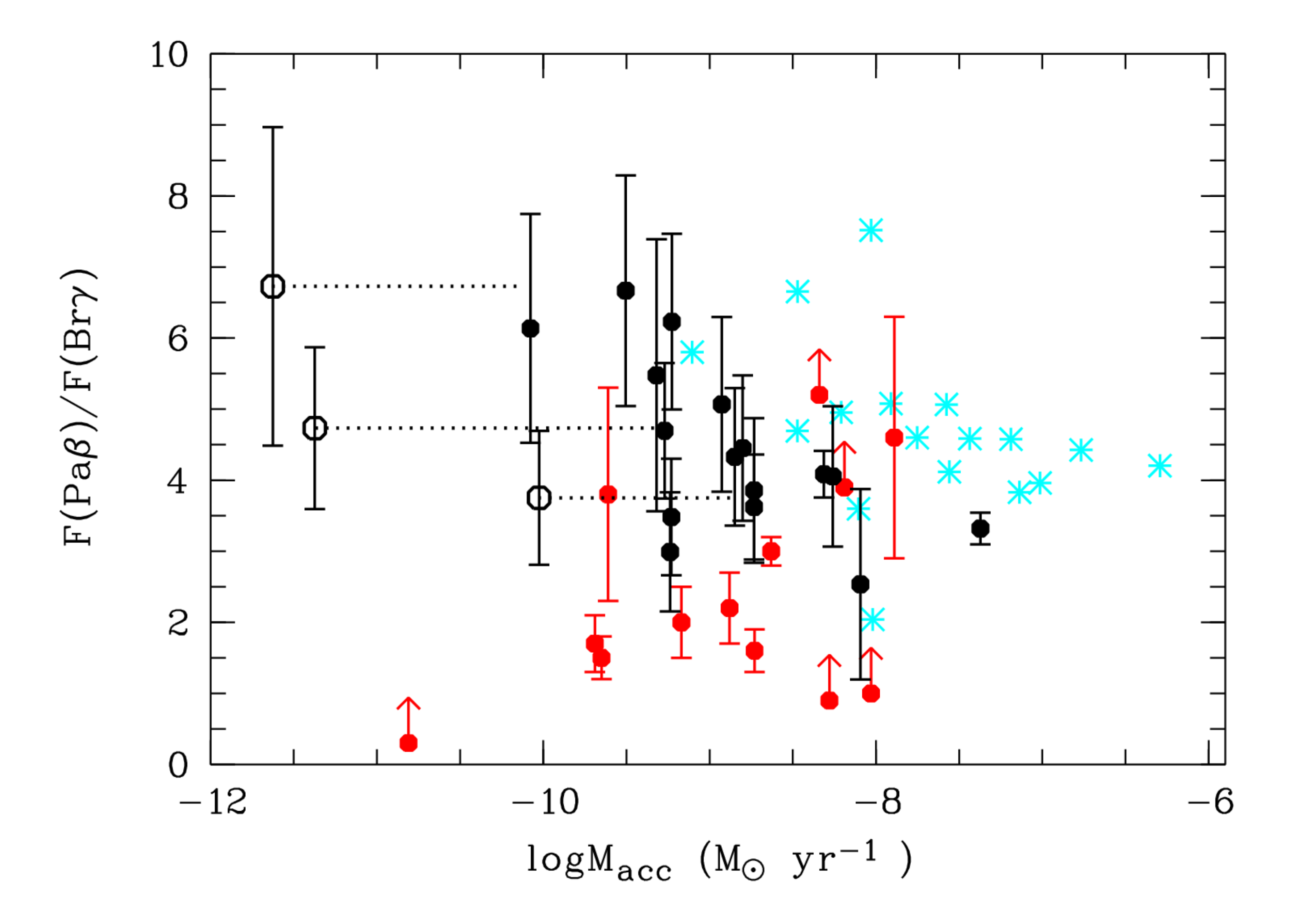}}
\caption{The \pab/\brg ~ratio for the 19 YSOs in which both lines are detected
(black dots and open circles).
The shift in \Macc ~to the right for the low-luminosity YSOs (open symbols), 
after correction for obscuration as explained in Section~\ref{subluminous}, 
is represented with the horizontal dotted lines. Objects in the
$\rho$-Oph cloud studied by \citet{gatti06} are represented with red dots, while
Taurus T Tauri stars from \citet{muzerolle01} are represented with cyan asterisks. 
	\label{PabBrg_Macc}}
\end{figure}

Another indication for similar physical conditions in the accreting gas 
comes from considering the relations between \Ll ~
and \Lacc ~derived in Section~\ref{correlations}. To zero order they are 
linear, with slopes varying between 0.99 and 1.18 for the hydrogen Balmer, 
Paschen and Brackett lines, all equal within the errors; similarly for 
the He\,{\sc i} lines, with slopes in the range 0.90--1.16. This means 
that, over a range of \Lacc ~of 5 orders of magnitude, a similar fraction 
of the accretion energy is emitted in each line, independently of their 
excitation potential and optical depth. Even if these fractions are in most 
cases very small (less than 1\% in the case of the hydrogen lines and 
less than 0.1\% for the He\,{\sc i} lines, with only \Ha ~ reaching  
$\sim$5\% of \Lacc), they are surprisingly stable across the examined 
stellar  parameters.  This, again, suggests that the physical conditions 
of the accreting gas are very similar in all objects, and that only the 
covering factor changes from object to object. 

To our knowledge, the suggestion by \citet{muzerolle01} that it is the 
geometry of the accretion flow  that controls the  rate at which the disc 
material accretes onto the central star has not been followed up by 
detailed models which include, for example, complex magnetic field 
configurations. 

\subsection{On the accretion properties of Lupus YSOs}

Using simplified stellar evolution calculations for stars subject to
time-dependent accretion, \citet{tilling08} conclude that the \Lacc--\Lstar 
~plane presents two main features namely, the 
\Lacc$\approx$\Lstar ~upper boundary and diagonal tracks, which can 
be ascribed to accretion rates as the stars descend the Hayashi tracks.
They show that the slope of such tracks on the \Lacc--\Lstar ~plane
is related to the power-law index of the \Macc ~vs. age relationship. 
Our sample is consistent with a \Lacc $\propto$ \Lstar$^{1.7}$ law, 
and under the \citet{tilling08} prescription, where 
\Lacc $\propto$ \Lstar$^{(\eta-0.3)/0.7}$ (see their equation 12), 
this would mean $\eta=$1.5, leading to \Macc~$\propto t^{-1.5}$.

A detailed observational study of the evolution of accretion 
requires complete and homogeneous samples of YSOs, with a
diversity of ages, and well determined \Macc ~\citep[c.f.][]{sciagui10}.
\citet{gatti08} find that the accretion rates are significantly 
lower in $\sigma$-Ori than in younger regions, such as $\rho$-Oph, 
consistently with viscous disc evolution. \citet{manara12} and 
\citet{rigliaco11a} discussed the evolution of accretion as a 
function of stellar mass in the ONC and the $\sigma$-Ori cluster, 
respectively.
The age range of our sample is too narrow to investigate in detail the
\Macc~$\propto t^{-\eta}$ relation, but selecting the objects in the 
mass range 0.4\Msun ~to 0.8\Msun, and excluding sub-luminous objects, 
an attempt to fit the relation yields $\eta = 1.3\pm0.3$, albeit with a 
large dispersion.
The average \Macc ~for objects in the same mass range is 
1.3$(^{+1.9}_{-0.8})\times$10$^{-9}$\,\Msun~yr$^{-1}$, which is consistent 
with the calculations of the evolution of viscous discs for 0.5\,\Msun ~YSOs
by \citet{hartmann98} at 3\,Myr (see their Figure~3). 
To investigate empirically the \Macc ~vs. age relationship, \citet{caratti12}  
normalised the \Macc ~to \Mstar$^2$ in order to account for the 
dependence of \Macc ~on mass. Following this approach for the Lupus 
YSOs, some number statistics is gained by including in the fit YSOs with 
mass up to $\sim$1\,\Msun. Although the dispersion is slightly 
reduced, the result is practically identical.

Since the stellar mass undergoes negligible changes during the Class-II 
phase, the \Macc ~vs. \Mstar ~relation represents a diagnostic 
tool for the evolution of \Macc ~\citep{clarke06}. From the theoretical
point of view, it has been suggested by \citet{vorobyov08} that the 
\Macc~$\propto$~\Mstar$^2$ relationship can be explained on the basis of 
self-regulated accretion by gravitational torques in self-gravitating discs. 
These authors argue that the relationship can be better described as 
a double power-law, with the break occurring at \Mstar$\approx$0.25\,\Msun, 
rather than a function with a single exponent \citep{vorobyov09}.
Interestingly, such kink occurs close to the value of mass where the 
techniques to calculate \Macc ~differ. Thus, they  also point out the different 
techniques used to determine \Macc ~as a possible cause for the apparent 
bi-modal power-law.

The power-law index of the \Macc--\Mstar ~relation for Lupus is also
$\sim$2, but the scatter is smaller than in previous data sets 
\citep[c.f. the standard deviation for the Lupus fit is a factor 2 less 
than for the Taurus sample in][]{HH08}. As pointed out in Section~\ref{Sec:spectra},
our sample represents about 50\% of the complete population of Class-II
YSOs in the Lupus-I and Lupus-III clouds. It is thus rather unlikely that 
the large range of \Macc ~($>$ 2\,dex) ~at a given mass observed in 
other data sets will show-up also in Lupus, if a more complete sample is 
investigated. However, it would be worthwhile to study the YSOs more 
massive than those in our sample, using spectra of the same quality as 
here. In addition, although our number statistics is low, there is no 
evidence for a double power-law in our sample, supporting the conclusion 
that the apparent bi-modal relations suggested in the literature between 
\Macc ~and other YSOs properties are probably the result of mixing up
\Macc's derived with different methods.

\subsection{The sub-luminous YSOs: evolution or edge-on discs ?}
\label{subluminous}
The underluminosity and strong emission lines in the optical spectrum of 
Par-Lup3-4 were first discussed by \citet{comeron03}, who found difficult 
explaining their observations in terms of either a photospheric continuum 
suppressed by an edge-on disc or an embedded source seen in scattered 
light. They favor instead the scenario in which the PMS evolution is 
significantly modified by the accretion process, as suggested in 
\citet[][]{baraffe10}. 
The analysis of the SED resulting in a disc inclination angle of 
$\sim$81$^\circ$ \citep{huelamo10} and the tiny difference in velocity 
between the red-shifted and blue-shifted components of the outflow 
\citep{bacciotti11, whelan13} provide evidence that we are seeing
the Par-Lup3-4 disc  almost edge-on. However, as in \citet{comeron03}, 
strong Ca\,{\sc ii}~IRT lines are also detected in our X-Shooter spectrum, 
meaning that not all the emission from the accretion-flows in the inner 
regions is suppressed by the optically thick disc.
  
\citet{comeron03} also pointed out Sz\,106 and Sz\,113 as underluminous, 
although not so extreme. Underluminosity is confirmed here for Sz\,106, 
while no evidence for such phenomenon is observed in 
Sz\,113\footnote{We noted some differences between our spectra
of Sz\,106 and Sz\,113 and those published by \citet{comeron03}. 
For instance, the \Ha ~equivalent width we measure for Sz\,106 is 
11.6\,\AA ~while \citet{comeron03} claim more than 100\,\AA. Also,  
lots of forbidden lines are seen in our spectrum of Sz\,113, while 
\citet{comeron03} detect none. Note, however, that the \citet{comeron03} 
spectra have much lower resolution than ours.}. We also find 
Lup\,706 and Sz\,123B underluminous, but not as much as Par-Lup3-4. 
  
As examined by \citet{baraffe10}, episodic strong accretion during PMS 
evolution of low-mass stars  produces objects  with smaller radius, 
higher central temperature, and lower luminosity compared to the non-accreting 
counterpart of the same mass and age, resulting in low-luminosity
objects. Should the PMS evolution of the 4 sub-luminous YSOs have been 
significantly altered by episodic strong accretion, their surface 
gravity should also appear larger than for objects of the same mass 
and age. One way to test this possibility is through independent estimates 
of the surface gravity.
We have done such estimates  while correcting the photospheric 
contribution of the  $\ion{Na}{i}$~D and  Ca\,{\sc ii}~IRT lines 
(see Section~\ref{linefluxes}), when using the gravity sensitive 
Na\,{\sc i} ($\lambda\lambda$818.3\,nm/819.48\,nm) and the 
K\,{\sc i} ($\lambda\lambda$766.50\,nm/769.90\,nm) doublets
\citep[see][for the method]{stelzer13b}, 
for which veiling is negligible in these objects.
The results of the analysis yield $\log{g}$ values (3.5, 3.7, 4.3, 
and 3.9 for Par-Lup3-4, Lup\,706, Sz\,106 and  Sz\,123B, respectively)
that are very similar as for the other YSOs in the sample, and lower 
than those expected for apparently more evolved objects. Hence, we 
reject the hypothesis that underluminosity of these objects is the 
result of modified PMS evolution by accretion episodes.

The scatter in several plots shown throughout the previous sections
is significantly reduced when \Lacc ~is normalised to the YSOs 
luminosity; the \Lacc/\Lstar ~values of the sub-luminous objects are
consistent with those for the other YSOs, suggesting that some process 
is in act that affects \Lacc ~ and \Lstar ~in the same way. As in the case 
of Par-Lup3-4, such a process may be gray circumstellar extinction. 
In the magnetospheric 
accretion model, the line and accretion luminosities originate in 
the inner parts of the star-disc system. Thus, under the assumption 
of gray circumstellar extinction due to disc obscuration both 
\Lacc ~and \Ll, as well as the 
YSO luminosity are dimmed by the same amount, which depends on disc 
inclination angle, disc flaring, and dust opacity. The effect thus 
cancels out when considering the \Lacc/\Lstar ~ratio and has no consequences
on the \Lacc ~vs. \Ll ~relationships, but may have an important impact
on the \Macc ~determinations for the most obscured objects. 

Under the hypothesis of a gray obscuration, it can be inferred from 
Figure~\ref{HRD} that the ''obscuration factors'' by which the luminosity 
of the four objects should be multiplied to fit the average trend of
the mass-luminosity relationship are $\sim$25, 10, 6, and 4 for Par-Lup3-4, 
Lup706, Sz\,123B, and Sz\,106, respectively. At a fixed mass, Equation~\ref{Macc} 
implies \Macc~$\propto$~\Lacc$\cdot$\Lstar$^{0.5}$, because the YSO radius 
scales with the square root of the luminosity. Assuming that the obscuration 
factor suppresses both \Lacc ~and \Lstar ~by the same amount, the \Macc 
~values for the low-luminosity YSOs  can be corrected as 
\Macc$(corrected)=(obscuration factor)^{1.5} \cdot$\Macc. 
When the obscuration factors are applied to \Lstar, \Lacc ~and \Macc, the four 
underluminous objects behave exactly as the other YSOs in the various plots.  
Detailed analysis of the spectral energy distribution of these YSOs is required 
to constrain the geometry of their disc.

\section{Summary and conclusions}
\label{summary}
Our study with X-Shooter@VLT is the first presenting UV-excess measures 
of accretion luminosity, simultaneously to intermediate-resolution spectroscopy
of a large number of emission line diagnostics, from $\sim$330\,nm to 2500\,nm, 
in a significant and homogeneous sample of very low-mass young stellar and
sub-stellar objects in Lupus. 
The quality of the spectra and the accuracy in flux calibration, both 
in the continuum and the lines, allowed the characterisation of the sample
and the computation of \Lacc ~vs. \Ll ~ relations for an unprecedentedly 
large number (39) of emission line diagnostics, as well as to study the 
accretion properties of the sample. The main results are summarised here.

The accretion emission in our sample is dominated by continuum emission
in the Balmer and Paschen continuum. 
For the vast majority of the YSOs the integrated  line luminosity 
totals less than one third of \Lacc. Most of the line luminosity is
due to Balmer lines, yet the contribution of the Paschen, Brackett
and other permitted lines in the strongest accretors may be comparable
to the total Balmer line luminosity.
The accretion level of all the YSOs studied here is well above the 
expected chromospheric contribution, even at the lowest values of \Lacc. 

The 39 empirical  \Lacc ~vs. \Ll ~ relationships computed here 
have in general a lower dispersion as compared to previous relationships 
in the literature. Our \Lacc ~vs. \Ll ~relationships are in good agreement 
with previous results, but systematic differences may exist at different 
mass regimes with respect to studies adopting other methodologies to measure 
\Lacc ~or \Macc. We confirm that for low-mass YSOs (\Mstar$<$0.3\Msun), 
\Ha ~line profile modelling may underestimate \Macc ~by 0.6 to 0.8\,dex 
with respect to \Macc ~derived from continuum-excess measures. 
  
The least scattered among our \Lacc--\Ll ~relationships are those for 
the Balmer lines n$>$3, the \brg ~line and the He\,{\sc i} 
lines. The \pab ~and the \brg ~relations are recommended because
less affected by chromospheric activity than the optical lines.
The most scattered relations are those of the Ca\,{\sc ii}~IRT. 
Likewise, the Ca\,{\sc ii}~IRT surface-flux relationships have large 
scatter ($\sim$0.6\,dex), mainly due to the blending and contribution 
of Paschen lines in strong accretors and to the uncertainties on stellar 
parameters. The previously suggested bi-modality with respect to the 
mass of the Ca\,{\sc ii}~IRT surface-flux relationships is most likely 
induced by the mix of different methodologies to derive \Macc.
More generally, we conclude that mixing mass-accretion rates calculated 
with different techniques may lead to a spurious bi-modality in 
the relationships between \Macc ~and YSOs properties.
The average \Lacc ~derived from several lines, measured simultaneously, 
has a significantly reduced error.

The accretion properties of the YSOs studied here are similar to 
those of other low-mass YSOs in regions like Taurus, $\rho$-Oph or 
$\sigma$-Ori. We derive \Macc  ~in the range from 
2$\times$10$^{-12}$ to 4$\times$10$^{-8}$\,\Msun~yr$^{-1}$ for objects
with masses from 0.03 to 1.2\,\Msun.  We conclude that \Macc~$\propto$~\Mstar$^{1.8(\pm0.2)}$
for the Lupus sample studied here, in agreement with most studies 
of the  \Macc -- \Mstar ~relationship. The scatter for the Lupus 
relationship is smaller than for other data sets.
The average \Macc ~for objects with mass between 0.4 and 0.8\,\Msun 
is consistent with the calculations of the evolution of \Macc ~in 
viscous discs for 3\,Myr old objects with 0.5\,\Msun ~\citep{hartmann98}.

We confirm, and extend over more than 5 orders of magnitude in \Macc, 
 some properties of the accretion emission, already known for a 
more limited range of \Macc. In particular,  that line ratios, 
as well as the fraction of \Lacc ~emitted in each line, are roughly 
independent of \Macc, and that the line luminosities increase almost 
linearly with \Macc  ~over the whole range. This suggests that
some inconsistencies between magnetospheric accretion models and
observations still prevail, but a number of properties (e.g. constant 
hydrogen line ratios, same accretion budget emitted in each line 
independently of optical depth, and linear correlation of the line 
luminosity  with \Macc, among other) suggest that the physical conditions 
of the accreting gas, over a large range of \Macc, are similar. Our data 
show that this properties are valid over a large range in  \Macc, 
extending down to the very low-mass regime.
We thus confirm the suggestion by \citet{muzerolle01} that it is the 
geometry of the accretion flows that controls the rate at which the 
disc material accretes onto the central star: larger mass accretion 
rates require larger emitting areas. Detailed magnetospheric
accretion models, incorporating complex magnetic field topologies, are 
needed to understand whether other physical parameters (e.g. magnetic 
field topology) play a role in the accretion physics.
   
\acknowledgements{We thank the anonymous referee for her/his 
 careful reading and for suggestions. JMA, EC, BS, and KB thank
 G. Attusino for stimulating discussions.
 We also thank B. Nisini, T. Giannini and S. Antoniucci for lively discussions.
 We thank V. D'Odorico, P. Goldoni and A. Modigliani for their help with the 
 X-Shooter pipeline, and F. Getman and G. Capasso for the installation of the
 different pipeline versions at Capodimonte. We also thank the ESO staff, in 
 particular F. Patat for suggestions in OB preparation and C. Martayan for 
 support during the observations. Financial support from INAF is also 
 acknowledged. This research made use of the SIMBAD database, operated at 
 the CDS (Strasbourg, France). This publication makes use of data products 
 from the Two Micron All Sky Survey, which is a joint project of the University 
 of Massachusetts and the Infrared Processing and Analysis Center/California 
 Institute of Technology, funded by NASA and the National Science Foundation. 
 This publication makes use of data products from the Wide-field Infrared Survey 
 Explorer, which is a joint project of the University of California, Los Angeles, 
 and the Jet Propulsion Laboratory/California Institute of Technology, funded by 
 the National Aeronautics and Space Administration.}


\bibliographystyle{/home/jmae/aa-package/bibtex/aa}

\begin{appendix}

\section{Correction for telluric bands} 
\label{appendix1}

The flux calibrated one-dimensional spectra resulting from
the X-Shooter pipeline or our MIDAS procedure are not corrected
for the contribution of telluric bands. Therefore, the telluric
standards were used to perform the correction using the 
IRAF task ''telluric''. The procedure basically consists on dividing 
the target spectrum by the telluric spectrum multiplied by an appropriate 
scaling factor. This factor depends on the ratio of the airmass of the
target and the telluric standard. Since the targets and their assigned 
telluric standards were observed at very similar air-masses in most cases, 
such factor is normally close to one. The procedure was applied 
independently in the VIS and NIR spectra, in a different way.

\subsection{Telluric correction in the VIS spectra}
The correction was done directly on the one-dimensional flux calibrated
VIS spectra of the targets using the assigned tellurics. However, in order 
to avoid the introduction of the spectral energy distribution of the 
telluric on the flux calibrated spectra, the tellurics were first normalized
to their continuum and their hydrogen lines and other stellar lines were removed 
by fitting combinations of Gaussian, Lorentzian and Voigt functions. The resulting 
normalised telluric spectra, free of photospheric lines were then used as 
input in the IRAF task ''telluric''. In Figure~\ref{tellcorr1} left panels 
examples of the telluric correction in the spectral range of the K\,{\sc i}
$\lambda\lambda$ 766.49, 769.90\,nm and the Na\,{\sc i} $\lambda\lambda$  
818.33, 819.481\,nm doublets are shown.

\subsection{Telluric correction in the NIR spectra}
For the NIR spectra, a pseudo-response function was first derived 
by  dividing the non flux-calibrated telluric by a black-body of 
the same effective temperature as the telluric. Such pseudo-response 
function was also cleaned for photospheric lines in the same way 
as in the VIS. The resulting pseudo-response function, free of stellar 
lines and containing the 
telluric bands, was then used as input in the IRAF task  "telluric". 
Note that the target spectrum to be used as input of the IRAF task
was the non-flux calibrated one-dimensional spectrum.
In this way, the telluric correction and the correction for 
the response function were done simultaneously in the NIR. Although the 
shape of the resulting spectrum after this procedure is correct, the 
flux calibration is only relative to the pseudo-response function. 
Therefore, a factor was applied in order to bring the NIR spectrum to 
the absolute flux scale. Such factor was estimated using the non-telluric
corrected, but flux-calibrated spectra resulting from both the X-Shooter 
pipeline and our MIDAS procedure. In Figure~\ref{tellcorr2} 
examples of the telluric correction of the YSO Lup713 are shown in two 
spectral ranges of the NIR arm. Note that in  the range 2000-2140\,nm 
the pipeline yields a bump with respect to the telluric corrected 
spectrum. This defect in the latter is corrected because 
the pseudo-response function contains such bump. 

\begin{figure}[h]
\resizebox{1.0\hsize}{!}{\includegraphics[]{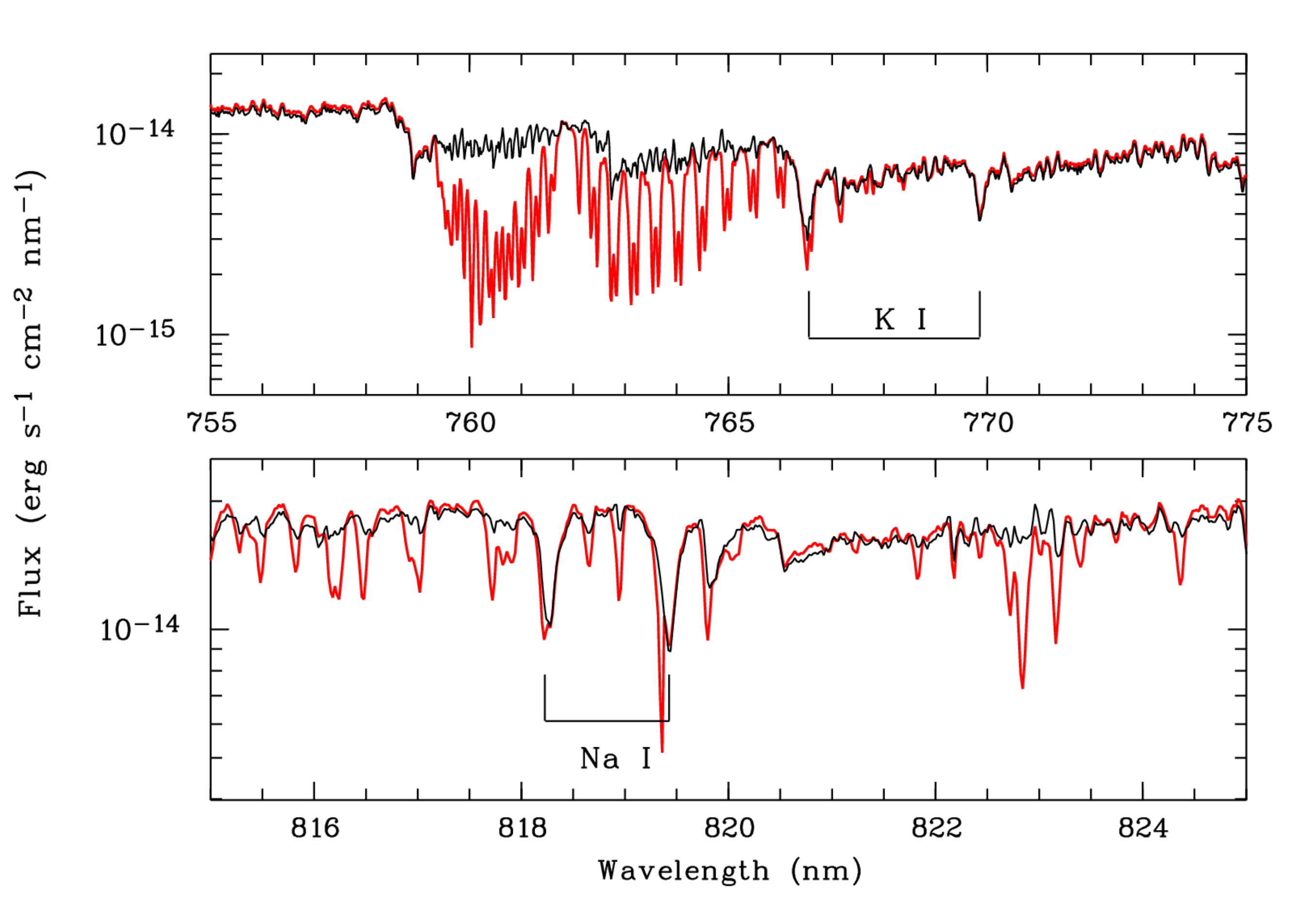}}
\caption{Example of telluric-corrected spectrum (black lines) of the YSO Lup713 in
       the range of the K\,{\sc i} $\lambda\lambda$ 766.49, 769.90\,nm (upper panel)
       and the Na\,{\sc i} $\lambda\lambda$ 818.33, 819.48\,nm (lower panel) 
       doublets. The non-corrected spectra are overlaid with red lines.
     \label{tellcorr1}}
\end{figure}

\begin{figure}[h]
\resizebox{1.0\hsize}{!}{\includegraphics[]{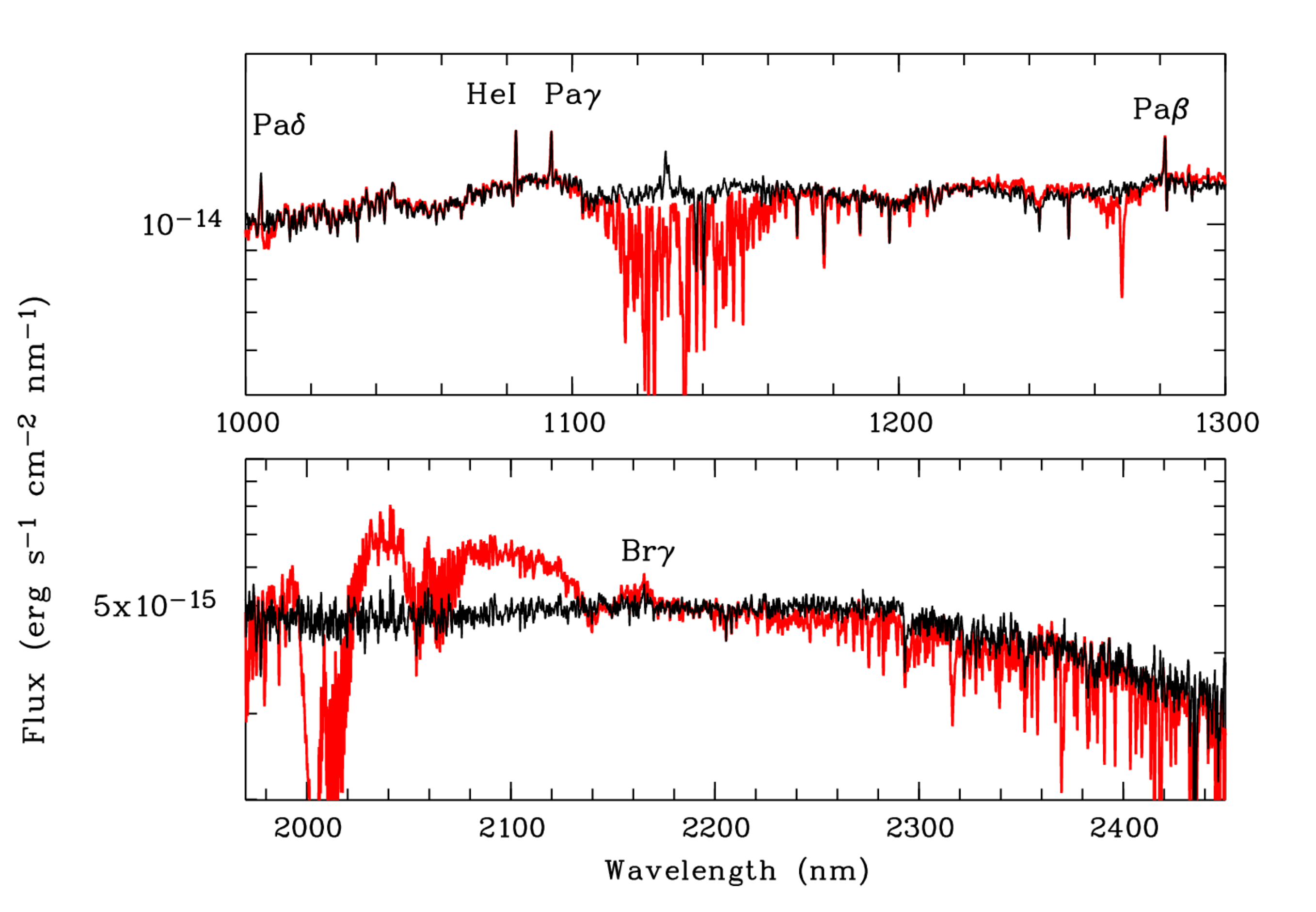}}
\caption{Examples of the telluric corrected spectrum of Lup713 in two spectral ranges 
       of the NIR arm. The non-corrected spectra are overlaid with red lines in both panels.
     \label{tellcorr2}}
\end{figure}

\section{Impact of veiling on spectral typing and extinction} 
\label{appendix2}

\subsection{Veiling and spectral indices}
Strong veiling may influence the spectral indices used to classify
the M-type YSOs. To investigate the impact of veiling on the spectral 
indices, we proceed as follows. A constant veiling was artificially 
added to the spectra of the Class-III templates in the spectral range 
between 740\,nm and 860\,nm, where the indices are computed. Then, the 
new indices and spectral type of the "veiled" spectrum were derived. 
The effect of veiling is to yield systematically earlier types, but 
in that spectral range it should be greater than one in order to 
induce a difference of about one spectral sub-class. 
As seen in Section~\ref{bjmodelling}, the strongest veiling, 
at $\sim$710\,nm, among the M-type YSOs is less than 0.6 and 
corresponds to Sz\,113. For this object we estimate that veiling
would introduce an uncertainty of 0.4 sub-class in spectral type.
For the other YSOs excess emission in the Paschen continuum is 
smaller (see Table~\ref{accretion}). For all the M-type YSOs 
veiling is estimated to influence the spectral type classification to 
less than 0.3 spectral sub-classes.

\subsection{Veiling and extinction}
The effect of strong veiling is to make YSOs objects intrinsically 
bluer than the templates we use to derive extinction. Thus, our method
may underestimate extinction if veiling is very strong. To investigate
by how much, we proceed as follows. The observed spectrum of a given
YSO, $F_{YSO}(\lambda)$, is the sum of the flux of the object, 
$F_{*}(\lambda)$, plus the flux of the continuum excess emission, 
$F_{cont}(\lambda)$, extincted with an extinction law, $A_{\lambda}/A_{V}$. 
So the flux we observe is $F_{YSO}(\lambda) = (F_{*}(\lambda) + F_{cont}(\lambda)) \cdot e^{-0.4 \cdot (A_{\lambda}/A_{V}) \cdot A_{V}}$.
In order to get the extinction, free of veiling effects,
we must apply our methods, not to $F_{YSO}(\lambda)$, but to the 
de-veiled spectrum:

\begin{equation}
F_{YSO}(\lambda) - F_{cont}(\lambda) \cdot e^{-0.4 \cdot (A_{\lambda}/A_{V}) \cdot A_{V}}
\label{deveiled}
\end{equation}

We can assume that the continuum flux is the model continuum excess 
emission derived in Section~\ref{bjmodelling}, hence from Equation~\ref{deveiled}
we can compute a "de-veiled" spectrum, provided we know $A_{V}$.
Since we do not know $A_{V}$ in advance, we  proceed in a two-step
fitting procedure. As first guess of $A_{V}$ we use the value derived
from our template fitting procedure applied to $F_{YSO}(\lambda)$,
and compute an de-veiled spectrum using Equation~\ref{deveiled}.  
A new $A_{V}$ is then derived from our  fitting procedure
applied to the first de-veiled spectrum. This spectrum
is used to compute a new $A_{V}$ value and to 
redden the $F_{cont}(\lambda)$ to be used again to calculate a 
second de-veiled spectrum, and so on until a  $A_{V}$ value 
is found which leads to the best match with the template.

In practice, we use as a test case Sz\,113, the most veiled M-Type 
YSO in our sample. The extinction we derive as "first guess" is 1\,mag.
Thus, the model continuum excess emission derived in Section~\ref{bjmodelling} 
was reddened at extinctions between 1\,mag and 2\,mag in steps of 
0.25\,mag. The results were then subtracted to the observed spectrum 
of Sz\,113. The several de-veiled spectra were then processed with our
 fitting procedure. The best compromise between reddened model 
continuum and de-reddened de-veiled spectra occurs at $A_{\rm V}=$1.5\,mag. 
We conclude that for our most veiled M-Type YSO extinction
is underestimated by less than 0.5\,mag.

\Online

\section{On-line material} 
\onecolumn

\begin{landscape}
\begin{flushleft}
\begin{longtable}{lcccccccc}
\caption{Ancillary optical and near-IR photometry}\label{optnirphot}\\
\hline \hline

Object  &  $B$ & $V$ & $R$ & $I$ & $Z$ & $J$ & $H$ & $K$ \\
        &  (mag)  & (mag)  &  (mag)  & (mag)   & (mag)  & (mag)  & (mag)   &  (mag) \\ 
\hline
  Sz66             	& 16.45$\pm$0.01 &  15.00$\pm$0.01  &  14.51$\pm$0.01  &  12.56$\pm$0.01  &  11.53$\pm$0.02  &  10.89$\pm$0.03  &   9.88$\pm$0.03  &   9.29$\pm$0.03  \\       
  AKC2006-19            & 18.38$\pm$0.05 &  17.74$\pm$0.05  &  17.04$\pm$0.01  &  14.91$\pm$0.01  &  14.09$\pm$0.01  &  12.98$\pm$0.02  &  12.41$\pm$0.02  &  12.13$\pm$0.02  \\
  Sz69                  & 17.41$\pm$0.06 &  16.25$\pm$0.06  &  14.84$\pm$0.01  &  13.63$\pm$0.01  &  12.28$\pm$0.01  &  11.18$\pm$0.03  &  10.16$\pm$0.03  &   9.41$\pm$0.03  \\
  Sz71             	& 13.80$\pm$0.01 &  13.54$\pm$0.01  &  13.70$\pm$0.01  &  11.83$\pm$0.01  &  11.28$\pm$0.02  &  10.07$\pm$0.02  &   9.18$\pm$0.02  &   8.63$\pm$0.02  \\       
  Sz72             	& 15.96$\pm$0.01 &  14.79$\pm$0.01  &  13.55$\pm$0.02  &  12.25$\pm$0.02  &	  ...	    &  10.57$\pm$0.03  &   9.77$\pm$0.03  &   9.33$\pm$0.02  \\        
  Sz73             	& 18.21$\pm$0.04 &  16.37$\pm$0.02  &  14.90$\pm$0.02  &  13.44$\pm$0.01  &	  ...	    &  10.74$\pm$0.03  &   9.53$\pm$0.03  &   8.83$\pm$0.02  \\ 	 
  Sz74             	& 15.96$\pm$0.02 &  14.18$\pm$0.01  &  12.98$\pm$0.02  &  11.44$\pm$0.01  &	  ...	    &	9.23$\pm$0.02  &   8.10$\pm$0.03  &   7.43$\pm$0.02  \\ 	 
  Sz83             	& 11.70$\pm$0.01 &  11.40$\pm$0.01  &  10.63$\pm$0.02  &   9.84$\pm$0.02  &	  ...	    &	8.73$\pm$0.03  &   7.82$\pm$0.04  &   7.14$\pm$0.02  \\ 	 
  Sz84             	& 17.54$\pm$0.03 &  16.16$\pm$0.01  &  14.53$\pm$0.04  &  12.94$\pm$0.02  &	  ...	    &  10.93$\pm$0.02  &  10.20$\pm$0.02  &   9.85$\pm$0.03  \\ 	 
  Sz130                 & 15.82$\pm$0.05 &  14.71$\pm$0.05  &  13.56$\pm$0.54  &  12.46$\pm$0.01  &  12.26$\pm$0.02  &  10.73$\pm$0.05  &   9.93$\pm$0.05  &   9.62$\pm$0.05  \\
  Sz88A (SW)       	& 13.67$\pm$0.01 &  13.02$\pm$0.01  &  13.06$\pm$0.01  &  11.91$\pm$0.01  &  11.21$\pm$0.01  &  10.20$\pm$0.03  &   9.36$\pm$0.03  &   8.76$\pm$0.02  \\       
  Sz88B (NE)       	& 17.17$\pm$0.01 &  15.72$\pm$0.01  &  15.32$\pm$0.01  &  13.33$\pm$0.01  &  12.57$\pm$0.01  &  11.53$\pm$0.03  &  10.87$\pm$0.03  &  10.51$\pm$0.02  \\       
  Sz91             	& 16.25$\pm$0.05 &  14.58$\pm$0.05  &  14.38$\pm$0.01  &  12.92$\pm$0.01  &  12.25$\pm$0.01  &  11.06$\pm$0.02  &  10.12$\pm$0.02  &   9.85$\pm$0.02  \\       
  Lup713                & 18.38$\pm$0.05 &   ... 	    &  17.77$\pm$0.02  &  15.68$\pm$0.02  &  14.06$\pm$0.01 &  13.24$\pm$0.03  &  12.57$\pm$0.03  &  12.13$\pm$0.03  \\ 
  Lup604s               & 16.94$\pm$0.05 &  17.01$\pm$0.05  &  16.51$\pm$0.01  &  14.31$\pm$0.01  &  13.34$\pm$0.01  &  12.15$\pm$0.03  &  11.45$\pm$0.03  &  11.07$\pm$0.02  \\
  Sz97                  & 16.18$\pm$0.05 &  14.61$\pm$0.05  &  14.77$\pm$0.01  &  12.92$\pm$0.01  &  12.24$\pm$0.01  &  11.24$\pm$0.02  &  10.55$\pm$0.02  &  10.22$\pm$0.02  \\ 
  Sz99                  & 15.49$\pm$0.05 &  16.0 $\pm$0.05  &  15.74$\pm$0.01  &  14.17$\pm$0.01  &  13.28$\pm$0.01  &  11.93$\pm$0.02  &  11.21$\pm$0.02  &  10.75$\pm$0.02  \\ 
  Sz100                 & 16.62$\pm$0.05 &  15.43$\pm$0.05  &  15.22$\pm$0.01  &  13.12$\pm$0.01  &  12.18$\pm$0.01  &  10.98$\pm$0.02  &  10.35$\pm$0.02  &   9.91$\pm$0.02  \\ 
  Sz103                 &   ...  	&  11.14$\pm$0.05  &  15.18$\pm$0.01  &  13.15$\pm$0.01  &  12.40$\pm$0.01  &  11.38$\pm$0.02  &  10.62$\pm$0.02  &  10.23$\pm$0.02  \\ 
  Sz104                 &   ...  	&  15.25$\pm$0.05  &  15.64$\pm$0.01  &  13.57$\pm$0.01  &  12.78$\pm$0.01  &  11.66$\pm$0.02  &  11.00$\pm$0.02  &  10.65$\pm$0.02  \\ 
  Lup706                &      ...	 &   ... 	   &  20.70$\pm$0.06  &  18.44$\pm$0.05  &  16.68$\pm$0.03  &  15.17$\pm$0.06  &  14.24$\pm$0.06  &  13.83$\pm$0.04  \\ 
  Sz106            	& 17.63$\pm$0.05 &  16.26$\pm$0.02  &  15.97$\pm$0.02  &  14.66$\pm$0.01  &  14.14$\pm$0.01 &	11.65$\pm$0.03  &  10.66$\pm$0.03  &  10.15$\pm$0.02  \\       
  Par-Lup3-3            & 16.76$\pm$0.05 &   ... 	    &  16.27$\pm$0.01  &  14.24$\pm$0.01  &  13.19$\pm$0.01 &  11.45$\pm$0.02  &  10.17$\pm$0.02  &   9.55$\pm$0.02  \\
  Par-Lup3-4            &   ...  	&  21.19:	   &  19.53$\pm$0.04  &  18.18$\pm$0.05  &  17.74$\pm$0.06  &  15.46$\pm$0.06  &  14.26$\pm$0.04  &  13.31$\pm$0.04  \\ 
  Sz110                 & 15.27$\pm$0.05 &  14.58$\pm$0.05  &  13.72$\pm$0.05  &  12.28$\pm$0.01  &  12.09$\pm$0.02  &  10.97$\pm$0.02  &  10.22$\pm$0.02  &   9.75$\pm$0.02  \\ 
  Sz111            	& 15.34$\pm$0.05 &  13.98$\pm$0.05  &  13.29$\pm$0.05  &  12.24$\pm$0.01  &	  ...	    &  10.62$\pm$0.02  &   9.80$\pm$0.02  &   9.54$\pm$0.02  \\        
  Sz112            	& 16.62$\pm$0.05 &  15.39$\pm$0.05  &  14.92$\pm$0.01  &  12.93$\pm$0.01  &  12.12$\pm$0.01  &  11.00$\pm$0.02  &  10.29$\pm$0.02  &   9.96$\pm$0.02  \\       
  Sz113                 & 17.46$\pm$0.05 &  16.5 $\pm$0.05  &  16.73$\pm$0.01  &  14.65$\pm$0.01  &  13.69$\pm$0.01  &  12.47$\pm$0.02  &  11.72$\pm$0.02  &  11.26$\pm$0.02  \\ 
2MASS J16085953-3856275 &   ...  	 &   ... 	   &  19.79$\pm$0.04  &  16.91$\pm$0.02  &  15.56$\pm$0.02  &  13.90$\pm$0.03  &  13.29$\pm$0.03  &  12.84$\pm$0.03  \\ 
SSTc2d160901.4-392512   & 16.92$\pm$0.05 &  15.31$\pm$0.05  &  15.23$\pm$0.01  &  13.49$\pm$0.01  &  13.23$\pm$0.01  &  11.61$\pm$0.02  &  10.68$\pm$0.02  &  10.29$\pm$0.02  \\ 
  Sz114                 & 15.33$\pm$0.04 &  14.12:	    &  14.73$\pm$0.01  &  12.54$\pm$0.01  &  11.58$\pm$0.01  &  10.41$\pm$0.02  &   9.70$\pm$0.02  &   9.32$\pm$0.02  \\ 
  Sz115            	& 16.88$\pm$0.05 &  15.48$\pm$0.05  &  15.17$\pm$0.01  &  13.12$\pm$0.01  &  12.44$\pm$0.01  &  11.33$\pm$0.03  &  10.65$\pm$0.03  &  10.45$\pm$0.03  \\       
  Lup818s               & 17.37$\pm$0.05 &   ... 	    &  17.58$\pm$0.02  &  15.28$\pm$0.01  &  14.24$\pm$0.01  &  13.01$\pm$0.03  &  12.39$\pm$0.03  &  11.99$\pm$0.03  \\ 
  Sz123A  (S)      	& 16.27$\pm$0.05 &  14.90$\pm$0.05  &  15.04$\pm$0.01  &  13.37$\pm$0.01  &  12.68$\pm$0.01  &  11.47$\pm$0.02  &  10.55$\pm$0.02  &  10.03$\pm$0.02  \\       
  Sz123B  (N)      	& 17.22$\pm$0.05 &  15.72$\pm$0.05  &  15.79$\pm$0.01  &  14.29$\pm$0.01  &  13.84$\pm$0.01  &  12.42$\pm$0.02  &  11.63$\pm$0.02  &  11.51$\pm$0.02  \\       
  SST-Lup3-1            & 18.17$\pm$0.05 &  16.55$\pm$0.05  &  16.25$\pm$0.01  &  14.26$\pm$0.01  &  13.46$\pm$0.01  &  12.20$\pm$0.02  &  11.51$\pm$0.03  &  11.20$\pm$0.02  \\

\hline
\end{longtable}
\end{flushleft}
\end{landscape}

\onecolumn

\setlength{\tabcolsep}{3pt}
\begin{landscape}
\begin{longtable}{lrrrrrrrrrr}
\caption{Available photometric fluxes, in milli Jy, from 3.4$\mu$m to 70.0$\mu$m }\label{mirphot}\\
\hline \hline

Object  & 3.4$\mu$m  & 3.6$\mu$m & 4.5$\mu$m & 4.6$\mu$m & 5.8$\mu$m & 8.0$\mu$m & 12.8$\mu$m  & 22.4$\mu$m & 24.0$\mu$m & 70.0$\mu$m  \\
        &  & & & & & & & & &  \\
\hline
  Sz66                  &   115.32$\pm$ 4.14   &  95.80$\pm$4.98 &  81.50$\pm$4.98  &   106.18$\pm$ 3.32   &  82.50$\pm$3.99  & 100.00$\pm$4.84 &    95.31$\pm$ 2.90   &  144.77$\pm$ 8.80 &  167.00$\pm$15.50  &    $<$ 50.00         \\ 
  AKC2006-19            &     5.46$\pm$ 0.12   &   5.48$\pm$0.27 &   4.06$\pm$0.27  &     4.03$\pm$ 0.08   &   3.29$\pm$0.16  &   3.73$\pm$0.18 &     3.84$\pm$ 0.15   &    7.54$\pm$ 0.89 &	3.44$\pm$ 0.38  &    $<$ 50.00         \\ 
  Sz69                  &   131.07$\pm$ 2.66   & 111.00$\pm$5.84 & 120.00$\pm$5.84  &   127.54$\pm$ 2.11   &  94.60$\pm$4.59  &  83.00$\pm$4.12 &    79.43$\pm$ 1.17   &  144.91$\pm$ 3.07 &  100.00$\pm$ 9.72  &    154.00$\pm$20.10  \\ 
  Sz71                  &   126.92$\pm$ 2.69   &	...	 &	    ...     &   104.53$\pm$ 2.02   &	  ...	      & 140.00 :	&   163.97$\pm$ 2.27   &  276.62$\pm$ 6.12 &  260.00 :  	&	  ...	       \\ 
  Sz72                  &   101.09$\pm$ 2.14   &	...	 &	    ...     &    92.31$\pm$ 1.79   &	  ...	      & 150.00 :	&   130.72$\pm$ 1.69   &  232.86$\pm$ 3.86 &  250.00 :  	&	  ...	       \\ 
  Sz73                  &   303.61$\pm$ 6.71   &	...	 &	    ...     &   361.43$\pm$ 6.66   &	  ...	      & 500.00 :	&   421.46$\pm$ 5.82   &  920.21$\pm$10.17 & 1200.00 :  	&	  ...	       \\   
  Sz74                  &   865.99$\pm$28.71   &	...	 &	    ...     &   897.90$\pm$20.68   &	  ...	      &      ...	&   759.20$\pm$10.49   & 1386.40$\pm$16.60 &	  ...		&	  ...	       \\   
  Sz83                  &  1260.99$\pm$60.39   &	...	 &	    ...     &  1739.52$\pm$59.28   &	  ...	      &  1500.00: 	&  2400.80$\pm$19.90   & 4465.67$\pm$41.13 & 3500.00 :  	&	  ...	       \\   
  Sz84                  &    42.10$\pm$ 0.89   &  42.00$\pm$0.04 &  29.00$\pm$0.04  &    32.30$\pm$ 0.63   &  20.10$\pm$0.03  &  12.00$\pm$0.03 &     6.95$\pm$ 0.21   &   24.54$\pm$ 1.20 &   20.90$\pm$0.41	&   244.00$\pm$73.20   \\   
  Sz130                 &    58.49$\pm$ 1.19   &  59.00$\pm$3.02 &  53.50$\pm$3.02  &    51.91$\pm$ 0.91   &  53.30$\pm$2.53  &  70.10$\pm$3.35 &    79.21$\pm$ 1.17   &  133.63$\pm$ 2.71 &  115.00$\pm$10.60  &    165.00$\pm$22.10  \\ 
  Sz88A (SW)            &   $<$ 292.90         &        ...      &	  ...	    &     $<$  264.99      & 	   ...        &  ...	        &  $<$   300.85        & $<$  363.99       &      ...	        &	      ...      \\	
  Sz88B (NE)            &   $<$ 292.90         &        ...      &	  ...	    &     $<$  264.99      & 	   ...        &  ...	        &  $<$   300.85        & $<$  363.99       &      ...	        &	      ...      \\	
  Sz91                  &    41.37$\pm$ 0.80   &  38.60$\pm$1.93 &  24.70$\pm$1.93  &    25.64$\pm$ 0.47   &  17.20$\pm$0.83  &  10.90$\pm$0.52 &     6.89$\pm$ 0.19   &   13.02$\pm$ 1.00 &   9.72 $\pm$ 0.98  &    502.00$\pm$53.20  \\ 
  Lup713                &     6.72$\pm$ 0.15   &   6.96$\pm$0.46 &   6.12$\pm$0.46  &     6.12$\pm$ 0.13   &   5.65$\pm$0.31  &   6.69$\pm$0.33 &     5.62$\pm$ 0.26   &    5.36$\pm$ 1.16 &	6.39$\pm$ 0.64  &     $<$ 50.00        \\ 
  Lup604s               &    16.49$\pm$ 0.35   &  16.30$\pm$0.90 &  15.10$\pm$0.90  &    13.72$\pm$ 0.25   &  12.80$\pm$0.64  &  14.40$\pm$0.69 &    14.02$\pm$ 0.30   &   21.22$\pm$ 1.15 &   16.60$\pm$ 1.57  &     $<$ 50.00        \\ 
  Sz97                  &    33.17$\pm$ 0.70   &  27.90$\pm$1.73 &  25.20$\pm$1.73  &    26.82$\pm$ 0.49   &  23.70$\pm$1.14  &  25.70$\pm$1.23 &    24.01$\pm$ 0.55   &   39.51$\pm$ 2.26 &   31.20$\pm$ 2.97  &     88.60$\pm$16.90  \\ 
  Sz99                  &    25.70$\pm$ 0.54   &  23.80$\pm$1.39 &  22.90$\pm$1.39  &    22.22$\pm$ 0.41   &  20.10$\pm$1.01  &  21.10$\pm$1.00 &    24.41$\pm$ 0.52   &   30.33$\pm$ 1.65 &   26.20$\pm$ 2.45  &    $<$ 50.00         \\ 
  Sz100                 &    59.58$\pm$ 1.32   &  53.10$\pm$3.21 &  50.40$\pm$3.21  &    51.86$\pm$ 1.00   &  48.70$\pm$2.48  &  61.70$\pm$3.04 &    67.98$\pm$ 1.13   &  140.44$\pm$ 3.62 &  130.00$\pm$12.10  &    223.00$\pm$28.50  \\ 
  Sz103                 &    40.28$\pm$ 0.93   &  32.10$\pm$1.97 &  28.10$\pm$1.97  &    33.89$\pm$ 0.69   &  30.00$\pm$1.81  &  33.80$\pm$1.77 &    39.77$\pm$ 1.21   &  151.04$\pm$ 4.31 &   81.20$\pm$ 7.66  &    317.00$\pm$40.00  \\ 
  Sz104                 & 	  ...	       &  25.50$\pm$1.47 &  22.50$\pm$1.47  &      ...  	   &  20.00$\pm$1.01  &  21.30$\pm$1.10 & 	  ...	       &	...	   &   47.90$\pm$ 4.62  &    441.00$\pm$53.20  \\ 
  Lup706                & 	  ...	       &   2.23$\pm$0.15 &   2.01$\pm$0.15  &      ...  	   &   1.77$\pm$0.11  &   1.93$\pm$0.11 & 	  ...	       &	...	   &	1.29$\pm$ 0.54  &     $<$ 50.00        \\ 
  Sz106                 &    67.53$\pm$ 1.49   &  60.10$\pm$4.42 &  81.30$\pm$4.42  &    60.65$\pm$ 1.12   &  72.10$\pm$3.58  &  73.80$\pm$3.86 &    54.90$\pm$ 1.01   &  106.83$\pm$ 2.66 &   69.50$\pm$ 6.56  &     $<$ 50.00        \\ 
  Par-Lup3-3            &    65.63$\pm$ 1.45   &  69.40$\pm$4.79 &  79.00$\pm$4.79  &    64.63$\pm$ 1.19   &  77.60$\pm$3.75  & 127.00$\pm$6.39 &   162.46$\pm$ 2.84   &  301.08$\pm$ 6.66 &  191.00$\pm$18.10  &     $<$ 50.00        \\ 
  Par-Lup3-4            &     2.77$\pm$ 0.07   &   2.80$\pm$0.15 &   2.88$\pm$0.15  &     3.39$\pm$ 0.08   &   2.24$\pm$0.13  &   1.73$\pm$0.10 &    10.40$\pm$ 0.47   &   69.93$\pm$ 2.83 &   26.60$\pm$ 2.55  &    492.00$\pm$53.60  \\ 
  Sz110                 &    70.85$\pm$ 1.44   &  69.90$\pm$3.61 &  67.80$\pm$3.61  &    63.51$\pm$ 1.17   &  58.80$\pm$2.82  &  70.50$\pm$3.45 &    99.62$\pm$ 1.56   &  193.33$\pm$ 4.27 &  171.00$\pm$16.20  &     $<$ 50.00        \\ 
  Sz111                 &    50.76$\pm$ 1.08   &	...	 &	  ...	    &    31.48$\pm$ 0.55   &	  ...	      &      ...	&     8.47$\pm$ 0.26   &   42.18$\pm$ 1.40 &   41.50$\pm$ 3.86  &     $<$ 50.00        \\ 
  Sz112                 &    49.42$\pm$ 1.14   &  48.70$\pm$2.96 &  38.00$\pm$2.96  &    39.38$\pm$ 0.69   &  30.40$\pm$1.49  &  24.80$\pm$1.22 &    38.33$\pm$ 0.64   &  152.72$\pm$ 3.66 &  124.00$\pm$11.70  &    120.00$\pm$25.20  \\ 
  Sz113                 &    25.09$\pm$ 0.56   &  15.40$\pm$0.89 &  15.00$\pm$0.89  &    26.16$\pm$ 0.43   &  14.90$\pm$0.72  &  20.50$\pm$0.99 &    32.39$\pm$ 0.51   &   59.69$\pm$ 1.76 &   56.10$\pm$ 5.35  &     88.30$\pm$16.00  \\ 
2MASS J16085953-3856275 &     3.42$\pm$ 0.07   &   3.89$\pm$0.21 &   3.69$\pm$0.21  &     3.66$\pm$ 0.07   &   3.77$\pm$0.23  &   4.31$\pm$0.21 &     4.94$\pm$ 0.18   &    8.37$\pm$ 0.95 &	5.40$\pm$ 0.57  &    $<$ 50.00         \\ 
SSTc2d160901.4-392512   &    43.48$\pm$ 0.84   &  44.70$\pm$2.95 &  34.00$\pm$2.95  &    33.73$\pm$ 0.62   &  30.60$\pm$1.61  &  25.80$\pm$1.29 &    23.40$\pm$ 0.45   &   58.98$\pm$ 2.23 &   42.20$\pm$ 3.96  &    114.00$\pm$15.00  \\ 
  Sz114                 &   118.34$\pm$ 2.51   &  95.70$\pm$5.43 & 101.00$\pm$5.43  &   124.17$\pm$ 2.40   &  97.20$\pm$4.66  & 121.00$\pm$6.02 &   170.28$\pm$ 2.35   &  374.53$\pm$ 8.28 &  347.00$\pm$32.30  &    257.00$\pm$30.50  \\ 
  Sz115                 &    25.94$\pm$ 0.55   &  22.20$\pm$1.36 &  18.40$\pm$1.36  &    18.79$\pm$ 0.38   &  14.80$\pm$0.72  &  12.40$\pm$0.59 &    10.59$\pm$ 0.51   &   12.53$\pm$ 1.66 &   10.50$\pm$ 1.01  &     $<$ 50.00        \\ 
  Lup818s               &     7.42$\pm$ 0.26   &   7.44$\pm$0.39 &   6.50$\pm$0.39  &     7.17$\pm$ 0.20   &   5.81$\pm$0.29  &   7.63$\pm$0.37 &     7.52$\pm$ 0.21   &   12.45$\pm$ 1.01 &   11.70$\pm$ 1.12  &     96.90$\pm$22.10  \\ 
  Sz123A  (S)           &  $<$  55.20          & $<$  59.90      & $<$  50.40       &   $<$  45.97         & $<$  42.60       & $<$  44.90      &  $<$   43.65         &  $<$  72.36       &  $<$  61.00        & $<$   327.00         \\ 
  Sz123B  (N)           &  $<$  55.20          & $<$  59.90      & $<$  50.40       &   $<$  45.97         & $<$  42.60       & $<$  44.90      &  $<$   43.65         &  $<$  72.36       &  $<$  61.00        & $<$   327.00         \\ 
  SST-Lup3-1            &    14.02$\pm$ 0.30   &  15.80$\pm$0.84 &  13.10$\pm$0.84  &    11.48$\pm$ 0.21   &  11.00$\pm$0.53  &  12.90$\pm$0.61 &    13.27$\pm$ 0.31   &   20.70$\pm$ 1.64 &   21.00$\pm$ 1.98  &     $<$ 50.00        \\ 
\hline
\end{longtable}
\footnotesize{Notes:
\begin{itemize}
\item fluxes at 3.4$\mu$m, 4.6$\mu$m, 12.8$\mu$m and 22.4$\mu$m are from the WISE catalog, 
while those at 3.6$\mu$m, 4.5$\mu$m, 5.8$\mu$m, 8.0$\mu$m, 24.0$\mu$m and 70.0$\mu$m are 
from Spitzer surveys.
\item the components of  Sz88 and Sz123  are not resolved in the mid-IR images. Thus, 
their fluxes are given as upper limits.
\item Spitzer fluxes marked with ":" were estimated from Spitzer spectra \citep{olofsson10}
\item the only YSO in our sample with 160$\mu$m data available is Sz\,111 with a flux of 
1678.00$\pm$335.60 mJy.
\end{itemize}
}
\end{landscape}

\begin{figure} 
\resizebox{0.9\hsize}{!}{\includegraphics[]{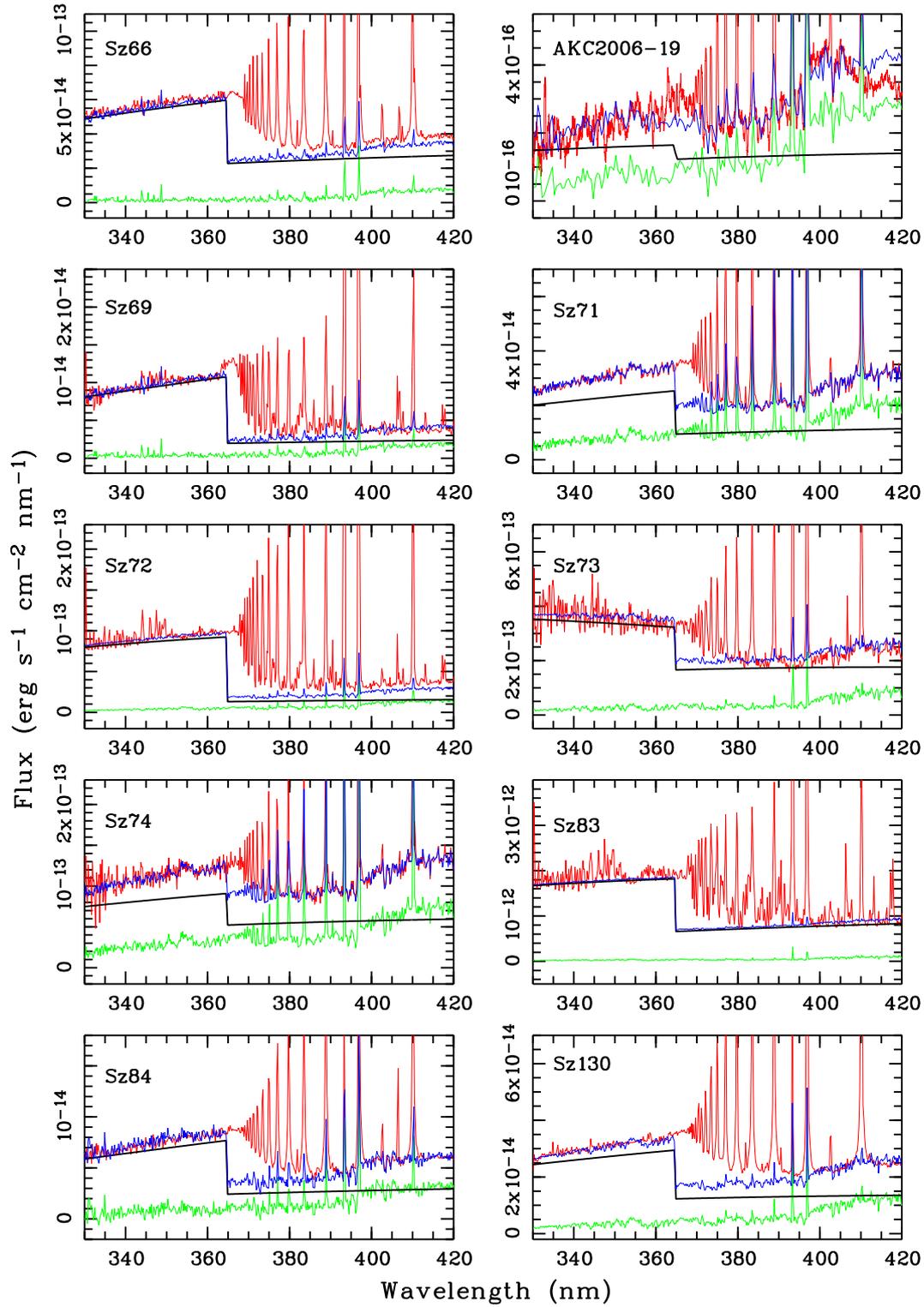}}
\caption{Extinction corrected spectra (red) fitted with a combination of a 
 photospheric template (green) and the synthetic continuum spectrum from
 a hydrogen slab (black). The total fit is represented with the blue line.
    \label{slab1}}
\end{figure}

\begin{figure} 
\resizebox{0.9\hsize}{!}{\includegraphics[]{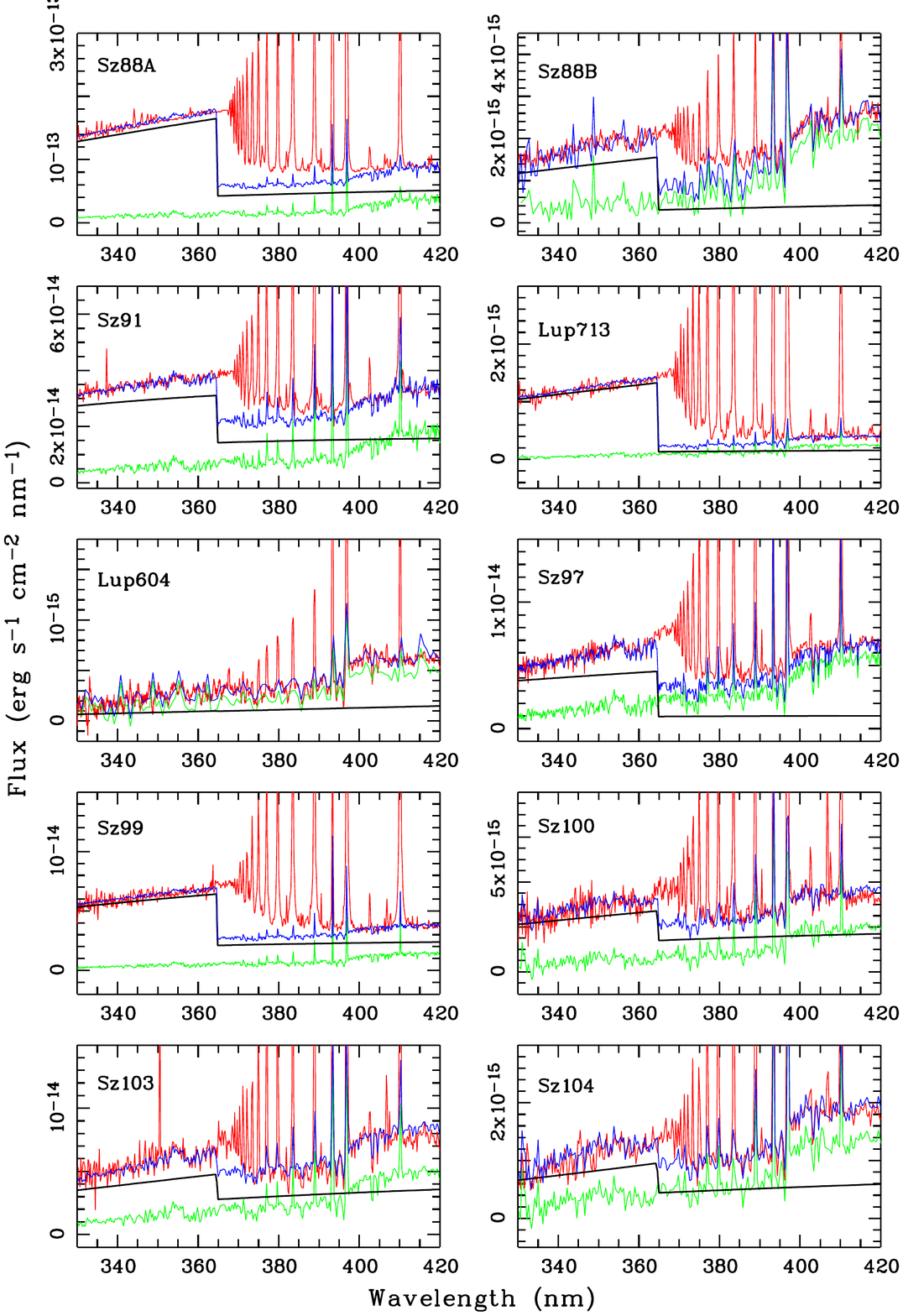}}
\caption{Extinction corrected spectra (red) fitted with a combination of a 
 photospheric template (green) and the synthetic continuum spectrum from
 a hydrogen slab (black). The total fit is represented with the blue line.
    \label{slab2}}
\end{figure}

\begin{figure} 
\resizebox{0.9\hsize}{!}{\includegraphics[]{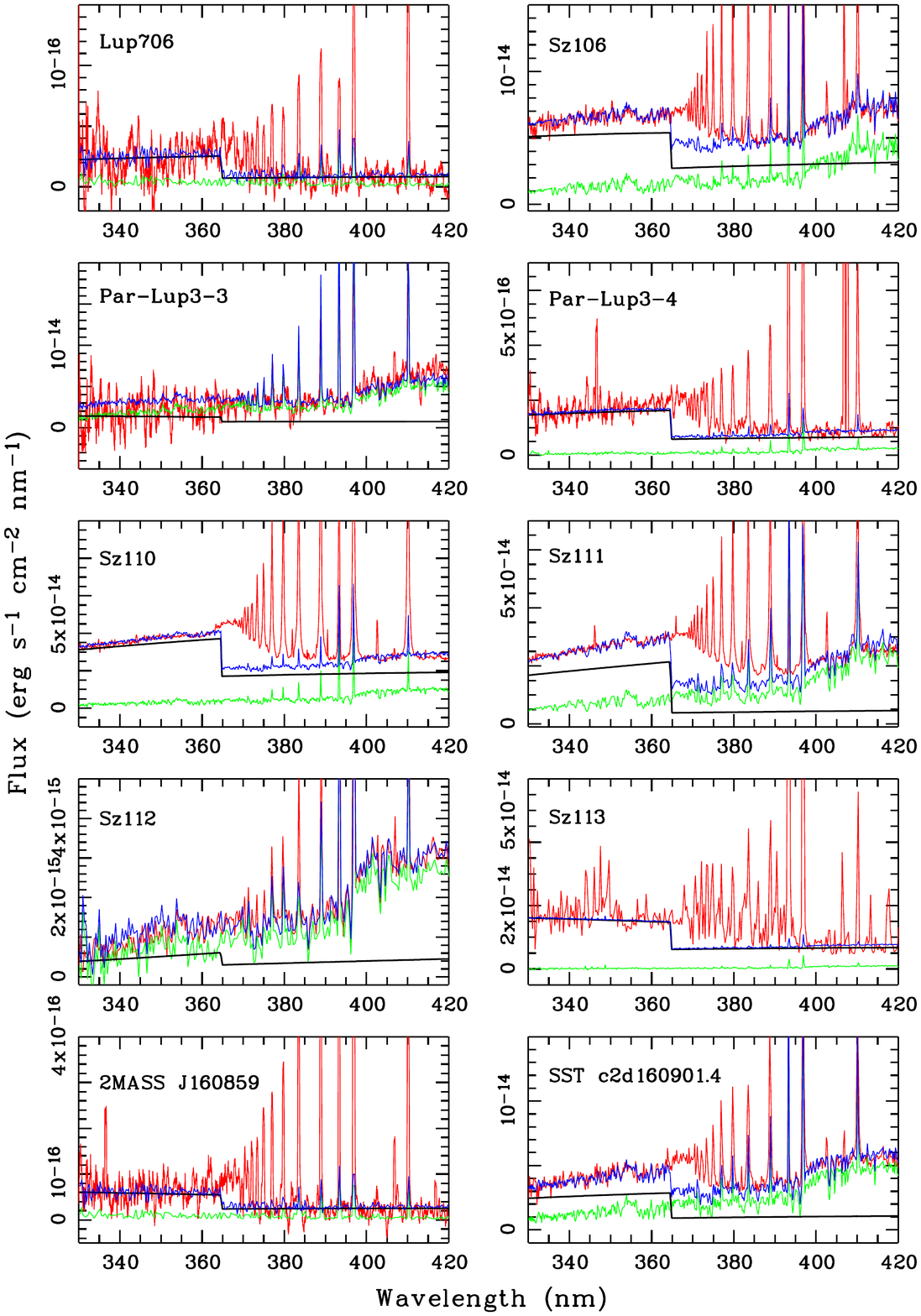}}
\caption{Extinction corrected spectra (red) fitted with a combination of a 
 photospheric template (green) and the synthetic continuum spectrum from
 a hydrogen slab (black). The total fit is represented with the blue line.
    \label{slab3}}
\end{figure}

\begin{figure} 
\resizebox{0.9\hsize}{!}{\includegraphics[]{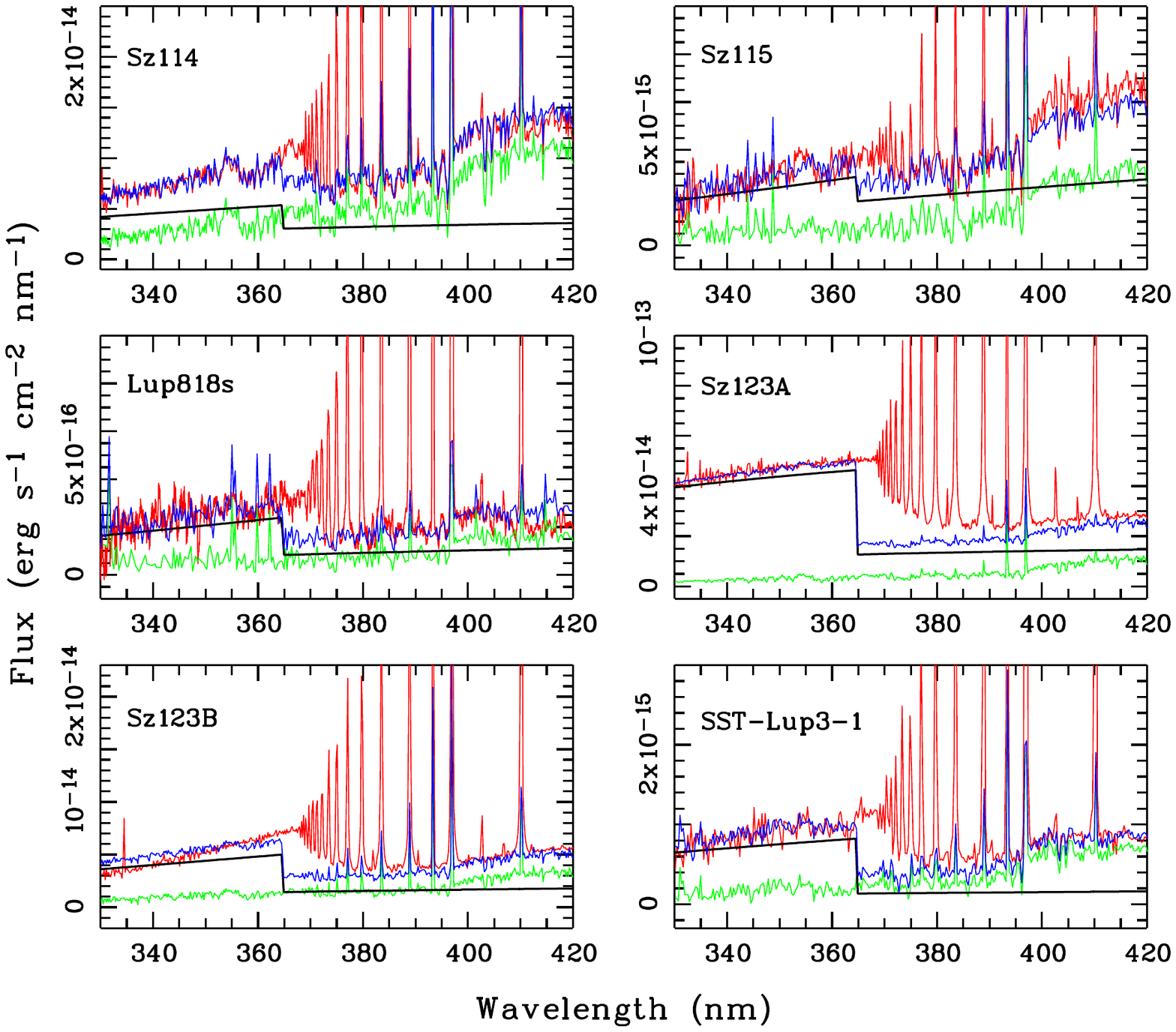}}
\caption{Extinction corrected spectra (red) fitted with a combination of a 
 photospheric template (green) and the synthetic continuum spectrum from
 a hydrogen slab (black). The total fit is represented with the blue line.
    \label{slab4}}
\end{figure}

\setlength{\tabcolsep}{4pt}

\begin{landscape}
\scriptsize
\begin{longtable}{l|c|r|c|r|c|r|c|r|c|r}
\caption[ ]{\label{tab:fluxes_EWs_Hae} Extinction corrected fluxes and equivalent widths of Balmer lines: H$\alpha$ to H$\epsilon$.}\\
\hline\hline
Object & $f_{\rm H\alpha}$ &  $EW_{\rm H\alpha}$&$f_{\rm H\beta}$& $EW_{\rm H\beta}$&$f_{\rm H\gamma}$& $EW_{\rm H\gamma}$&$f_{\rm H\delta}$& $EW_{\rm H\delta}$&$f_{\rm H\epsilon}$& $EW_{\rm H\epsilon}$\\
    & (erg\,s$^{-1}$\,cm$^{-2}$)&  (nm)	     &(erg\,s$^{-1}$\,cm$^{-2}$)& (nm)		&(erg\,s$^{-1}$\,cm$^{-2}$)& (nm)  	   &(erg\,s$^{-1}$\,cm$^{-2}$)& (nm)	      &(erg\,s$^{-1}$\,cm$^{-2}$)& (nm)		 \\
\hline
&&&&&&&&&&\\
\hline							  
\end{longtable}
\end{landscape}

\setlength{\tabcolsep}{5pt}

\begin{landscape}
\scriptsize
\begin{longtable}{l|c|r|c|r|c|r|c|r}
\caption[ ]{\label{tab:fluxes_EWs_H812} Extinction corrected fluxes and equivalent widths of Balmer: H8 to H11.}\\
\hline\hline
Object & $f_{\rm H8}$ &  $EW_{\rm H8}$&$f_{\rm H9}$& $EW_{\rm H9}$&$f_{\rm H10}$& $EW_{\rm H10}$&$f_{\rm H11}$& $EW_{\rm H11}$  \\
    & (erg\,s$^{-1}$\,cm$^{-2}$)&  (nm)	     &(erg\,s$^{-1}$\,cm$^{-2}$)& (nm)		&(erg\,s$^{-1}$\,cm$^{-2}$)& (nm)  	   &(erg\,s$^{-1}$\,cm$^{-2}$)& (nm)	      		 \\
\hline
&&&&&&&\\
\hline							  
\end{longtable}
\end{landscape}

\setlength{\tabcolsep}{5pt}

\begin{landscape}
\scriptsize
\begin{longtable}{l|c|r|c|r|c|r|c|r}
\caption[ ]{\label{tab:fluxes_EWs_H1315} Extinction corrected fluxes and equivalent widths of Balmer lines: H12 to H15.}\\
\hline\hline
Object & $f_{\rm H12}$ &  $EW_{\rm H12}$ & $f_{\rm H13}$ &  $EW_{\rm H13}$ &$f_{\rm H14}$& $EW_{\rm H14}$&$f_{\rm H15}$& $EW_{\rm H15}$                                \\
    & (erg\,s$^{-1}$\,cm$^{-2}$)&  (nm)	   & (erg\,s$^{-1}$\,cm$^{-2}$)&  (nm)	     &(erg\,s$^{-1}$\,cm$^{-2}$)& (nm)		&(erg\,s$^{-1}$\,cm$^{-2}$)& (nm)  	   \\
\hline
 & & & & & & & \\
\hline							  
\end{longtable}
\end{landscape}

\setlength{\tabcolsep}{5pt}

\begin{landscape}
\scriptsize
\begin{longtable}{l|c|c|c|c|c|c|c|c}
\caption[ ]{\label{tab:fluxes_EWs_PabPa8} Extinction corrected fluxes and equivalent widths of Paschen lines: Pa$\beta$ to Pa8.}\\
\hline\hline
Object & $f_{\rm Pa\beta}$ &  $EW_{\rm Pa\beta}$&$f_{\rm Pa\gamma}$& $EW_{\rm Pa\gamma}$&$f_{\rm Pa\delta}$& $EW_{\rm Pa\delta}$&$f_{\rm Pa8}$& $EW_{\rm Pa8}$  \\
    & (erg\,s$^{-1}$\,cm$^{-2}$)&  (nm)	     &(erg\,s$^{-1}$\,cm$^{-2}$)& (nm)		&(erg\,s$^{-1}$\,cm$^{-2}$)& (nm)  	   &(erg\,s$^{-1}$\,cm$^{-2}$) \\
\hline
 & & & & & & & \\
\hline			    			     	  
\end{longtable}
\end{landscape}

\setlength{\tabcolsep}{5pt}

\begin{landscape}
\scriptsize
\begin{longtable}{l|c|c|c|c|c|c|c|c}
\caption[ ]{\label{tab:fluxes_EWs_Pa9Br8} Extinction corrected fluxes and equivalent widths of Pa9, Pa10, Br$\gamma$, and Br8.}\\
\hline\hline
Object & $f_{\rm Pa9}$ &  $EW_{\rm Pa9}$   &  $f_{\rm Pa10}$ &  $EW_{\rm Pa10}$ & $f_{\rm Br\gamma}$& $EW_{\rm Br\gamma}$&$f_{\rm Br8}$& $EW_{\rm Br8}$\\
    & (erg\,s$^{-1}$\,cm$^{-2}$)&  (nm)	 & (erg\,s$^{-1}$\,cm$^{-2}$)&  (nm)	     &(erg\,s$^{-1}$\,cm$^{-2}$)& (nm)		&(erg\,s$^{-1}$\,cm$^{-2}$)& (nm) 	\\
\hline
 & & & & & & & \\
\hline			    			     	  
\end{longtable}
\end{landscape}

\setlength{\tabcolsep}{5pt}

\begin{landscape}
\scriptsize
\begin{longtable}{l|c|r|c|r|c|r|c|r|c|r}
\caption[ ]{\label{tab:fluxes_EWs_Helium1} Extinction corrected fluxes and equivalent widths of Helium lines.}\\
\hline\hline
Object &$f_{\rm \ion{He}{i}~\lambda402.6}$&$EW_{\rm \ion{He}{i}~\lambda402.6}$&$f_{\rm \ion{He}{i}~\lambda447.1}$&$EW_{\rm \ion{He}{i}~\lambda447.1}$&$f_{\rm \ion{He}{i}~\lambda471.3}$& $EW_{\rm \ion{He}{i}~\lambda471.3}$&$f_{\rm \ion{He}{i}~\lambda501.6}$&$EW_{\rm \ion{He}{i}~\lambda501.6}$&$f_{\rm \ion{He}{i}~\lambda587.6}$& $EW_{\rm \ion{He}{i}~\lambda587.6}$\\
   & (erg\,s$^{-1}$\,cm$^{-2}$)&  (nm)	     &(erg\,s$^{-1}$\,cm$^{-2}$)& (nm)		&(erg\,s$^{-1}$\,cm$^{-2}$)& (nm)  	   &(erg\,s$^{-1}$\,cm$^{-2}$)& (nm)	      &(erg\,s$^{-1}$\,cm$^{-2}$)& (nm)		 \\
\hline
&&&&&&&&&&\\
\hline							  
\end{longtable}
\end{landscape}

\setlength{\tabcolsep}{5pt}

\begin{landscape}
\scriptsize
\begin{longtable}{l|c|r|c|r|c|r|c|r|c|r}
\caption[ ]{\label{tab:fluxes_EWs_Helium2} Extinction corrected fluxes and equivalent widths of Helium lines.}\\
\hline\hline
Object &$f_{\rm \ion{He}{i}~\lambda667.8}$&$EW_{\rm \ion{He}{i}~\lambda667.8}$&$f_{\rm \ion{He}{i}~\lambda706.5}$&$EW_{\rm \ion{He}{i}~\lambda706.5}$&$f_{\rm \ion{He}{i}~\lambda1083.0}$& $EW_{\rm \ion{He}{i}~\lambda1083.0}$&$f_{\rm \ion{He}{i}+\ion{Fe}{i}~\lambda492.2}$&$EW_{\rm \ion{He}{i}+\ion{Fe}{i}~\lambda492.2}$&$f_{\rm \ion{He}{ii}~\lambda468.5}$& $EW_{\rm \ion{He}{ii}~\lambda468.5}$\\
   & (erg\,s$^{-1}$\,cm$^{-2}$)&  (nm)	     &(erg\,s$^{-1}$\,cm$^{-2}$)& (nm)		&(erg\,s$^{-1}$\,cm$^{-2}$)& (nm)  	   &(erg\,s$^{-1}$\,cm$^{-2}$)& (nm)	      &(erg\,s$^{-1}$\,cm$^{-2}$)& (nm)		 \\
\hline
&&&&&&&&&&\\
\hline							  
\end{longtable}
\end{landscape}

\setlength{\tabcolsep}{5pt}
\begin{landscape}
\scriptsize
\begin{longtable}{l|r|r|r|r|r|r|r|r|r|r}
\caption[ ]{\label{tab:fluxes_EWs_CaII} Extinction corrected fluxes and equivalent widths of \ion{Ca}{ii} H\&K and IRT lines.}\\
\hline\hline
Object & $f_{\rm \ion{Ca}{ii}~\lambda393.4}$ & $EW_{\rm \ion{Ca}{ii}~\lambda393.4}$&$f_{\rm \ion{Ca}{ii}~\lambda396.8}$& $EW_{\rm \ion{Ca}{ii}~\lambda396.8}$&$f_{\rm \ion{Ca}{ii}~\lambda849.8}$& $EW_{\rm \ion{Ca}{ii}~\lambda849.8}$&$f_{\rm \ion{Ca}{ii}~\lambda854.2}$& $EW_{\rm \ion{Ca}{ii}~\lambda854.2}$&$f_{\rm \ion{Ca}{ii}~\lambda866.2}$& $EW_{\rm \ion{Ca}{ii}~\lambda866.2}$\\
    & (erg\,s$^{-1}$\,cm$^{-2}$)&  (nm)	     &(erg\,s$^{-1}$\,cm$^{-2}$)& (nm)		&(erg\,s$^{-1}$\,cm$^{-2}$)& (nm)  	   &(erg\,s$^{-1}$\,cm$^{-2}$)& (nm)	      &(erg\,s$^{-1}$\,cm$^{-2}$)& (nm)		 \\
\hline
&&&&&&&&&&\\
\hline							  
\end{longtable}
\end{landscape}

\setlength{\tabcolsep}{5pt}

\begin{landscape}
\scriptsize
\begin{longtable}{l|c|r|c|r}
\caption[ ]{\label{tab:fluxes_EWs_NaI} Extinction corrected fluxes and equivalent widths of the $\ion{Na}{i}$~D lines }\\
\hline\hline
Object & $f_{\rm \ion{Na}{i}~\lambda588.99}$ &  $EW_{\rm \ion{Na}{i}~\lambda588.99}$ & $f_{\rm \ion{Na}{i}~\lambda589.59}$ &  $EW_{\rm \ion{Na}{i}~\lambda589.59}$ \\
    & (erg\,s$^{-1}$\,cm$^{-2}$)&  (nm)	   & (erg\,s$^{-1}$\,cm$^{-2}$)&  (nm)	      \\
\hline
 & & & &  \\
\hline							  
\end{longtable}
\end{landscape}


\begin{figure} 
\resizebox{1.0\hsize}{!}{\includegraphics[]{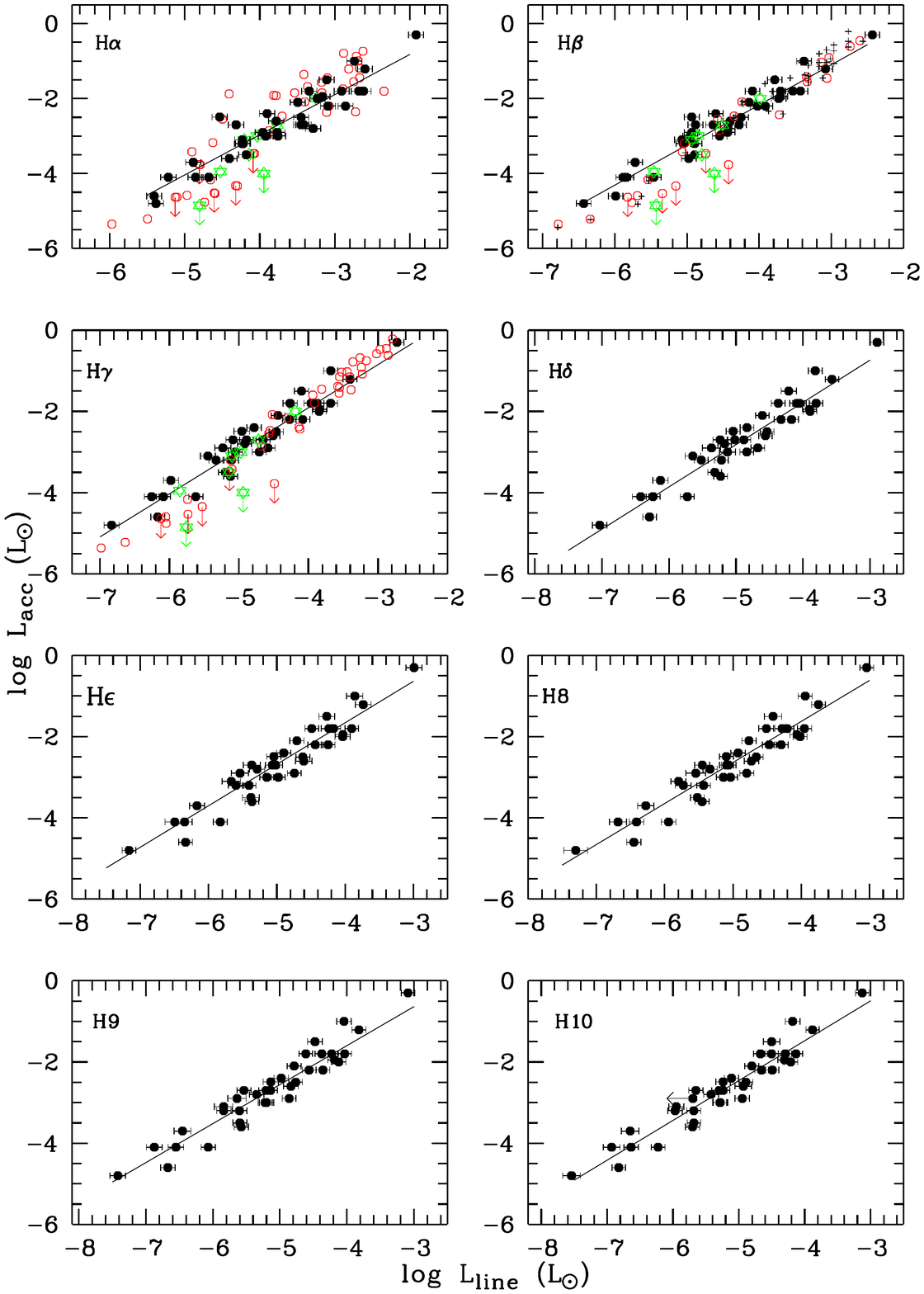}}
\caption{Relationships between accretion luminosity and line luminosity for 
     the several diagnostics as labelled in each panel. The Lupus YSOs are
      represented with black dots. The \Ha, \Hb ~and \Hg ~ data available in 
      literature for YSOs in Taurus \citep{HH08}, and the $\sigma$-Ori cluster 
      \citep{rigliaco12}, are overlaid with open circles and star symbols, 
      respectively.
    \label{correl1}}
\end{figure}

\begin{figure} 
\resizebox{1.0\hsize}{!}{\includegraphics[]{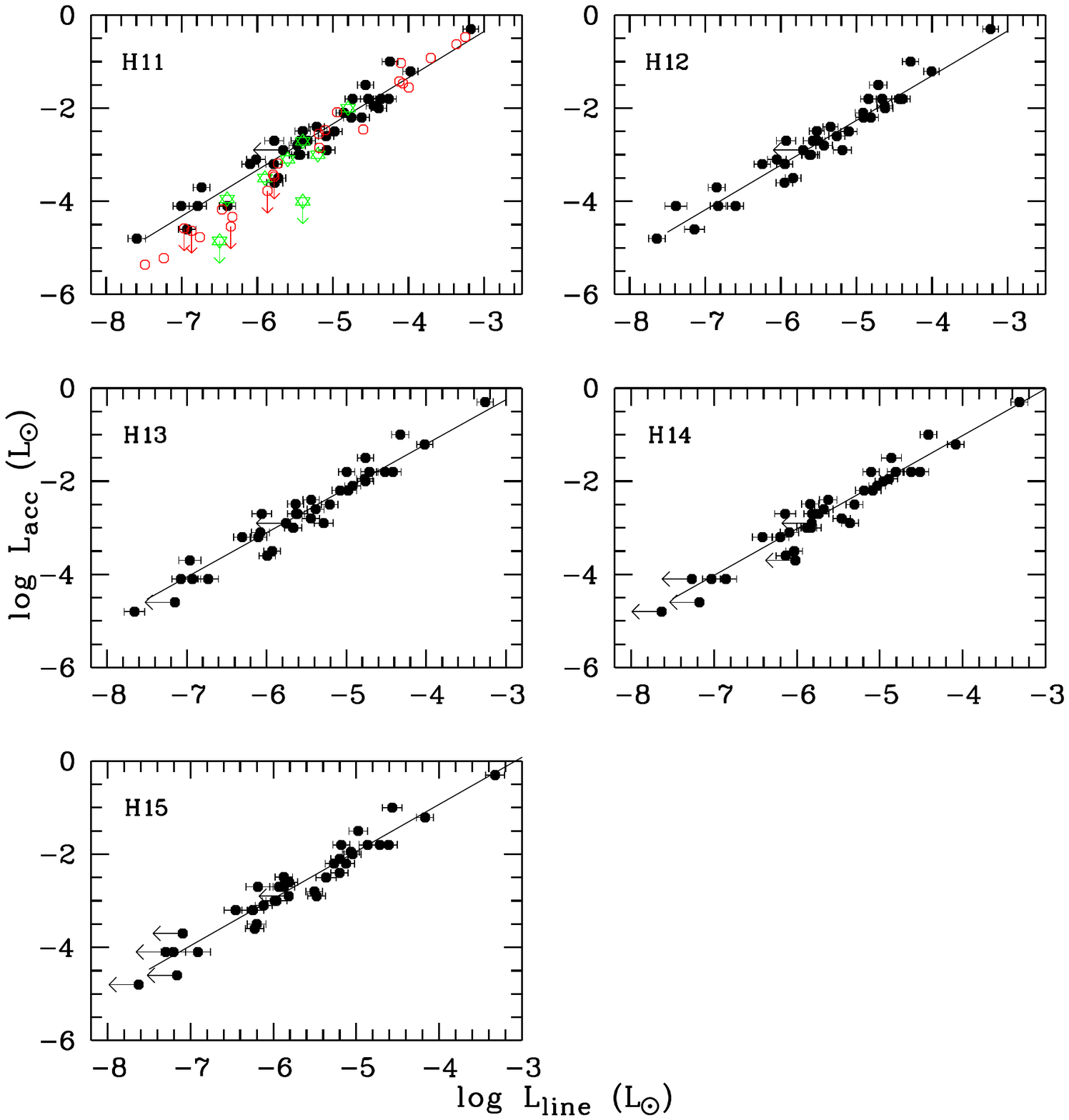}}
\caption{Relationships between accretion luminosity and line luminosity for 
     the several diagnostics as labelled in each panel. The Lupus YSOs are
      represented with black dots. The H11 data available in 
      literature for YSOs in Taurus \citep{HH08} and the $\sigma$-Ori cluster 
      \citep{rigliaco12}, are overlaid with open circles and star symbols, 
      respectively.
    \label{correl2}}
\end{figure}

\begin{figure} 
\resizebox{1.0\hsize}{!}{\includegraphics[]{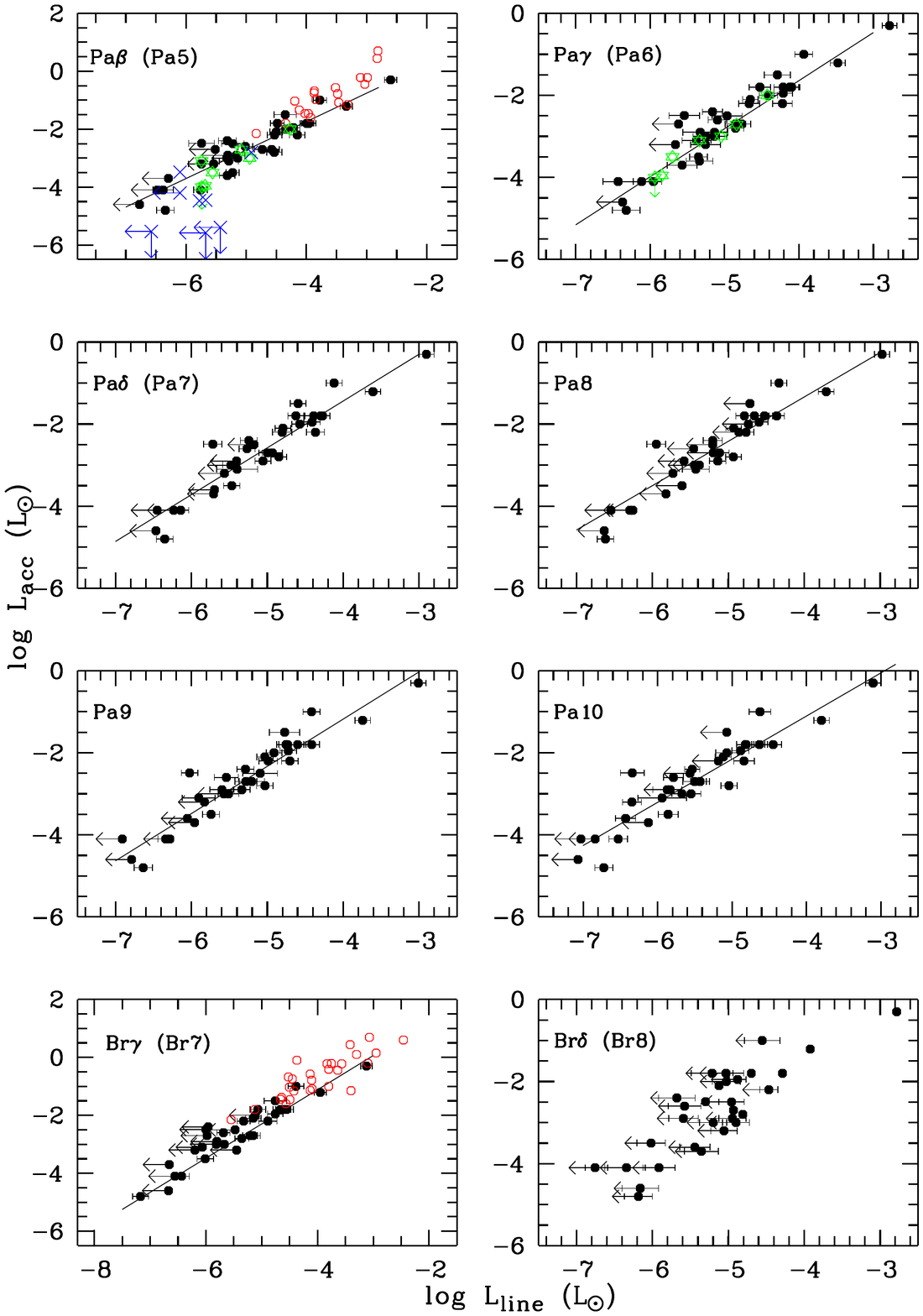}}
\caption{Relationships between accretion luminosity and line luminosity for 
      the several diagnostics as labelled in each panel. The Lupus YSOs are
      represented with black dots. The \pab, \pag ~and \brg ~data available in 
      literature for YSOs in Taurus \citep{muzerolle98, calvet00, calvet04},  
      $\rho$-Oph and Chamaeleon \citep{natta04} and the $\sigma$-Ori cluster 
      \citep{rigliaco12}, are overlaid with open circles, $\times$-symbols
       and star symbols, respectively.
    \label{correl3}}
\end{figure}

\begin{figure} 
\resizebox{1.0\hsize}{!}{\includegraphics[]{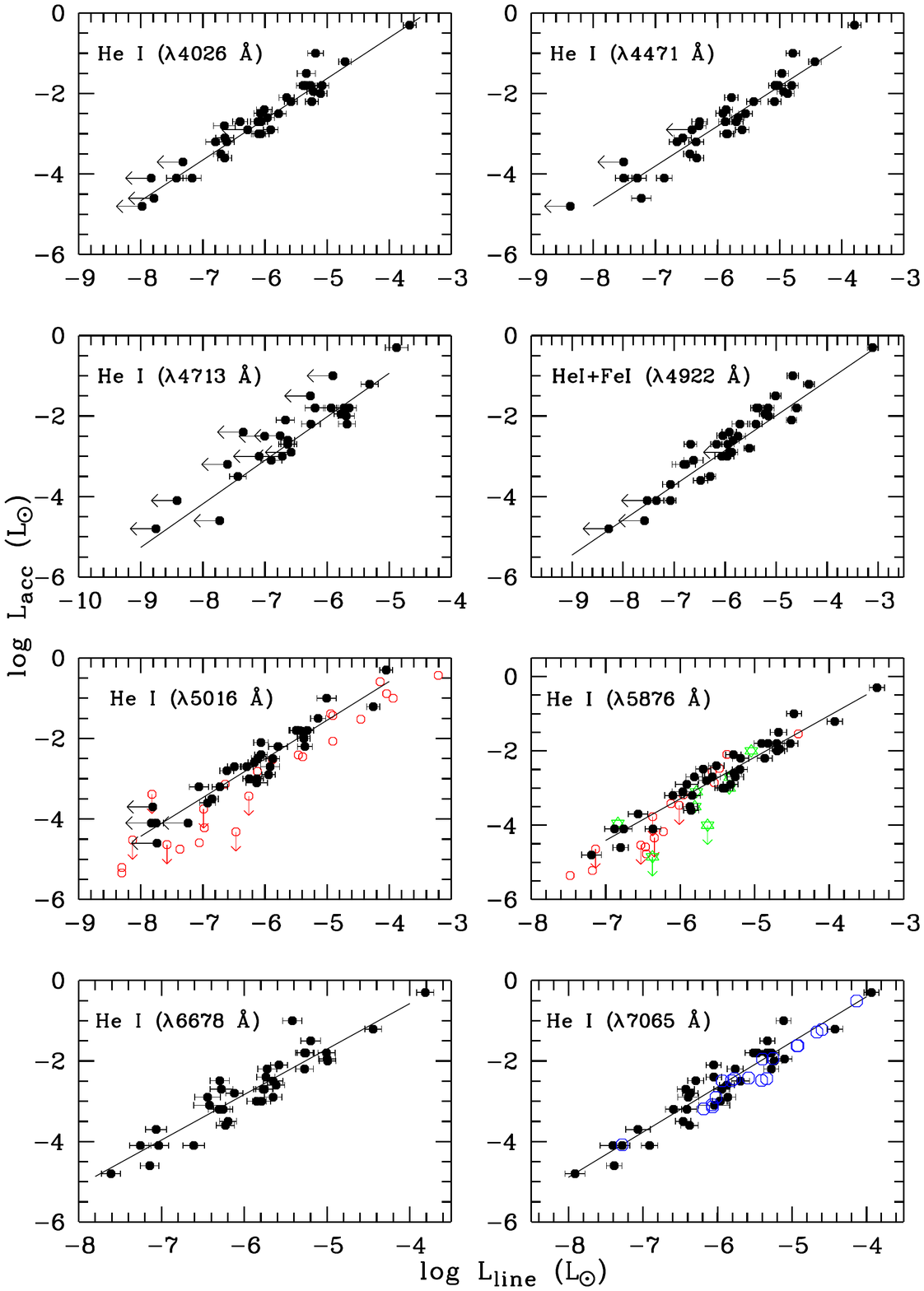}}
\caption{Relationships between accretion luminosity and line luminosity for 
      the several diagnostics as labelled in each panel. The Lupus YSOs are
      represented with black dots. The $\ion{He}{i}$ ($\lambda$5016\AA),
      $\ion{He}{i}$ ($\lambda$5876\AA), and $\ion{He}{i}$ ($\lambda$7065\AA)  
      ~data available in literature for YSOs in Taurus \citep{HH08},
      the $\sigma$-Ori cluster \citep{rigliaco12}, and the Cha-II 
      cloud \citep{biazzo12}, are overlaid with small open circles, 
      star symbols and big open circles, respectively.
    \label{correl4}}
\end{figure}


\begin{figure} 
\resizebox{1.0\hsize}{!}{\includegraphics[]{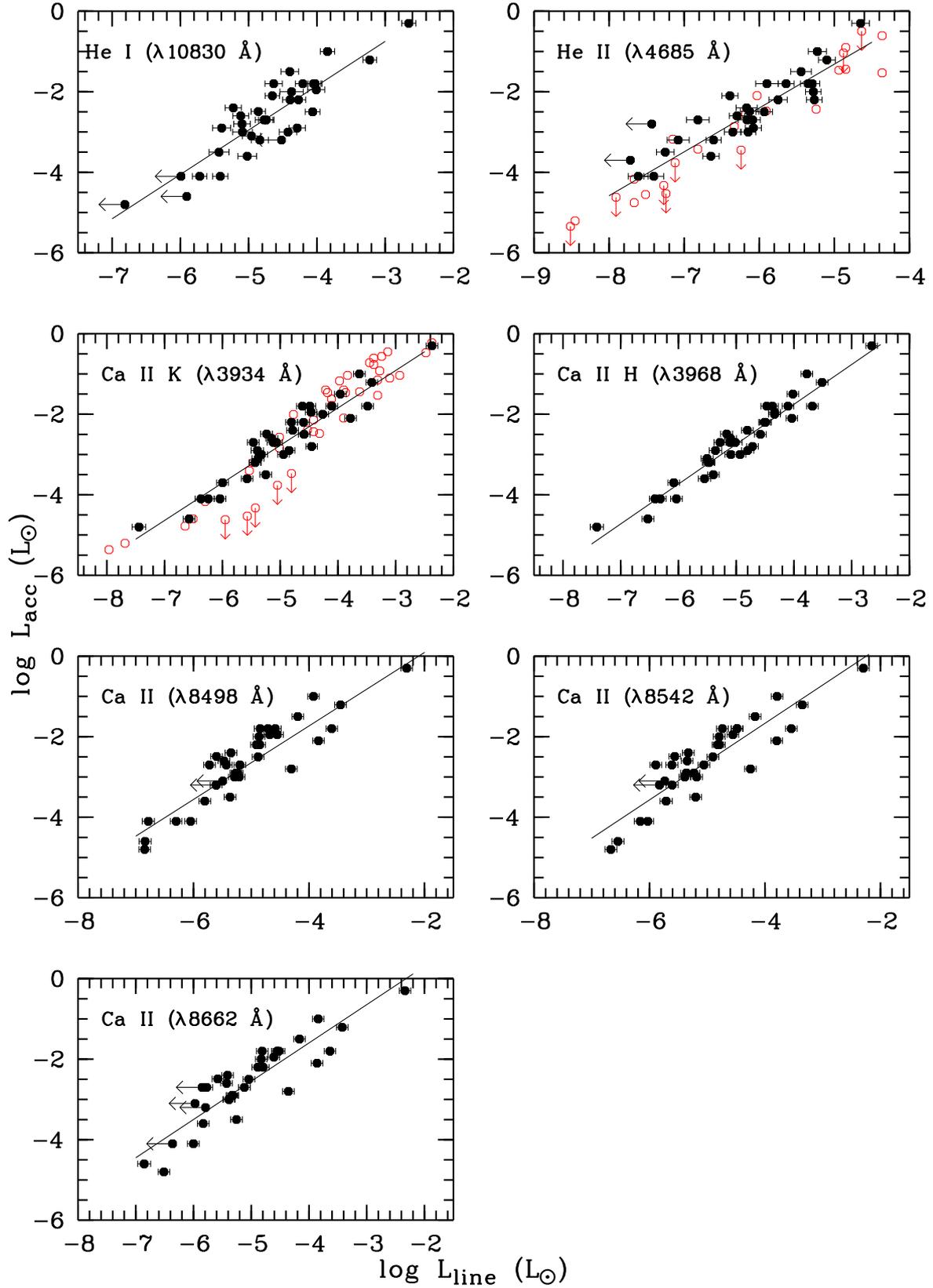}}
\caption{Relationships between accretion luminosity and line luminosity for 
     the several diagnostics as labelled in each panel. The Lupus YSOs are
      represented with black dots. The $\ion{He}{ii}$ ($\lambda$4685\AA) 
      and $\ion{Ca}{ii}$~K data available in literature for YSOs in 
      Taurus \citep{HH08} are overlaid with open circles.
    \label{correl5}}
\end{figure}


\begin{figure} 
\resizebox{1.0\hsize}{!}{\includegraphics[]{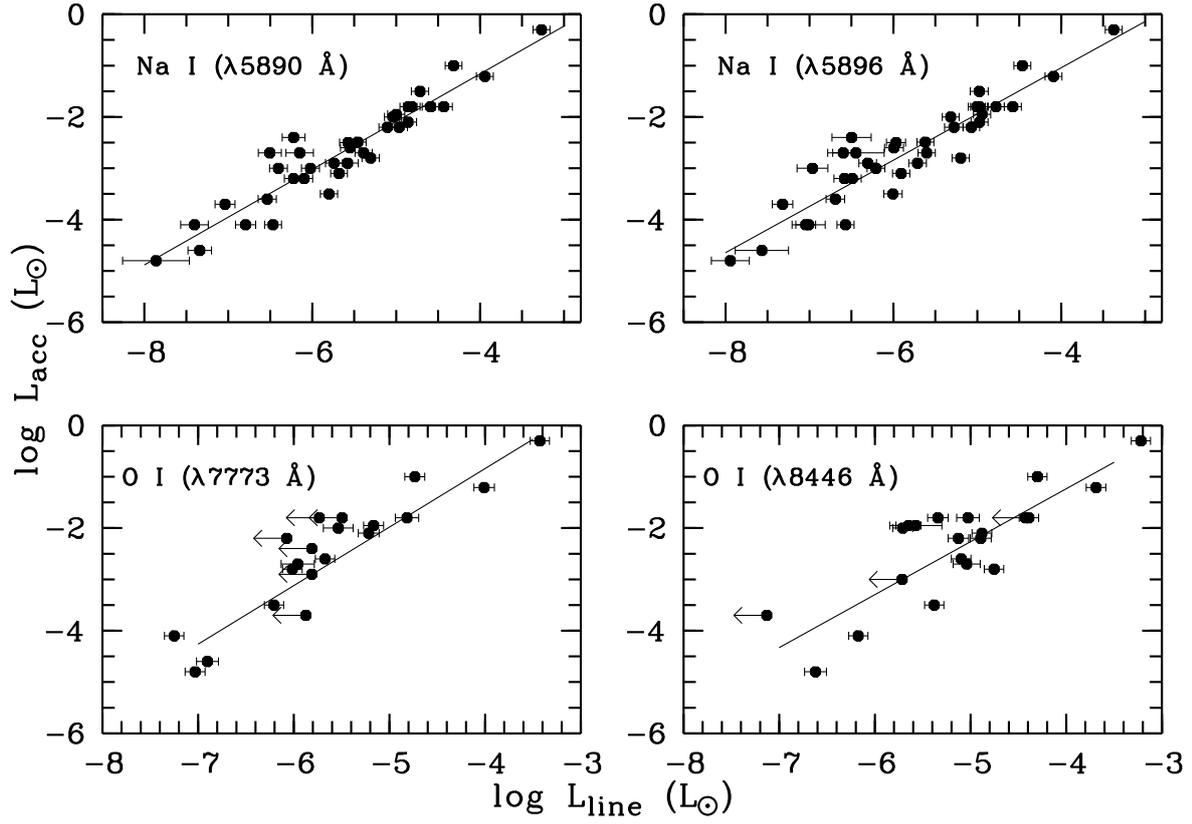}}
\caption{Relationships between accretion luminosity and line luminosity for 
     the several diagnostics as labelled in each panel. The Lupus YSOs are
      represented with black dots.
    \label{correl6}}
\end{figure}

\end{appendix}

\end{document}